\documentclass[showpacs,preprintnumbers,amsmath,amssymb,floatfix]{revtex4}

\usepackage{epsf}
 \newcommand{\insertplot}[5]{\begin{figure}
 \hfill\hbox to 0.05in{\vbox to #5in{\vfill
 \inputplot{#1}{#4}{#5}}\hfill}
 \hfill\vspace{-.1in}
 \caption{#2}\label{#3}
 \end{figure}}
 \newcommand{\inputplot}[3]{
 \special{ps: plotfile #1}
}

\usepackage[german, english]{babel}
\usepackage{ifthen}
\usepackage{epsfig}
\newcounter{fig}   \newcommand{\lbfig}[1]{\refstepcounter{fig}
\label{#1} }

\newcommand{\vphi}{\varphi}

\begin{document}
\title{Properties of Charged Rotating Electroweak Sphaleron-Antisphaleron Systems}
\author{
{\bf Rustam Ibadov}
}
\affiliation{Department of Theoretical Physics and Computer Science,\\
Samarkand State University, Samarkand, Uzbekistan}
\author{
{\bf Burkhard Kleihaus, Jutta Kunz and Michael Leissner}
}
\affiliation{
{Institut f\"ur Physik, Universit\"at Oldenburg, 
D-26111 Oldenburg, Germany}
}
\date{\today}

\begin{abstract}
We perform a systematic study of 
stationary sphaleron-antisphaleron systems
of Weinberg-Salam theory
at the physical value of the weak mixing angle.
These systems include rotating sphaleron-antisphaleron pairs,
chains and vortex rings.
We show that the angular momentum of these solutions
is proportional to their electric charge.
We study the dependence of their energy and magnetic moment
on their angular momentum.
We also investigate the influence of their
angular momentum on their local properties,
in particular on their energy density 
and on the node structure of their Higgs field configuration.
Furthermore, we discuss the equilibrium condition
for these solutions.
\end{abstract}

\maketitle



\section{Introduction}

It came as a surprise when 't Hooft \cite{'tHooft:1976up}
observed in 1976 that
because of the Adler-Bell-Jackiw anomaly
the standard model does not absolutely conserve
baryon and lepton number.
The process 't Hooft considered was
spontaneous fermion number violation due to instanton
induced transitions.
Later Ringwald \cite{Ringwald:1989ee} argued,
that such tunnelling transitions between
topologically distinct vacua might 
be observable at high energies at future accelerators.

The presence of baryon and lepton number
violating processes in the standard model
was considered by Manton \cite{Manton:1983nd}
from another point of view.
He investigated the topological structure
of the configuration space of Weinberg-Salam theory
and found the existence of noncontractible loops.
From these he predicted the existence of a static, unstable solution
of the bosonic field equations,
representing the top of the energy barrier between
topologically distinct vacua.
Because of its instability
this classical electroweak solution was termed sphaleron by Klinkhamer
and Manton \cite{Klinkhamer:1984di}.

At finite temperature the energy barrier between
topologically distinct vacua can be overcome
due to thermal fluctuations of the fields,
and baryon number violating
vacuum to vacuum transitions
involving changes of baryon and lepton number
can occur.
The rate for such baryon number violating processes
is largely determined by a Boltzmann factor,
containing the height of the barrier at a given
temperature and thus the energy of the sphaleron
\cite{McLerran:1993rv,Rubakov:1996vz,Klinkhamer:2003hz,Dine:2003ax}.
Entailing baryon number violating processes,
the sphaleron itself carries baryon number $Q_{\rm B}=1/2$ 
\cite{Klinkhamer:1984di}.

The energy of the sphaleron increases with increasing Higgs mass 
and ranges roughly between 7 and 13 TeV 
\cite{Klinkhamer:1984di,Dashen:1974ck,Boguta:1983xs}.
The energy is that high,
since its scale is not set by $M_W$ but by $M_W/\alpha_{\rm w}$.
For the physical value of the weak mixing angle the sphaleron energy 
is only slightly decreased as compared 
to vanishing mixing angle \cite{Kleihaus:1991ks,Kunz:1992uh}.
However, the configuration is no longer spherically symmetric,
and retains only axial symmetry.
At the same time, one finds a large value for the
magnetic dipole moment of the sphaleron
$\mu_{\rm S} \approx 1.8 e/(\alpha_{\rm w} M_W) $ 
\cite{Klinkhamer:1984di,Kleihaus:1991ks,Kunz:1992uh}.

Whereas the static electroweak sphaleron 
does not carry electric charge,
it was argued before \cite{Saffin:1997ae}
and demonstrated recently in nonperturbative studies
\cite{Radu:2008ta,Kleihaus:2008cv},
that the addition of electric charge 
leads to a non-vanishing Poynting vector 
and thus a finite angular momentum density of the system. 
Consequently,
a branch of electrically charged sphalerons arises,
that carry at the same time angular momentum.
In particular, their angular momentum and charge are proportional.
Since these charged sphalerons
carry non-vanishing baryon number as well,
they can also entail baryon number violating processes.

Beside the sphaleron,
the non-trivial topology of the configuration space of Weinberg-Salam theory
gives rise to further unstable classical solutions.
A superposition of $n$ sphalerons, for instance,
can lead to static axially symmetric solutions, multisphalerons, 
which carry baryon number $Q_{\rm B}=n/2$ and
whose energy density is torus-like 
\cite{Brihaye:1994ib,Kleihaus:1994yj,Kleihaus:1994tr}.
A superposition of a sphaleron and an antisphaleron,
on the other hand, can give rise to a bound
sphaleron-antisphaleron system, in which a sphaleron and an
antisphaleron are located at an equilibrium distance
on the symmetry axis 
\cite{Klinkhamer:1985ki,Klinkhamer:1990ik,Klinkhamer:1993hb}.
Such a sphaleron-antisphaleron pair has vanishing baryon number,
$Q_{\rm B}=0$, since the antisphaleron carries $Q_{\rm B}=-1/2$.
The sphaleron-antisphaleron pair therefore does not mediate
baryon number violating processes.

Recently, 
the sphaleron-antisphaleron pair solutions have been generalized,
leading to sphaleron-antisphaleron chains,
where $m$ sphalerons and antisphalerons are located
on the symmetry axis in static equilibrium
\cite{Kleihaus:2008gn},
in close analogy to the monopole-antimonopole chains
encountered in the Georgi-Glashow model 
\cite{Kleihaus:2003nj}.
When systems of multisphalerons and -antisphalerons
are considered, instead of the anticipated pairs and chains
a new type of solutions arises,
when $n \ge 3$ \cite{Kleihaus:2008gn}.
In these vortex ring solutions 
the Higgs field vanishes not (only) on isolated points
on the symmetry axis but (also) on one or more rings,
centered around the symmetry axis 
\cite{Kleihaus:2003xz,Kleihaus:2004is}.

In this paper we perform a systematic study of multisphalerons
and sphaleron-antisphaleron systems
endowed with electric charge 
\cite{Ibadov:2010ei}.
As for the simple sphaleron,
the non-vanishing Poynting vector leads to a finite angular momentum
density for these configurations.
Thus branches of rotating electrically charged sphaleron-antisphaleron systems
emerge from the respective static electrically neutral
configurations.
We construct these solutions explicitly 
for $m,n \le 6$ and discuss their properties.
We demonstrate that
the angular momentum and the electric charge of the solutions
are proportional 
\cite{VanderBij:2001nm,Radu:2008ta,Kleihaus:2008gn,Ibadov:2010ei}.

In section 2 we present the action, the Ansatz for
the stationary axially symmetric configurations, and the
boundary conditions. 
We then consider the
relevant physical properties and, in particular,
derive the linear relation
between angular momentum and electric charge.
We present and discuss the numerical results in section 3.
These include global properties of the solutions,
such as their energy, their angular momentum, their charge
and their magnetic moments,
but also local properties, such as their
energy density, their angular momentum density
and the modulus of their Higgs field.
Moreover, we discuss the equilibrium condition
for these solutions. 
We give our conclusions in section 4.

\section{Action, Ansatz and Properties}

\subsection{Weinberg-Salam Lagrangian}

We consider the bosonic sector of Weinberg-Salam theory
\begin{equation}
{\cal L} = -\frac{1}{2} {\rm Tr} (F_{\mu\nu} F^{\mu\nu})
-  \frac{1}{4}f_{\mu \nu} f^{\mu \nu}                                           
- (D_\mu \Phi)^{\dagger} (D^\mu \Phi) 
- \lambda (\Phi^{\dagger} \Phi - \frac{v^2}{2} )^2 
\  
\label{lag1}
\end{equation}
with su(2) field strength tensor
\begin{equation}
F_{\mu\nu}=\partial_\mu V_\nu-\partial_\nu V_\mu
            + i g [V_\mu , V_\nu ]
 , \end{equation}
su(2) gauge potential $V_\mu = V_\mu^a \tau_a/2$,
u(1) field strength tensor
\begin{equation}
f_{\mu\nu}=\partial_\mu A_\nu-\partial_\nu A_\mu 
 , \end{equation}
and covariant derivative of the Higgs field
\begin{equation}
D_{\mu} \Phi = \Bigl(\partial_{\mu}
             +i g  V_{\mu} 
             +i \frac{g'}{2} A_{\mu} \Bigr)\Phi
 , \end{equation}
where $g$ and $g'$ denote the $SU(2)$ and $U(1)$ gauge coupling constants,
respectively,
$\lambda$ denotes the strength of the Higgs self-interaction and
$v$ the norm of the vacuum expectation value of the Higgs field.

The gauge symmetry is spontaneously broken 
due to the non-vanishing vacuum expectation
value of the Higgs field
\begin{equation}
    \langle \Phi \rangle = \frac{v}{\sqrt2}
    \left( \begin{array}{c} 0\\1  \end{array} \right)   
 , \label{Higgs} \end{equation}
leading to the boson masses
\begin{equation}
    M_W = \frac{1}{2} g v   , \ \ \ \ 
    M_Z = \frac{1}{2} \sqrt{(g^2+g'^2)} v  \, , \ \ \ \ 
    M_H = v \sqrt{2 \lambda} \,  . 
\end{equation}
$ \tan \theta_{\rm w} = g'/g $ determines
the weak mixing angle $\theta_{\rm w}$,
defining the electric charge $e = g \sin \theta_{\rm w}$.  
We also denote the weak fine structure constant $\alpha_{\rm W}=g^2/4\pi$.

\subsection{Stationary axially symmetric Ansatz} 

To obtain stationary rotating solutions of the bosonic sector
of Weinberg-Salam theory,
we employ the time-independent axially symmetric Ansatz
\begin{equation}
V_\mu\, dx^\mu
  = \left( B_1\, \frac{\tau^{(n,m)}_r}{2g} 
         + B_2\, \frac{\tau^{(n,m)}_\theta}{2g} \right) \, dt 
            -n\sin\theta\left(H_3 \frac{\tau^{(n,m)}_r}{2g}
            + H_4 \frac{\tau^{(n,m)}_\theta}{2g}\right)\, d\varphi
+\left(\frac{H_1}{r}\, dr +(1-H_2)\, d\theta \right)
  \frac{\tau^{(n)}_\varphi}{2g}
  , \label{a1} \end{equation}
\begin{equation}
A_\mu\, dx^\mu = \left( a_1\, dt + a_2\, \sin^2 \theta \, d\varphi \right)/g'
  , \end{equation}
and
\begin{equation}
\Phi = i\, 
      \left( \phi_1 \, \tau^{(n,m)}_r 
           + \phi_2  \tau^{(n,m)}_\theta \right)
    \frac{v}{\sqrt2} \left( \begin{array}{c} 0\\1  \end{array} \right)
  , \end{equation}
where
\begin{eqnarray}          
\tau^{(n,m)}_r & = & \sin m\theta (\cos n\vphi \tau_x + \sin n\vphi \tau_y) 
           + \cos m\theta \tau_z \ , \ \ 
\nonumber \\       
\tau^{(n,m)}_\theta & = & \cos m\theta (\cos n\vphi \tau_x + \sin n\vphi \tau_y) 
           - \sin m\theta \tau_z \ , \ \ 
\nonumber \\       
\tau^{(n)}_\vphi & = & (-\sin n\vphi \tau_x + \cos n\vphi \tau_y) 
\ , \ \ \nonumber 
\end{eqnarray}
$n$ and $m$ are integers,
and $\tau_x$, $\tau_y$ and $\tau_z$ denote the Pauli matrices.

The two integers $n$ and $m$ determine the type of configuration,
that is put into rotation.
For $n=m=1$ the solutions correspond to rotating sphalerons.
Rotating multisphaleron configurations arise for $n>1$ and $m=1$.
For $n=1$ and $m>1$ rotating sphaleron-antisphaleron
pairs ($m=2$) or sphaleron-antisphaleron chains arise,
and for $n \ge 3$ rotating vortex ring solutions are obtained.

The ten functions $B_1$, $B_2$, $H_1,\dots,H_4$, $a_1$, $a_2$,
$\phi_1$, and $\phi_2$ depend on $r$ and $\theta$, only. 
With this Ansatz the full set of field equations reduces to a system 
of ten coupled partial differential equations in the independent variables 
$r$ and $\theta$. A residual $U(1)$ gauge degree of freedom is 
fixed by the condition $r\partial_r H_1 - \partial_\theta H_2=0$ 
\cite{Kleihaus:1991ks}.

\subsection{Boundary conditions}

Requiring regularity and finite energy, we impose 
for odd $m$ configurations the boundary conditions 
\begin{eqnarray}
r=0: &  & 
B_1 \sin m \theta + B_2 \cos m \theta =0  , \
\partial_r \left( B_1 \cos m \theta - B_2 \sin m \theta\right) =0  , \
H_1=H_3=H_4=0 , \ H_2=1 , \ 
\nonumber \\
&  & 
\partial_r a_1 =0  , \ a_2=0  , \
\phi_1=0 , \ \phi_2 = 0
\nonumber \\
r\rightarrow \infty: &  &  
B_1 = \gamma \cos m \theta  , \ B_2 = \gamma \sin m \theta  , \
H_1=H_3=0 , \ H_2=1-2m  ,  \ H_4=\frac{2\sin m\theta}{\sin\theta} , \ 
\nonumber \\[-0.7ex]
& & 
a_1=\gamma  , \ a_2 = 0  , \
\phi_1=1 , \ \phi_2 = 0 , \ 
 {\rm where } \ \gamma = const.
\nonumber \\[0.5ex]
\theta = 0: &  & 
\partial_\theta B_1 =0  , \ B_2 =0  , \
H_1=H_3=0 , \ \partial_\theta H_2=\partial_\theta H_4=0 , \ 
\partial_\theta a_1 = \partial_\theta a_2 = 0  , \
\partial_\theta \phi_1 = 0  , \ \phi_2=0  ,
\end{eqnarray}
where the latter hold also at $\theta=\pi/2$,
except for $B_1=0$ and $\partial_\theta B_2 =0$.
For even $m$ configurations the same set of boundary conditions
holds except for
\begin{eqnarray}
r=0: &  & 
\phi_1 \sin m \theta + \phi_2 \cos m \theta =0  , \
\partial_r \left( \phi_1 \cos m \theta - \phi_2 \sin m \theta\right) =0  
\nonumber \\
\theta=\pi/2: & &
\partial_\theta B_1=0 , \ B_2 = 0 \ , \partial_\theta H_3 = 0 , \ H_4=0 .
\end{eqnarray}


\subsection{Mass, angular momentum and charge}

We now address the global charges of the sphaleron-antisphaleron systems,
their energy, their angular momentum, their electric charge, 
and their baryon number.
The energy $E$ and angular momentum $J$ are defined in terms of
volume integrals of the respective components
of the energy-momentum tensor. The mass is obtained from
\begin{equation}
E =- \int T_t^t d^3 r , 
\label{Mint}
\end{equation}
while the angular momentum
\begin{equation}
J = \int T_\varphi^t d^3 r 
= \int \left[ 2 {\rm Tr} \left( F^{t\mu} F_{\varphi\mu} \right)
 + f^{t\mu} f_{\varphi\mu} + 2 \left( D^t \Phi \right)^\dagger
 \left( D_\varphi \Phi \right)   \right] d^3 r
\label{Jint}
\end{equation}
can be reexpressed with help of the
equations of motion and the symmetry properties of the Ansatz
\cite{VanderBij:2001nm,Kleihaus:2002tc,Volkov:2003ew,Radu:2008pp}
as a surface integral at spatial infinity 
\begin{equation}
J =  \int_{S_2} \left\{ 
 2 {\rm Tr} \left( \left(V_\varphi - \frac{n\tau_z}{2g}\right) F^{r t} \right) 
 + 
 \left(A_\varphi - \frac{n}{g'}\right) f^{r t}
 \right\} r^2 \sin \theta d \theta d \varphi .
\label{Joint}
\end{equation}
The power law fall-off of the $U(1)$ field
of a charged solution allows for a finite flux
integral at infinity and thus a finite angular momentum.
Insertion of the asymptotic expansion for the $U(1)$ field
\begin{eqnarray}
a_1= \gamma - \frac{\chi}{r} + O \left(\frac{1}{r^2}\right) , \nonumber\\
a_2= \frac{\zeta}{r} + O \left(\frac{1}{r^2}\right) ,
\label{asymp}
\end{eqnarray}
and of the analogous expansion for the $SU(2)$ fields
then yields for the angular momentum
\begin{equation}
\frac{J}{4 \pi} = 
\frac{n\chi}{g^2} + \frac{n\chi}{{g'}^2}
= \frac{n\chi}{g^2 \sin^2 \theta_{\rm w}} = \frac{n\chi}{e^2} .
\label{Joint2}
\end{equation}

The field strength tensor ${\cal F}_{\mu\nu}$ of the 
electromagnetic field ${\cal A}_{\mu}$,
\begin{equation}
{\cal A}_{\mu}=\sin \theta_{\rm w} V^3_\mu + \cos \theta_{\rm w} A_\mu ,
\label{emfield}
\end{equation}
as given in a gauge where the Higgs field asymptotically tends to
Eq.~(\ref{Higgs}),
then defines the electric charge $\cal Q$
\begin{equation}
{\cal Q} =  \int_{S_2}
  {^*}{\cal F}_{\theta\varphi} d\theta d\varphi
 =4\pi\left\{ 
 \frac{\sin \theta_{\rm w} \chi}{g} + \frac{\cos \theta_{\rm w} \chi}{g'}
 \right\}
 = 4\pi\frac{\chi}{e}
  , \label{Qel} \end{equation}
where the integral is evaluated at spatial infinity.
%
Comparison of Eqs.~(\ref{Joint2}) and (\ref{Qel}) then
yields a linear relation between the angular momentum $J$
and the electric charge $\cal Q$ \cite{Radu:2008ta,Kleihaus:2008cv}
\begin{equation}
 J = \frac{n \cal Q}{e}
\label{JQrel}
 . \end{equation}
This relation 
corresponds to the relation for
monopole-antimonopole systems without magnetic charge 
\cite{Kleihaus:2007vf}.
The magnetic moment $\mu$ is obtained from 
the asymptotic expansion Eq.~(\ref{asymp}), analogously to the 
electric charge,
\begin{equation}
\mu= \frac{4\pi\zeta}{e} \ .
\end{equation}

\subsection{Baryon number}

Addressing finally the baryon number $Q_{\rm B}$,
its rate of change is given by
\begin{equation}
 \frac{d Q_{\rm B}}{dt} = \int d^3 r \partial_t j^0_{\rm B}
= \int d^3 r \left[ \vec \nabla \cdot \vec j_{\rm B}
 + \frac{1}{32 \pi^2} \epsilon^{\mu\nu\rho\sigma} \, \left\{
 g^2 {\rm Tr} \left(F_{\mu\nu} F_{\rho\sigma} \right) 
 + \frac{1}{2} {g'}^2 f_{\mu\nu} f_{\rho\sigma} \right\} \right]  . 
\end{equation}
Starting at time $t=-\infty$ at the vacuum with $Q_{\rm B}=0$,
one obtains the baryon number of a sphaleron solution at
time $t=t_0$ \cite{Klinkhamer:1984di},
\begin{equation}
 Q_{\rm B} = 
\int_{-\infty}^{t_0} dt \int_S \vec K \cdot d \vec S
+  \int_{t=t_0} d^3r K^0 
  , \end{equation}
where the $\vec \nabla \cdot \vec j_{\rm B}$ term is neglected,
and the anomaly term is reexpressed in terms of the
Chern-Simons current
\begin{equation}
 K^\mu=\frac{1}{16\pi^2}\varepsilon^{\mu\nu\rho\sigma} 
 \left\{ g^2 {\rm Tr}\left( F_{\nu\rho}V_\sigma
 - \frac{2}{3} i g V_\nu V_\rho V_\sigma \right)
 + \frac{1}{2} {g'}^2 f_{\nu\rho}A_\sigma \right\}
  . \end{equation}
In a gauge, where
\begin{equation}
V_\mu \to \frac{i}{g} \partial_\mu \hat{U} \hat{U}^\dagger   , \ \ \ 
\hat{U}(\infty) = 1   , 
\end{equation}
$\vec K$ vanishes at infinity. 
Subject to the above Ansatz and boundary conditions
the baryon charge
of the sphaleron solution \cite{Kleihaus:1994tr,Kleihaus:2003tn} is then
\begin{equation}
Q_{\rm B}= \int_{t=t_0} d^3r K^0  = \frac{n \ (1-(-1)^m)}{4} \, .
\label{Q_B}
\end{equation}

\section{Results and Discussion}

\subsection{Numerical technique}

We have solved the set of ten coupled non-linear
elliptic partial differential equations numerically 
subject to the above boundary conditions.
We have employed the compactified dimensionless coordinate
\begin{equation}
x = \tilde r/(1+ \tilde r) \ , \ \ \
\tilde r = gvr \ ,
\label{xbar}
\end{equation}
instead of $\tilde r$,
to map spatial infinity to the finite value ${x}=1$.

The numerical calculations are performed with help of the program
FIDISOL \cite{FIDISOL}.
The equations are discretized on a non-equidistant
grid in ${x}$ and  $\theta$.
Typical grids used have sizes in the range $100 \times 60$ to
$120 \times 80$,
covering the integration region
$0\leq {x}\leq 1$ and $0\leq\theta\leq\pi/2$.

The numerical method is based on the Newton-Raphson
method, an iterative procedure to find a good approximation to
the exact solution.
The iteration stops when the Newton residual
is smaller than a prescribed tolerance.
Thus it is essential to have a good
first guess, to start the iteration procedure.
Our strategy therefore is to use a known solution as guess
and then vary some parameter to obtain the next solution.

Restricting to $M_H=M_W$,
and employing the physical value for the mixing angle $\theta_{\rm w}$,
we have performed a systematic study
of the rotating sphaleron-antisphaleron systems
with $1 \le m \le 6$ and $1 \le n \le 6$.
We have also obtained samples of solutions for $M_H=2 M_W$,
showing that the basic features of these solutions
do not depend on the particular value of the Higgs mass 
(in this range of masses).

Starting from a given static neutral solution
for a sphaleron-antisphaleron system characterized by
the integers $n$ and $m$,
we have constructed the corresponding branch of
rotating solutions, by slowly increasing
the value of the parameter
$\tilde \gamma = \gamma/gv$,
which specifies the boundary conditions for the
time components of the gauge fields.

The rotating branch ends 
when the limiting value $\tilde \gamma_{\rm max}=1/2$
is reached. 
In the asymptotic expansion, the exponential decay is determined by a decay constant proportional to $\sqrt{1-4\tilde \gamma^2}$.
Beyond $\tilde\gamma_{\rm max}$ some of the gauge field functions
would no longer decay exponentially,
precluding localized solutions 
for larger values of $\tilde \gamma$.
Consequently, at $\tilde \gamma_{\rm max}$ the 
respective solution has maximal angular momentum, maximal charge and 
maximal energy.

We have used the linear relation
(\ref{JQrel}) between the charge $\cal Q$ and
the angular momentum $J$ as
a check of the accuracy of the solutions.
According to this relation,
we should obtain a single straight line,
when exhibiting the charge
versus the scaled angular momentum $J/n$.

We demonstrate this in Fig.~\ref{f-1} for 
the sets of solutions with $m=2$ and $m=6$
by exhibiting the charge parameter $\chi$
(which is proportional to the charge $\cal Q$)
versus the scaled angular momentum $J/n$.
We indeed observe a single straight line,
which extends the further the greater $n$.
Since the charge parameter has been extracted from the
asymptotic fall-off of the $U(1)$ function $a_1$,
whereas the angular momentum has been obtained from
the volume integral of the angular momentum density $T^t_\varphi$,
this agreement reflects the good numerical quality of the
solutions.

\begin{figure}[h!]
\lbfig{f-1}
\begin{center}
\hspace{0.0cm} (a)\hspace{-0.6cm}
\includegraphics[height=.25\textheight, angle =0]{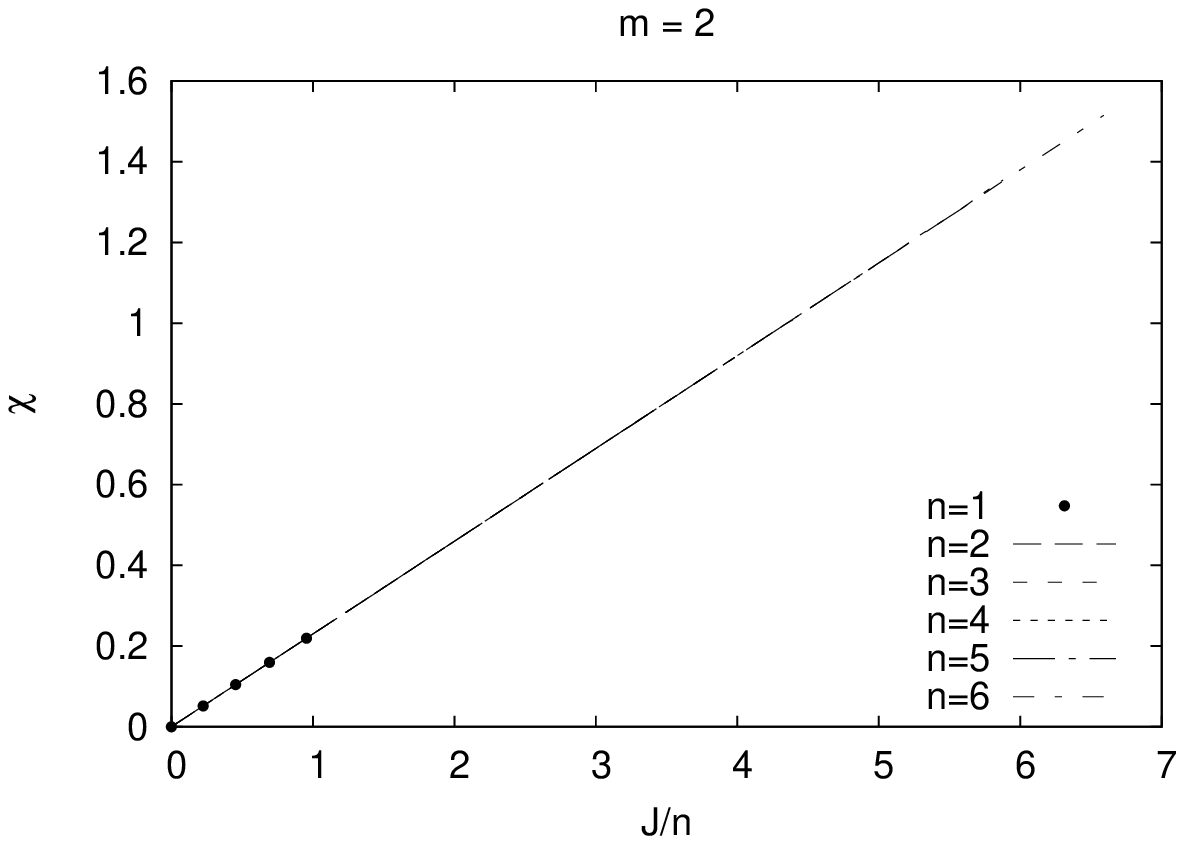}
\hspace{0.5cm} (b)\hspace{-0.6cm}
\includegraphics[height=.25\textheight, angle =0]{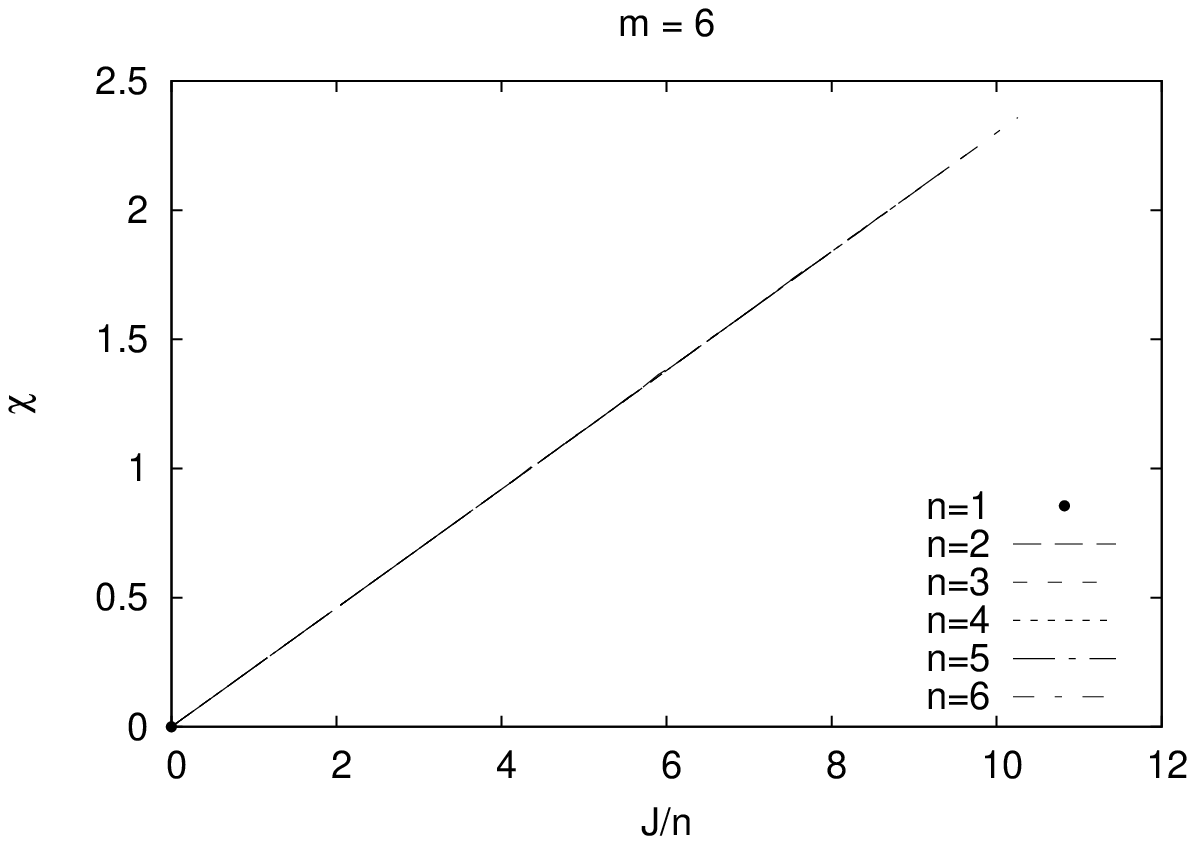}
\end{center}
\caption{
Quality of the solution sets:
the asymptotic value of the $U(1)$ field $\tilde \gamma = \gamma/gv$
versus the scaled angular momentum $J/n$
(in units of $J_0=4 \pi/g^2$)
(a) $m=2$, $n=1,\dots,6$,
(b) $m=6$, $n=1,\dots,6$.
}
\end{figure}

\subsection{Global properties}

We first address the domain of existence of these solutions.
For that purpose we exhibit in Fig.~\ref{f-2}
the asymptotic gauge field parameter
$\tilde \gamma = \gamma/gv$ versus the
scaled angular momentum of the solutions $J/n$,
which corresponds to the charge ${\cal Q}/e$
(choosing units of $J_0=4 \pi/g^2$).
As $\tilde \gamma$ increases from zero to
its maximal value $\tilde \gamma_{\rm max}=1/2$,
the angular momentum increases monotonically.
Consequently the solutions have maximal angular momentum
at $\tilde \gamma_{\rm max}$.

\begin{figure}[p!]
\lbfig{f-2}
\begin{center}
\hspace{0.0cm} (a)\hspace{-0.6cm}
\includegraphics[height=.25\textheight, angle =0]{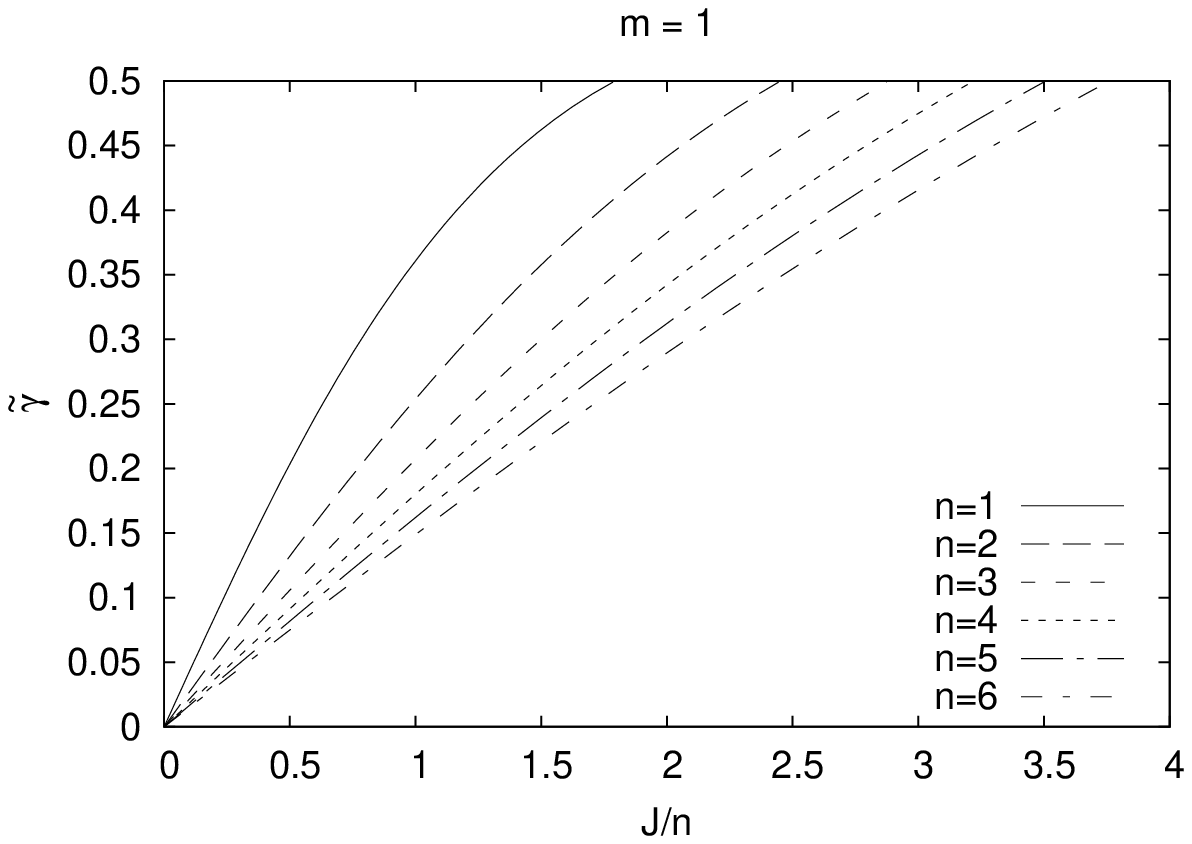}
\hspace{0.5cm} (b)\hspace{-0.6cm}
\includegraphics[height=.25\textheight, angle =0]{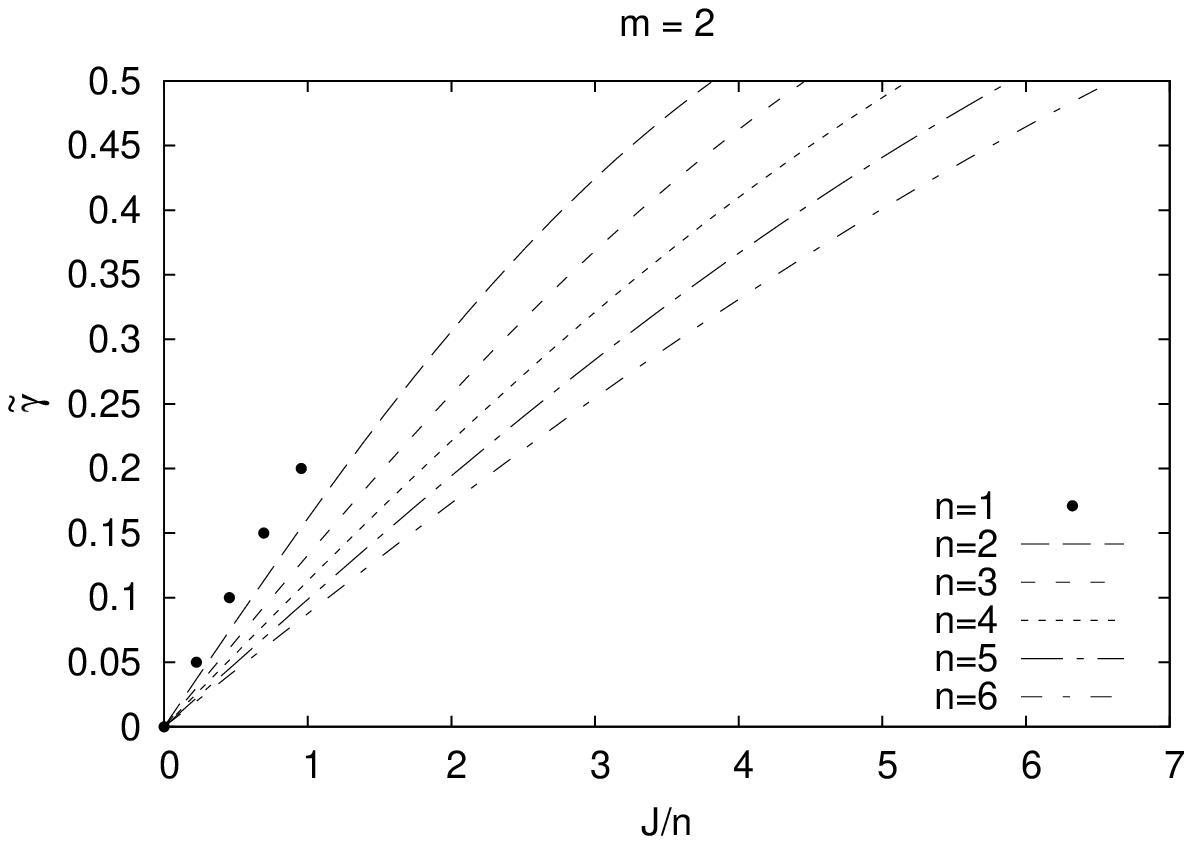}
\\
\hspace{0.0cm} (c)\hspace{-0.6cm}
\includegraphics[height=.25\textheight, angle =0]{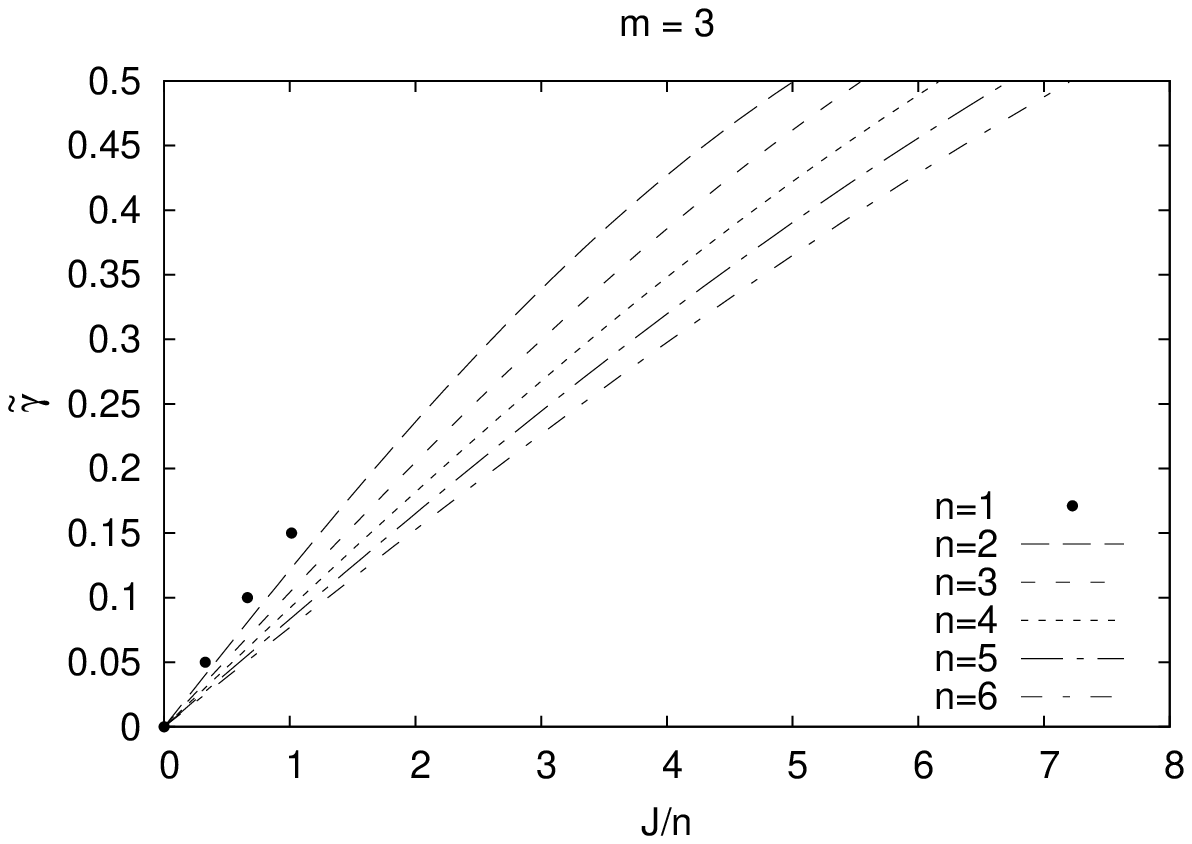}
\hspace{0.5cm} (d)\hspace{-0.6cm}
\includegraphics[height=.25\textheight, angle =0]{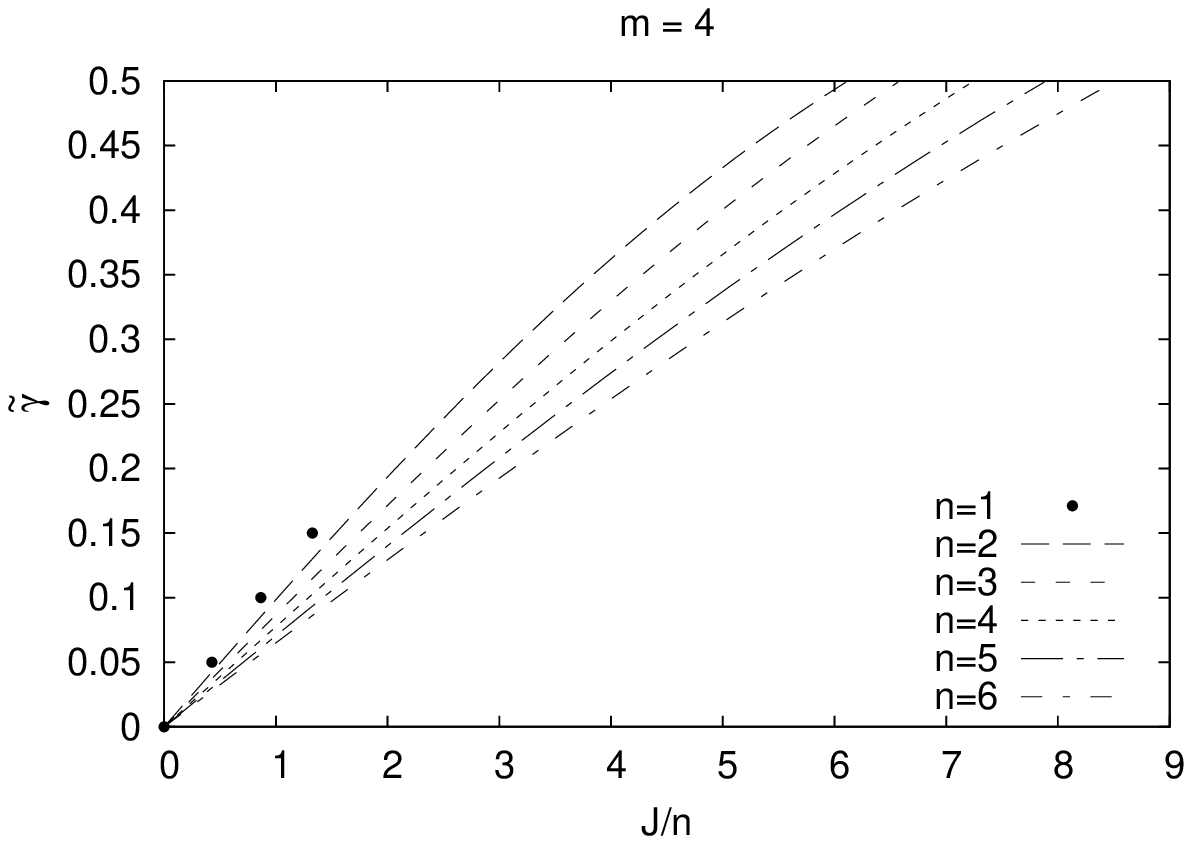}
\\
\hspace{0.0cm} (e)\hspace{-0.6cm}
\includegraphics[height=.25\textheight, angle =0]{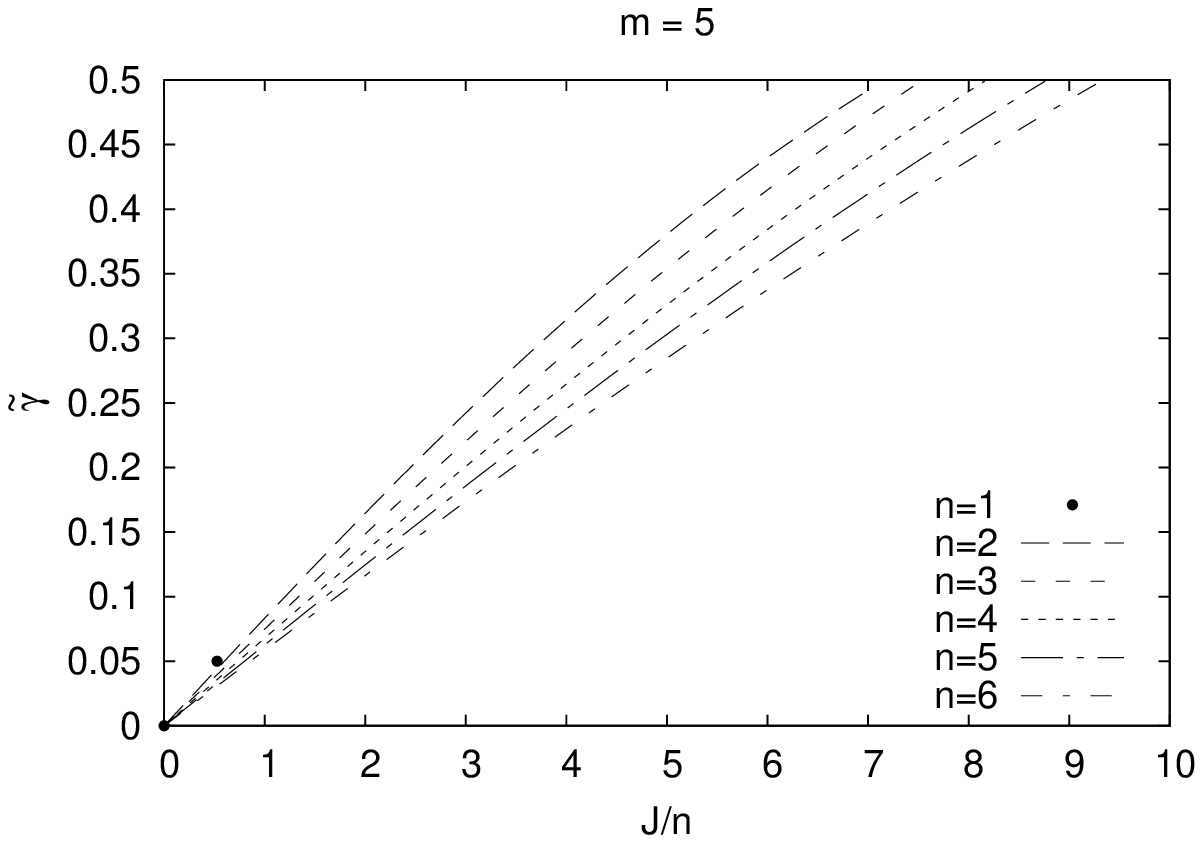}
\hspace{0.5cm} (f)\hspace{-0.6cm}
\includegraphics[height=.25\textheight, angle =0]{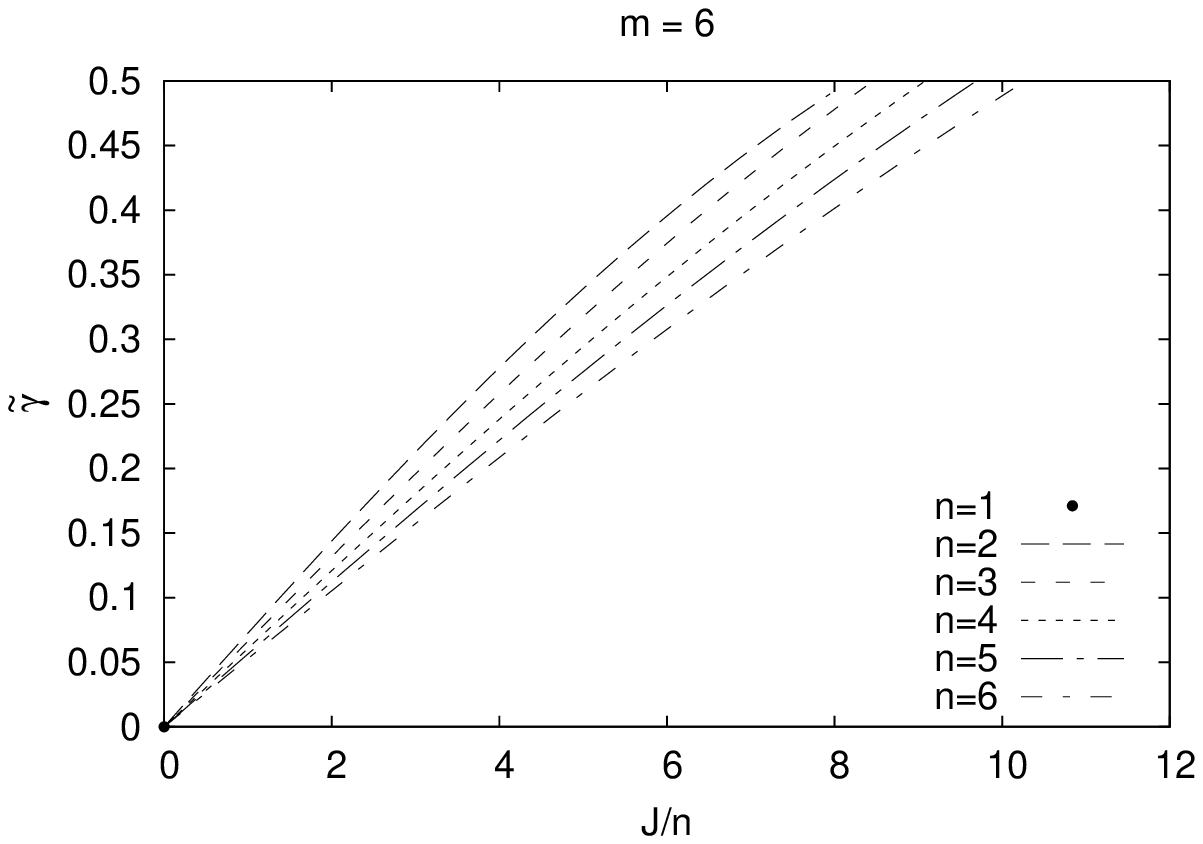}
\end{center}
\caption{
Domain of existence of the solutions:
the asymptotic value of the $U(1)$ field $\tilde \gamma = \gamma/gv$
versus the scaled angular momentum $J/n$
(in units of $J_0=4 \pi/g^2$)
(a) $m=1$, $n=1,\dots,6$,
(b) $m=2$, $n=1,\dots,6$,
(c) $m=3$, $n=1,\dots,6$,
(d) $m=4$, $n=1,\dots,6$,
(e) $m=5$, $n=1,\dots,6$,
(f) $m=6$, $n=1,\dots,6$.
}
\end{figure}

We observe that for fixed $m$ the maximal value of the
scaled angular momentum $J/n$ (respectively charge ${\cal Q}/e$)
increases with $n$.
Thus the value of the maximal angular momentum $J_{\rm max}$
increases faster than linearly with $n$.
For $m=1$ the solutions are multisphaleron solutions,
with $n$ sphalerons superimposed at the origin.
We thus see that the more sphalerons a configuration
consists of, the more angular momentum the constituents
can carry.
We reach the analogous conclusion by considering a fixed 
value of $n$ and varying $m$.
The maximal value of the scaled angular momentum
$J_{\rm max}/n$ increases with $m$.
Thus the higher the number of constituents of a configuration
(encoded in the product $mn$),
the more angular momentum each of the constituents
can carry.

\begin{figure}[p!]
\lbfig{f-3}
\begin{center}
\hspace{0.0cm} (a)\hspace{-0.6cm}
\includegraphics[height=.25\textheight, angle =0]{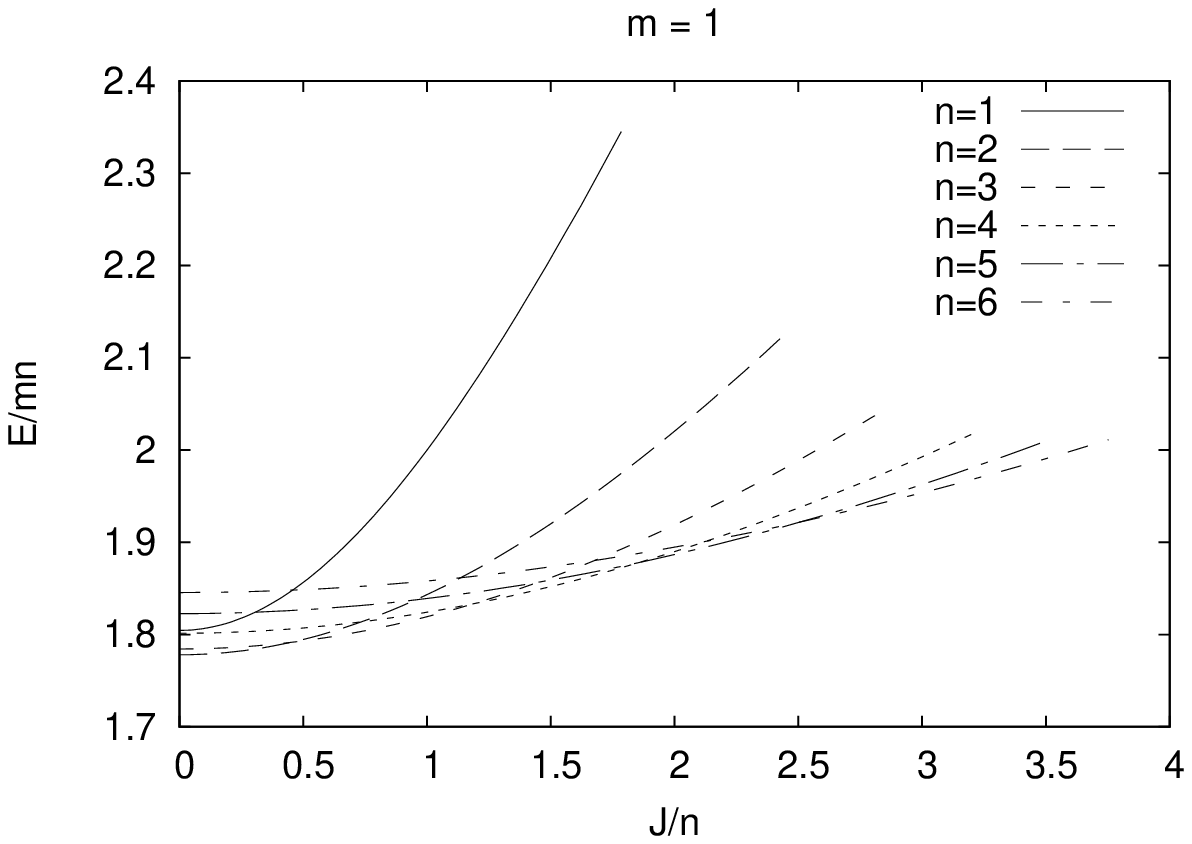}
\hspace{0.5cm} (b)\hspace{-0.6cm}
\includegraphics[height=.25\textheight, angle =0]{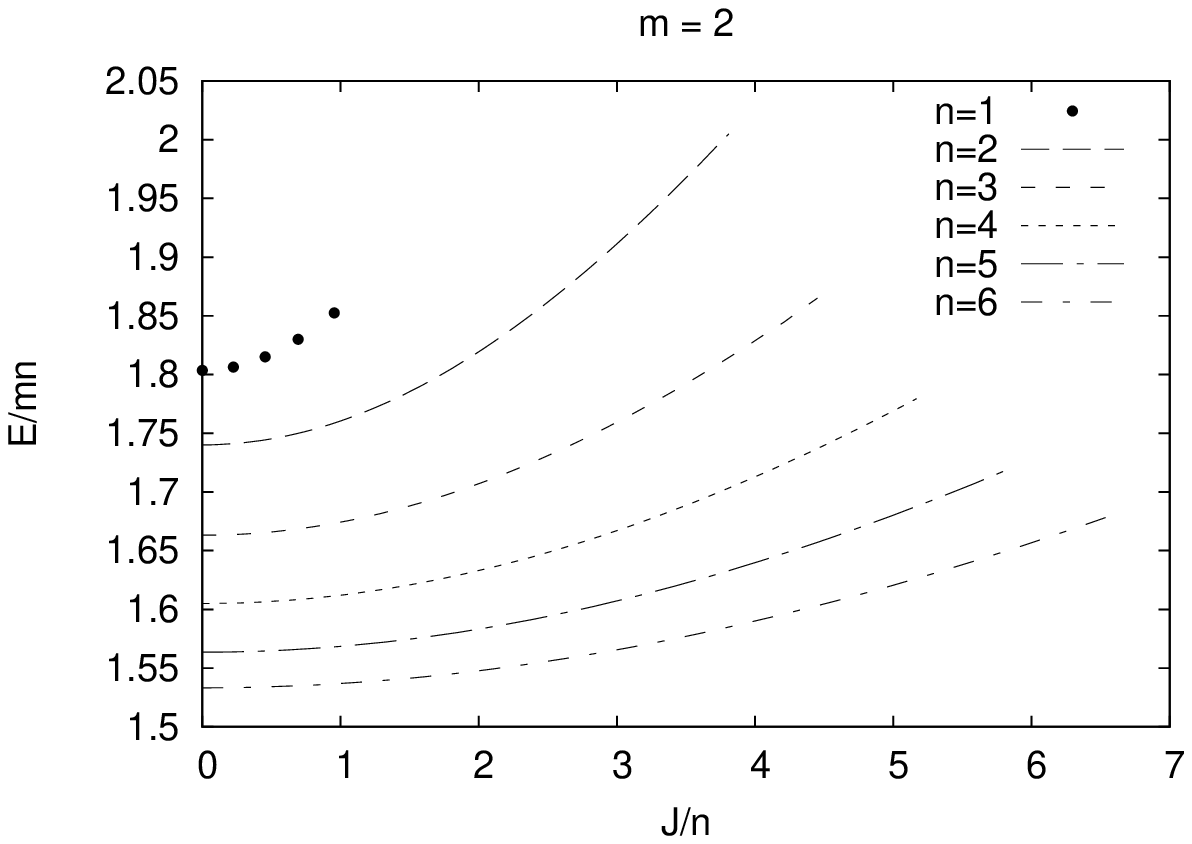}
\\
\hspace{0.0cm} (c)\hspace{-0.6cm}
\includegraphics[height=.25\textheight, angle =0]{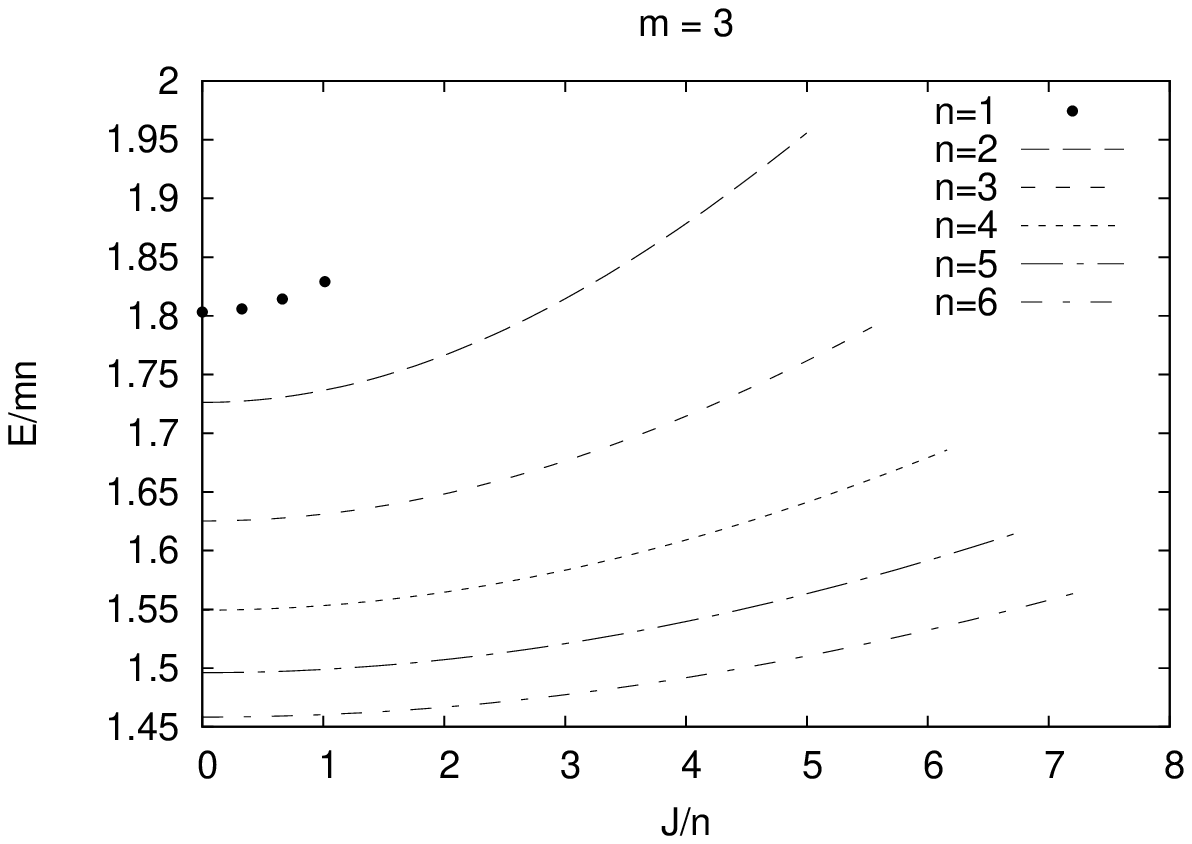}
\hspace{0.5cm} (d)\hspace{-0.6cm}
\includegraphics[height=.25\textheight, angle =0]{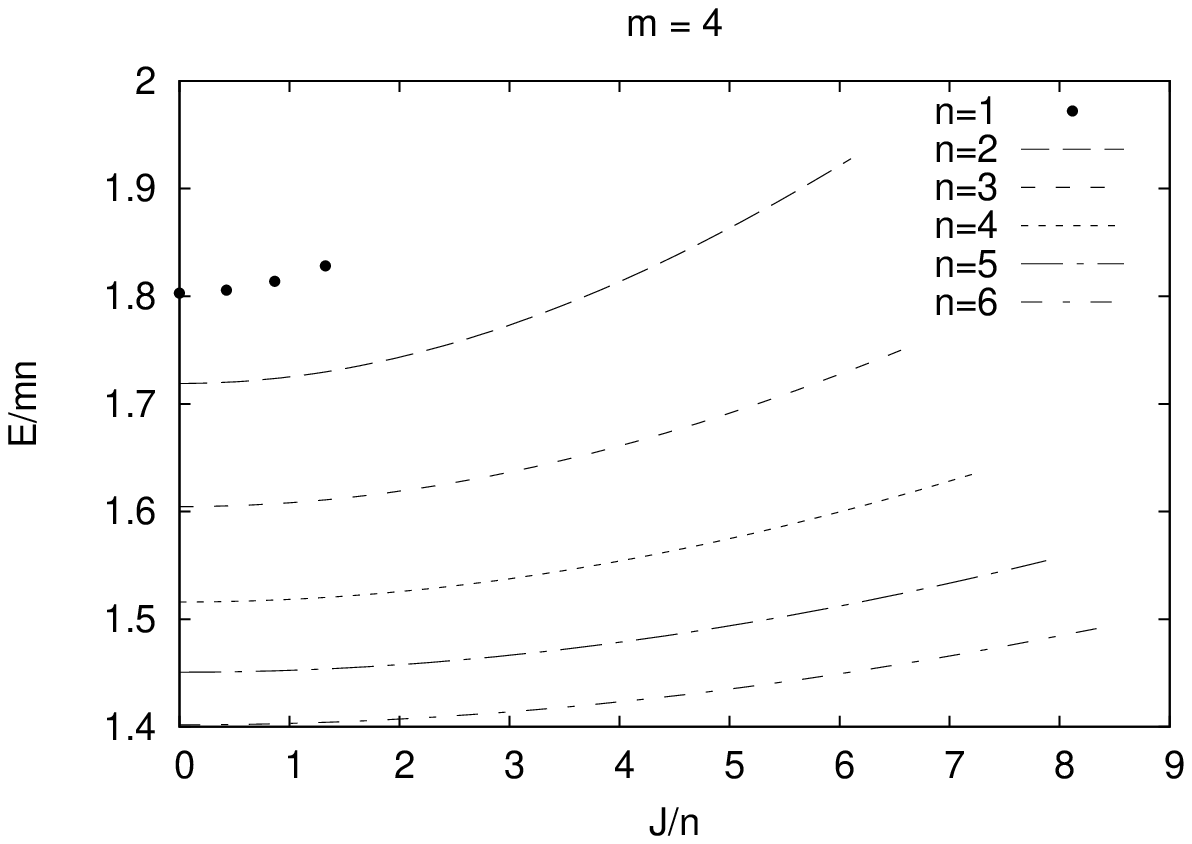}
\\
\hspace{0.0cm} (e)\hspace{-0.6cm}
\includegraphics[height=.25\textheight, angle =0]{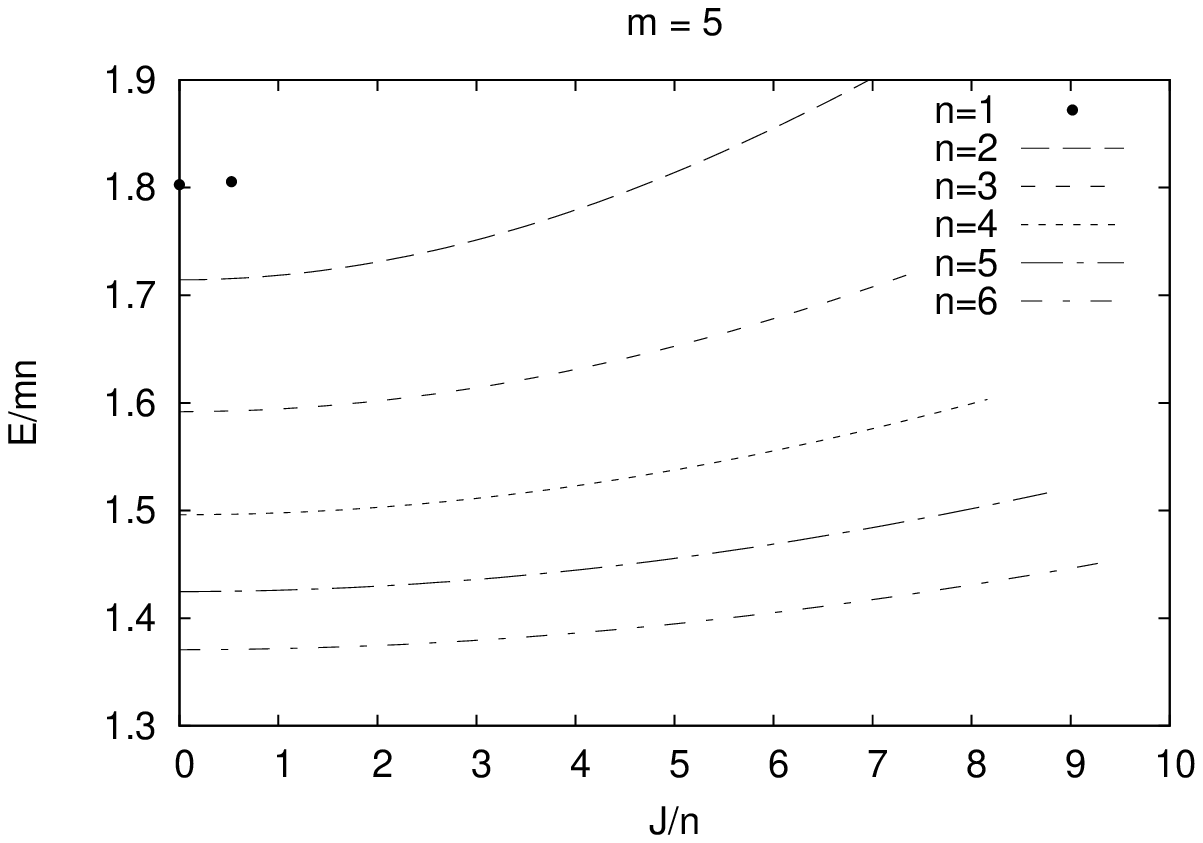}
\hspace{0.5cm} (f)\hspace{-0.6cm}
\includegraphics[height=.25\textheight, angle =0]{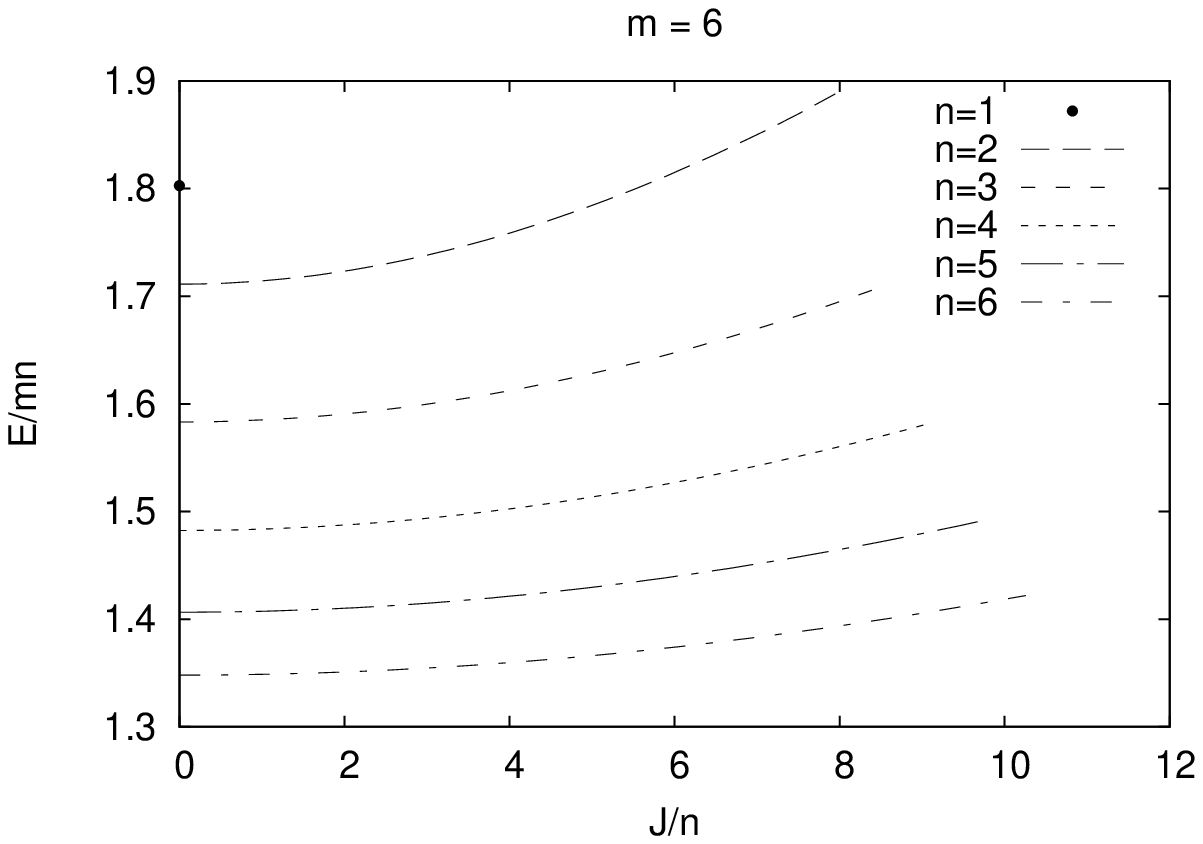}
\end{center}
\caption{
Properties of the solutions:
the scaled energy $E/mn$ (in units of $E_0=4\pi v/g$)
versus the scaled angular momentum $J/n$
(in units of $J_0=4 \pi/g^2$)
(a) $m=1$, $n=1,\dots,6$,
(b) $m=2$, $n=1,\dots,6$,
(c) $m=3$, $n=1,\dots,6$,
(d) $m=4$, $n=1,\dots,6$,
(e) $m=5$, $n=1,\dots,6$,
(f) $m=6$, $n=1,\dots,6$.
}
\end{figure}

\begin{figure}[p!]
\lbfig{f-4}
\begin{center}
\hspace{0.0cm} (a)\hspace{-0.6cm}
\includegraphics[height=.25\textheight, angle =0]{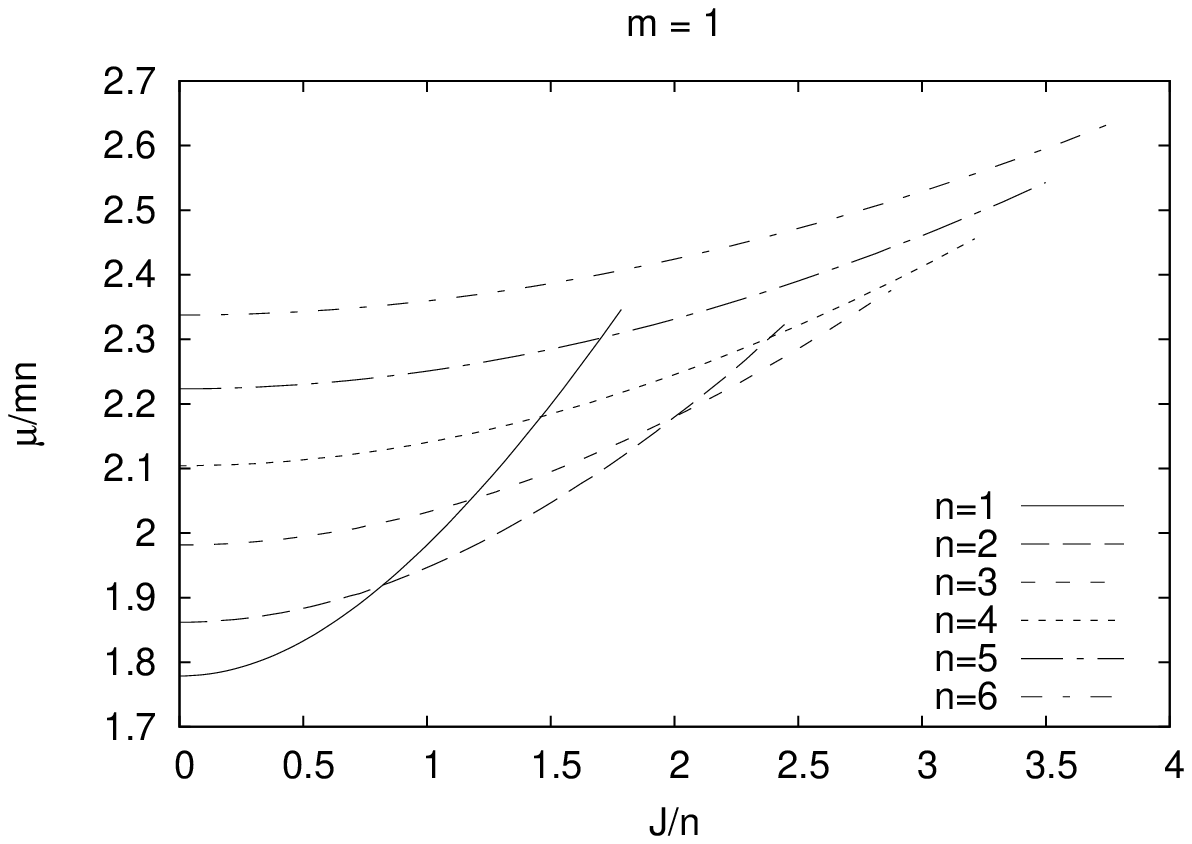}
\hspace{0.5cm} (b)\hspace{-0.6cm}
\includegraphics[height=.25\textheight, angle =0]{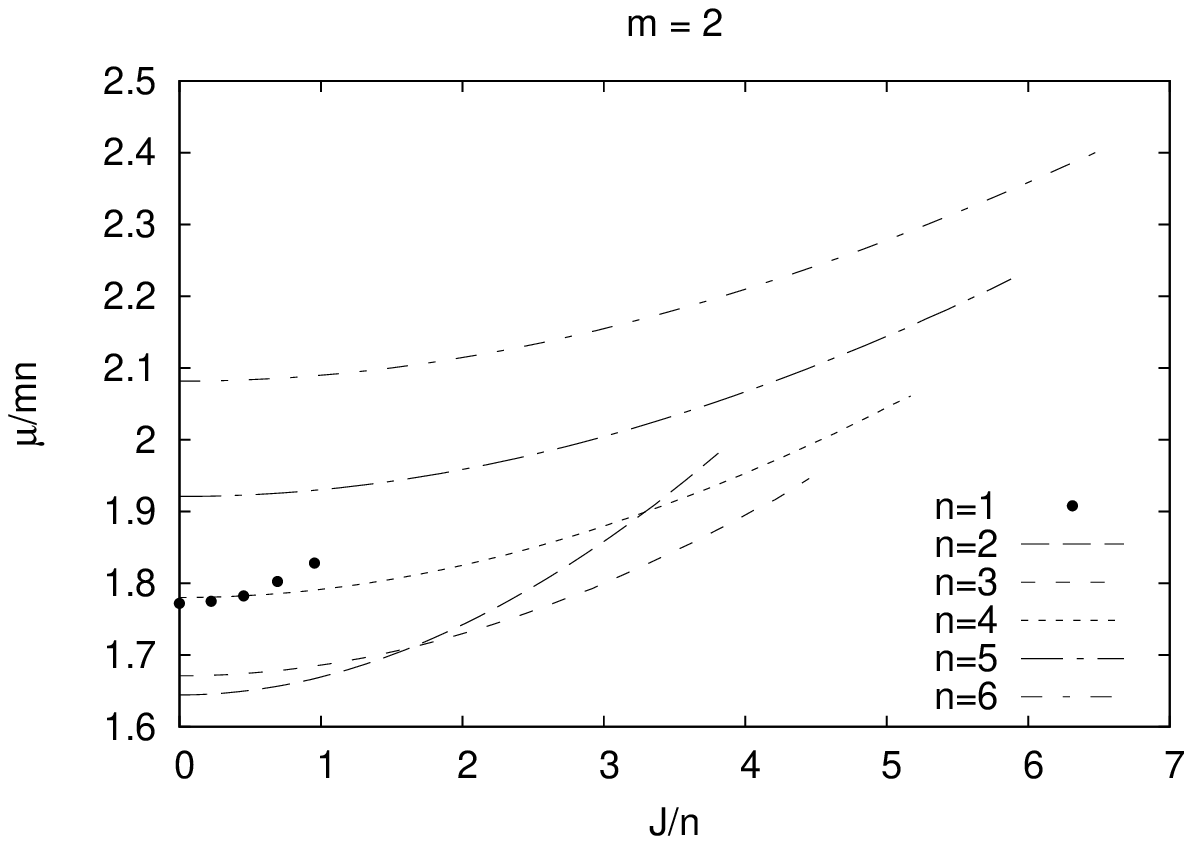}
\\
\hspace{0.0cm} (c)\hspace{-0.6cm}
\includegraphics[height=.25\textheight, angle =0]{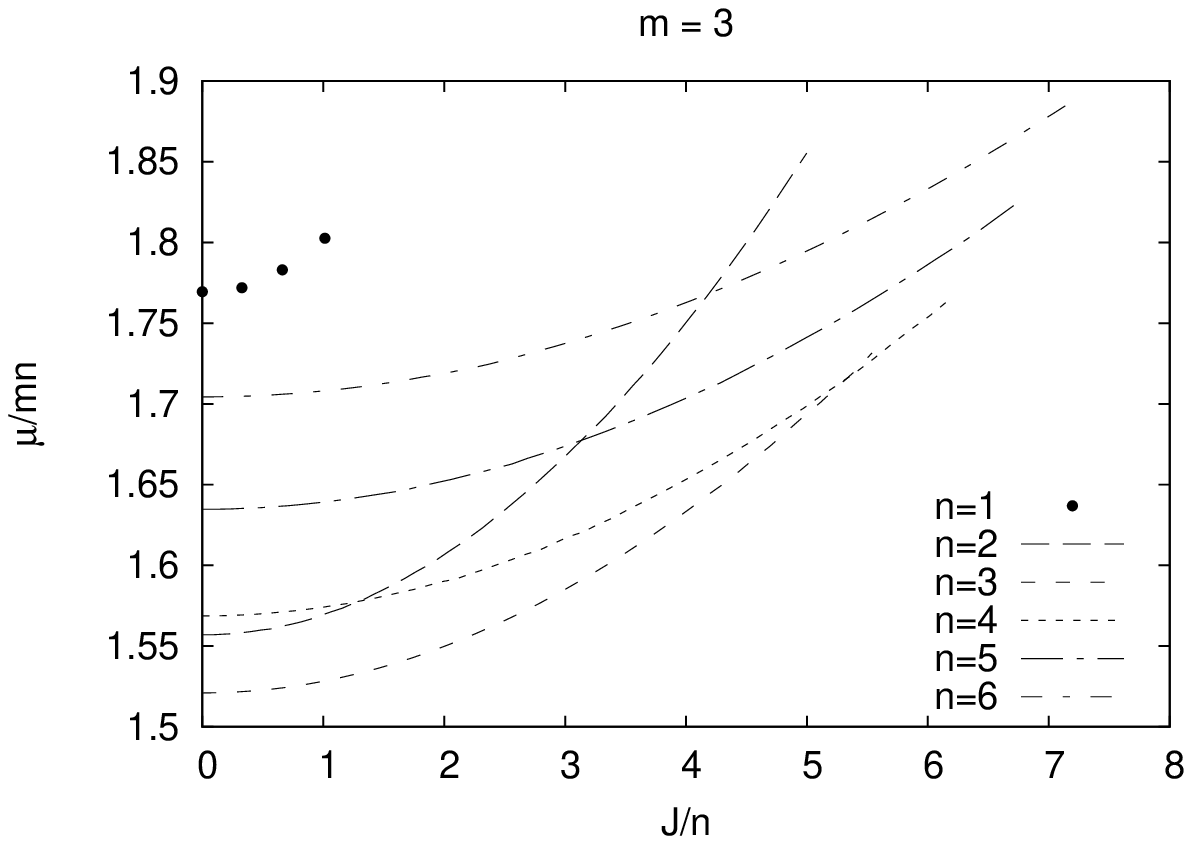}
\hspace{0.5cm} (d)\hspace{-0.6cm}
\includegraphics[height=.25\textheight, angle =0]{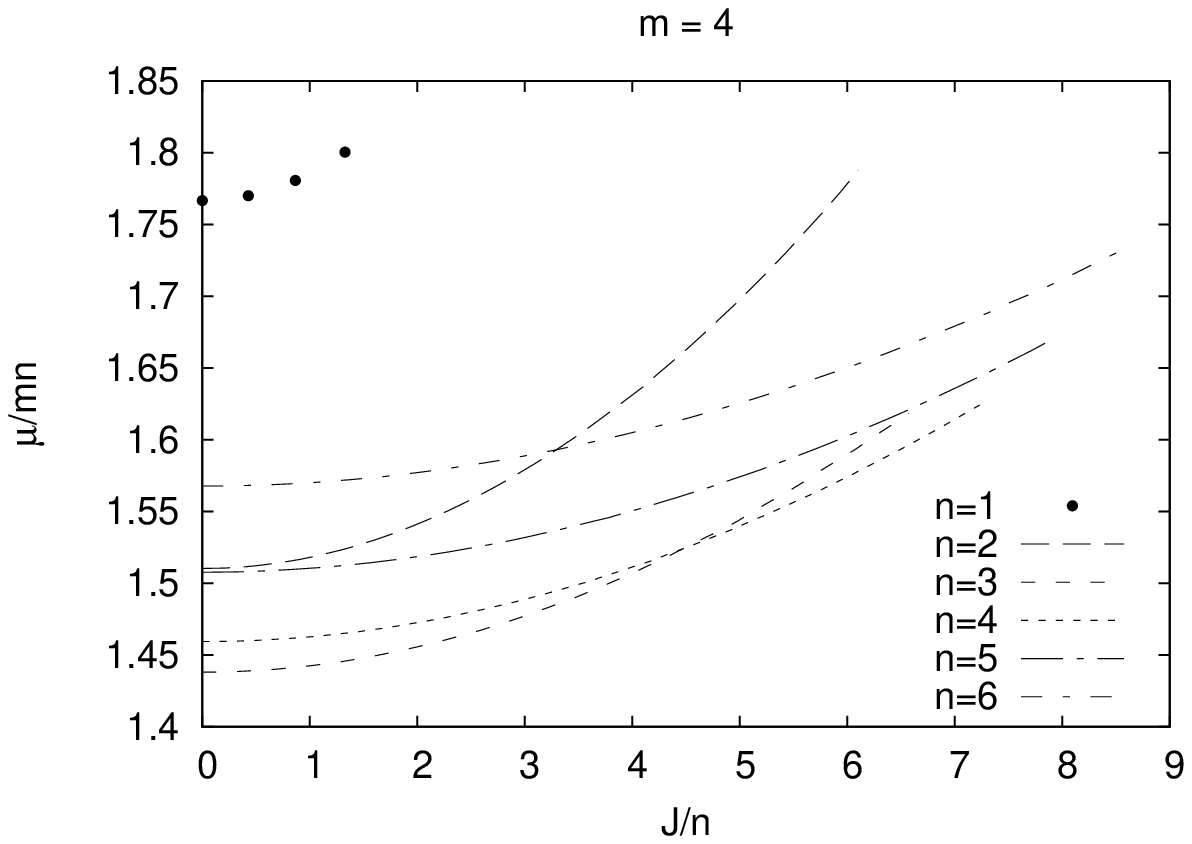}
\\
\hspace{0.0cm} (e)\hspace{-0.6cm}
\includegraphics[height=.25\textheight, angle =0]{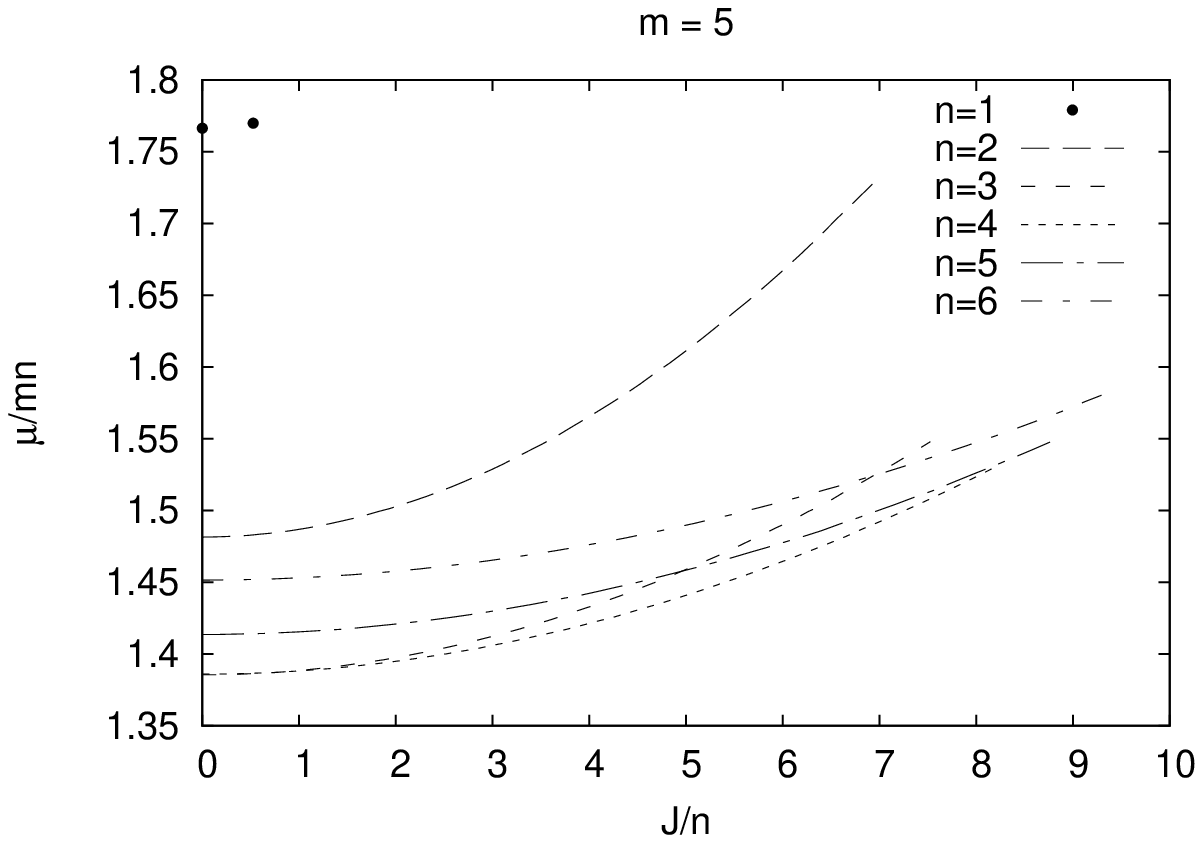}
\hspace{0.5cm} (f)\hspace{-0.6cm}
\includegraphics[height=.25\textheight, angle =0]{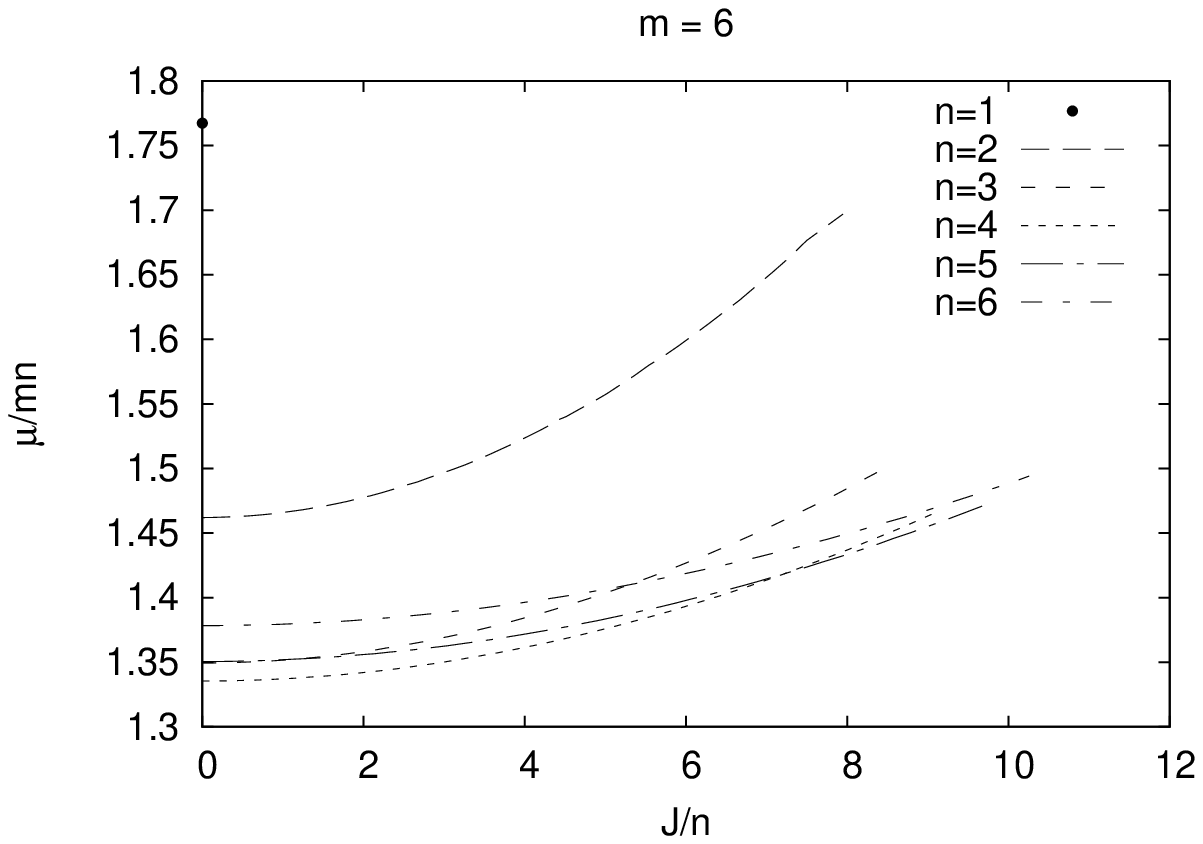}
\end{center}
\caption{
Properties of the solutions:
the scaled magnetic moment $\mu/n$
(in units of $\mu_0=e/\alpha_{\rm W} M_{\rm W}$)
versus the scaled angular momentum $J/n$
(in units of $J_0=4 \pi/g^2$)
(a) $m=1$, $n=1,\dots,6$,
(b) $m=2$, $n=1,\dots,6$,
(c) $m=3$, $n=1,\dots,6$,
(d) $m=4$, $n=1,\dots,6$,
(e) $m=5$, $n=1,\dots,6$,
(f) $m=6$, $n=1,\dots,6$.
}
\end{figure}

We exhibit in Fig.~\ref{f-3} the energy of the sets of solutions.
In Fig.~\ref{f-3}a
the energy of the multisphaleron solutions is shown.
For multisphalerons consisting of $n$ sphalerons 
the energy is on the order of $n$ times
the energy of a single sphaleron,
thus $E/n$ is roughly constant.
The deviations of the energy per sphaleron $E/n$
from the energy of a single sphaleron
can be attributed to the interaction of the $n$ sphalerons
and therefore be interpreted in terms of the binding energy
of these multisphaleron configurations.
For the employed value of the Higgs mass
the static solutions with $n=2-4$ represent bound states,
since $E/n$ is smaller than the energy of a single sphaleron,
whereas the static solutions with $n>4$ are slightly unbound
\cite{Kleihaus:1994tr}.
Since the binding energy is, however, sensitive to the
value of the Higgs mass, bound configurations may
turn into unbound configurations, when the value
of the Higgs mass is sufficiently changed.
\
When charge is added to these static multisphaleron configurations and the
solutions begin to rotate, their energy increases monotonically
with their angular momentum.
The increase of the energy per sphaleron $E/n$
with the angular momentum per sphaleron $J/n$ is strongest
for the branch of single sphaleron solutions.
The more sphalerons a multisphaleron configuration consists of,
the weaker is the increase of its energy per sphaleron $E/n$
with increasing angular momentum per sphaleron $J/n$.
Thus charge and rotation contribute relatively less
to the total energy for these ``many sphaleron'' configurations
(e.g.~only 8\% for $n=6$ as compared to 30\% for $n=1$).
Consequently, the rotating multisphaleron configurations
turn into bound states beyond some critical value
of the angular momentum.

We next address the energy of the general sphaleron-antisphaleron systems,
which we like to think of as consisting of the number $mn$ of constituents.
We therefore exhibit the scaled energy $E/mn$,
i.e., the energy per constituent,
in Figs.~\ref{f-3}b-f.
We see, that for all $m>1$ the energy per constituent $E/mn$
is of the same order of magnitude.
The deviations of $E/mn$
from the energy of a single sphaleron
are attributed to the interaction of the sphalerons
and antisphalerons in the system
(as long as these can be discerned)
and can again be interpreted in terms of their binding energy.
We note that the binding energy increases with 
increasing number of constituents.
Charge and rotation contribute therefore relatively less
to the total energy in the ``many constituents'' configurations.

Finally, we exhibit in Fig.~\ref{f-4} 
the magnetic moment of the sets of solutions.
Sphalerons possess a large magnetic moment $\mu$.
For multisphalerons consisting of $n$ sphalerons one
expects from the superposition picture
that the magnetic moment should be roughly $n$ times
the magnetic moment of a single sphaleron.
As seen in Fig.~\ref{f-4}a, where we exhibit
the magnetic moment per sphaleron $\mu/n$
of the multisphaleron configurations
versus the angular momentum per sphaleron $J/n$,
this guess is not that good for the static multisphalerons configurations.
In fact, for them the interaction between the sphalerons
gives rise to an almost linear increase of
the magnetic moment per sphaleron $\mu/n$ with
the number of sphalerons.
When charge and thus angular momentum
is added to these multisphaleron configurations,
their magnetic moment increases monotonically
with increasing angular momentum.
This increase is strongest
for the branch of single sphaleron solutions,
and the more sphalerons a configuration consists of,
the weaker is the increase.

Addressing finally the magnetic moment $\mu$ of the
sphaleron-antisphaleron systems, 
we exhibit in Fig.~\ref{f-4}b-f
the magnetic moment per constituent $\mu/mn$.
Interestingly, for the chain configurations with $n=1$,
the magnetic moment per constituent $\mu/m$
is almost independent of $m$.
However, generically the interaction between the constituents
leads to a decrease of the magnetic moment per constituent 
$\mu/mn$ with increasing $m$.

\subsection{Local properties}

Having discussed the global properties of the
sphaleron-antisphaleron systems, 
we now turn to their local properties.
In particular, we address the effect of
the presence of charge and rotation
on the energy density $-T^t_t$,
and on the modulus of the Higgs field $|\Phi|$.
We also consider the angular momentum density $T^t_\varphi$
and the component of the stress-energy density $T^z_z$,
relevant for equilibrium.

\begin{figure}[h!]
\lbfig{f-5}
\begin{center}
\hspace{0.0cm} (a1)\hspace{-0.6cm}
\includegraphics[width=.45\textwidth, angle =0]{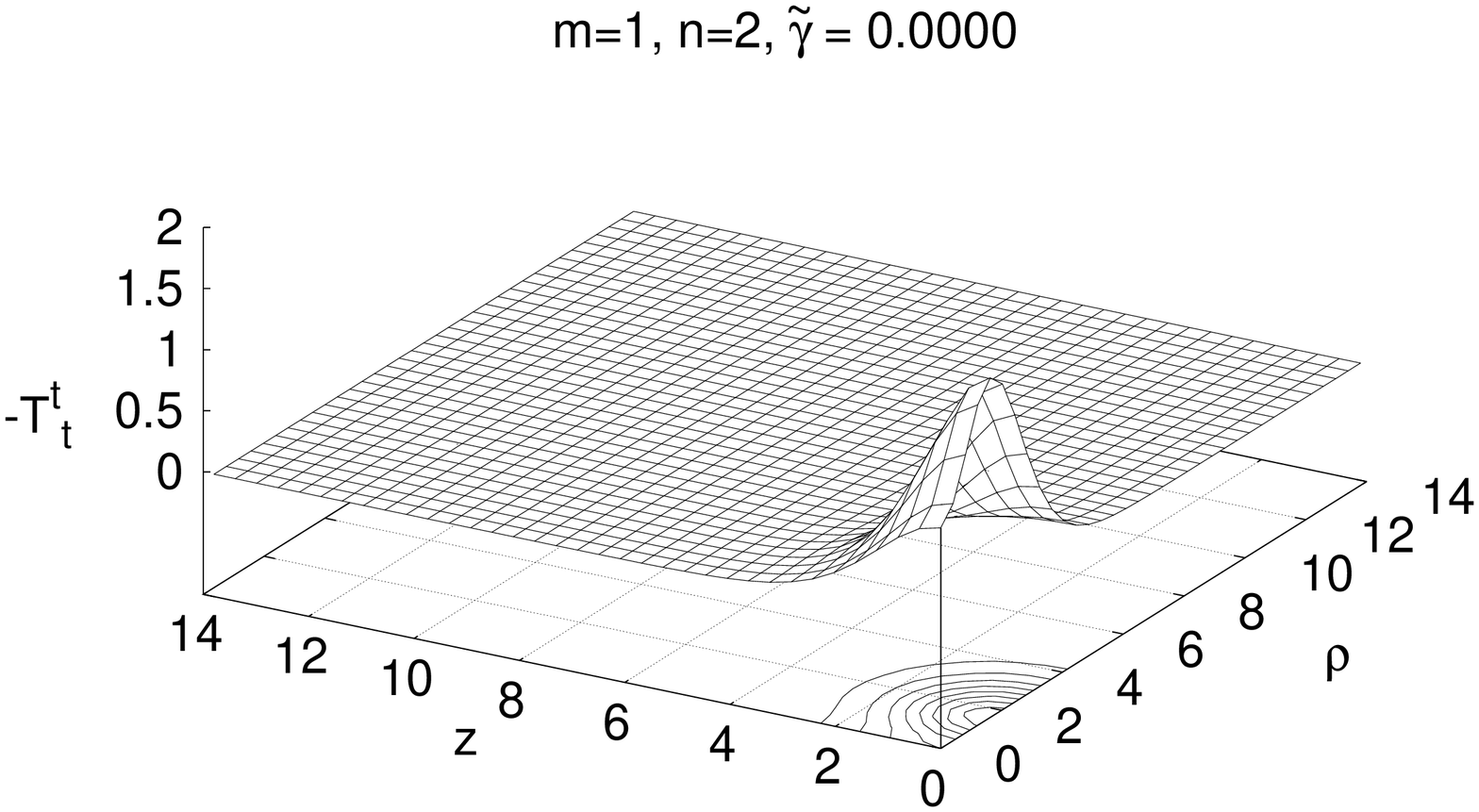}
\hspace{0.0cm} (a2)\hspace{-0.6cm}
\includegraphics[width=.45\textwidth, angle =0]{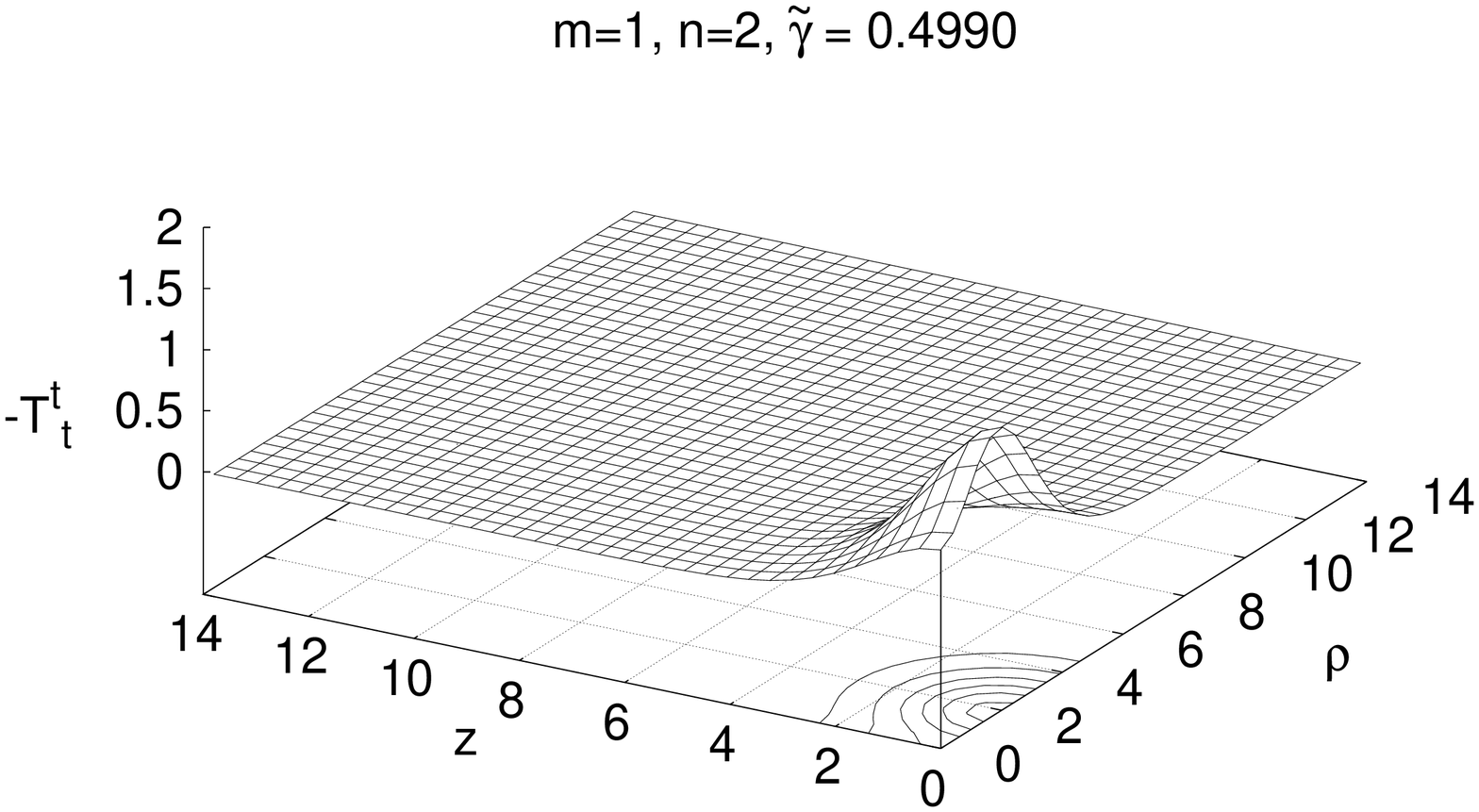}\\
\hspace{0.0cm} (b1)\hspace{-0.6cm}
\includegraphics[width=.45\textwidth, angle =0]{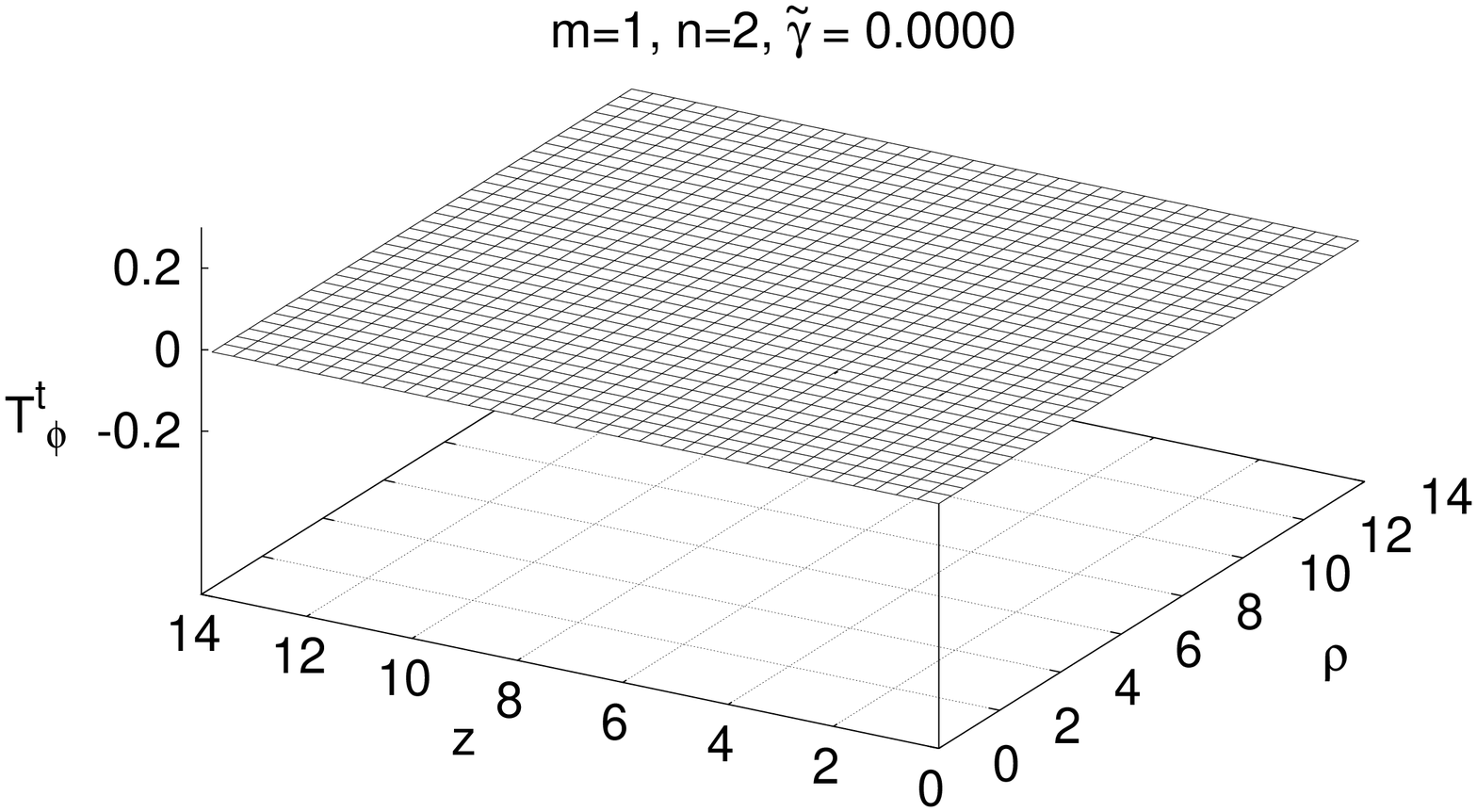}
\hspace{0.0cm} (b2)\hspace{-0.6cm}
\includegraphics[width=.45\textwidth, angle =0]{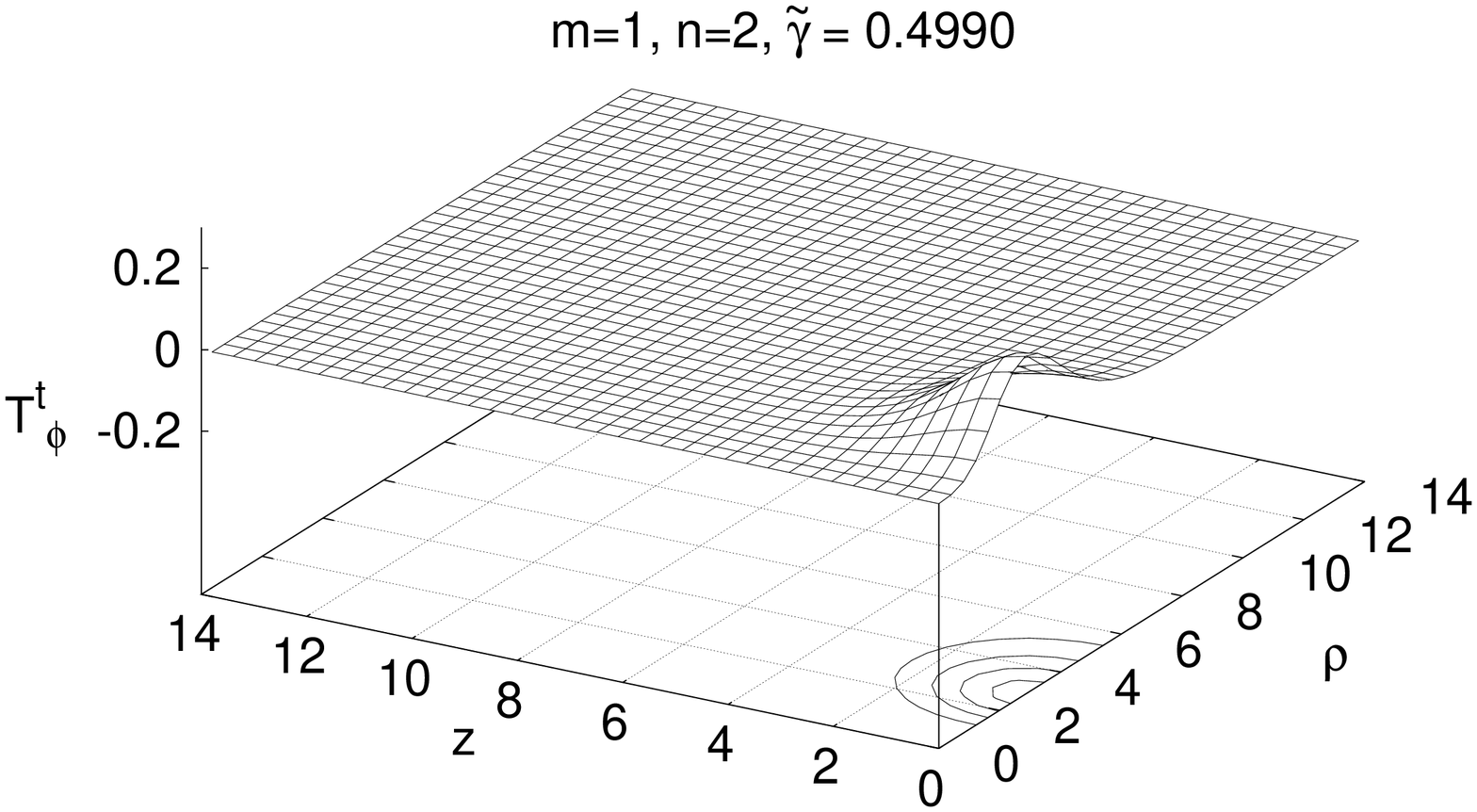}\\
\hspace{0.0cm} (c1)\hspace{-0.6cm}
\includegraphics[width=.45\textwidth, angle =0]{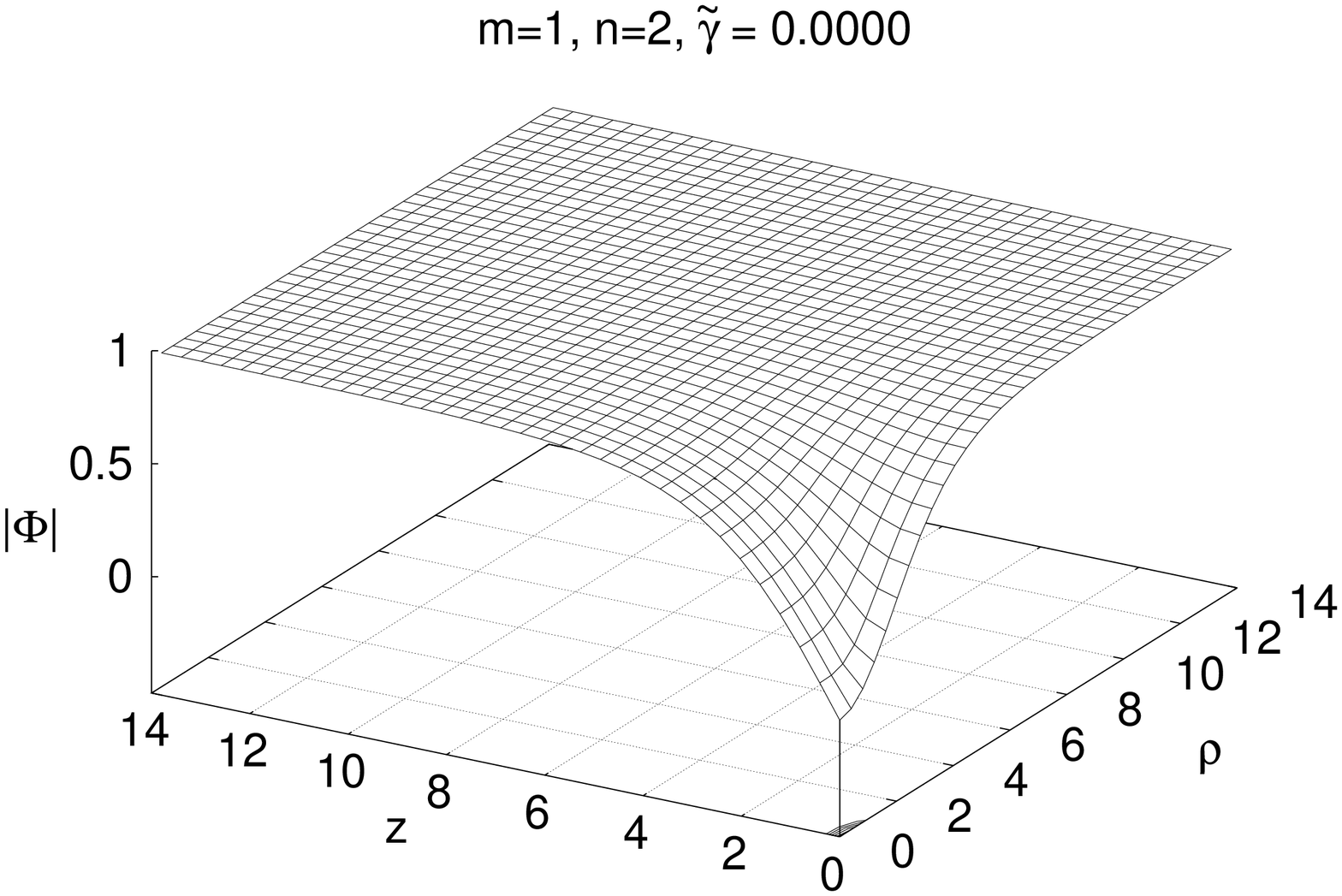}
\hspace{0.0cm} (c2)\hspace{-0.6cm}
\includegraphics[width=.45\textwidth, angle =0]{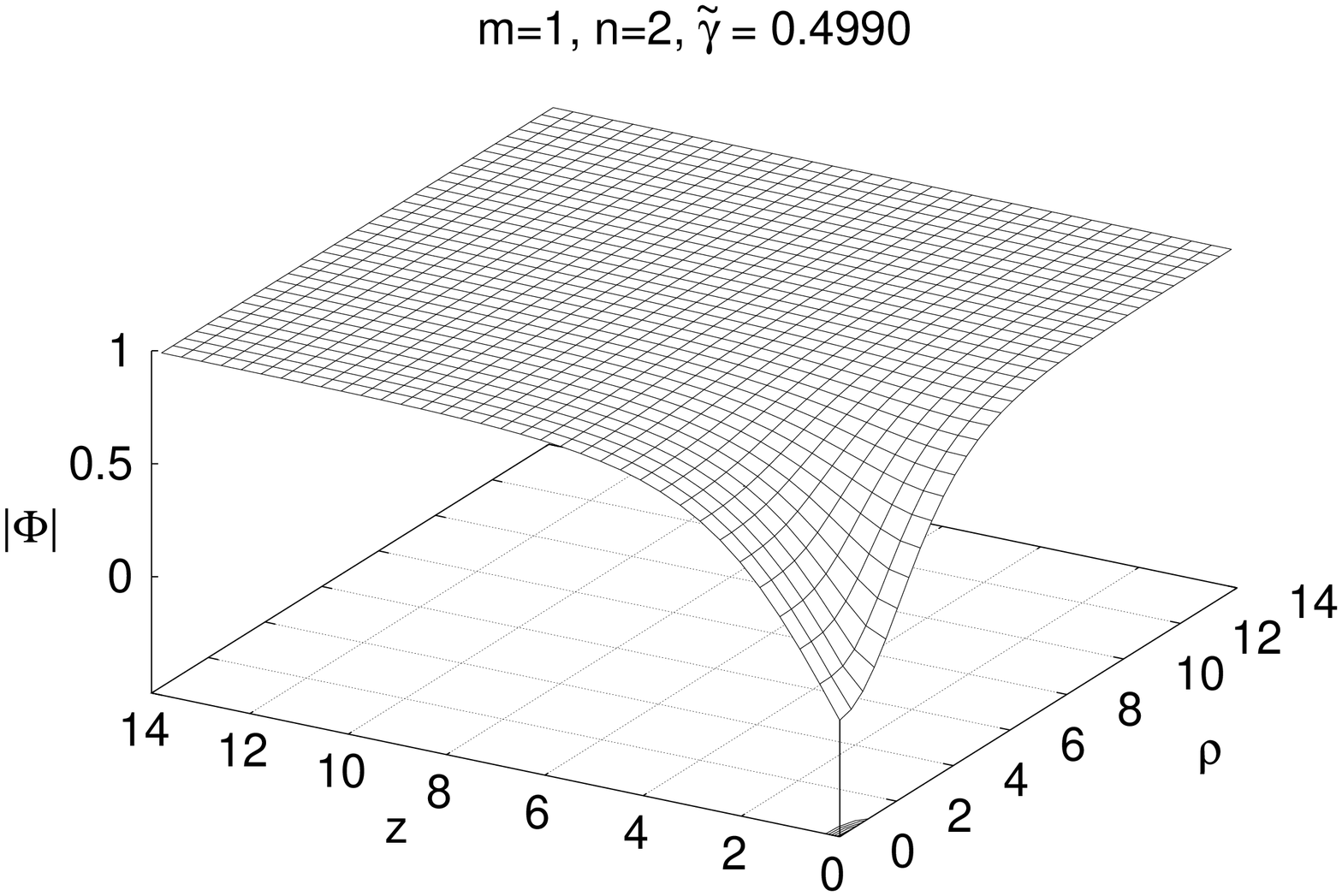}\\
\hspace{0.0cm} (d1)\hspace{-0.6cm}
\includegraphics[width=.45\textwidth, angle =0]{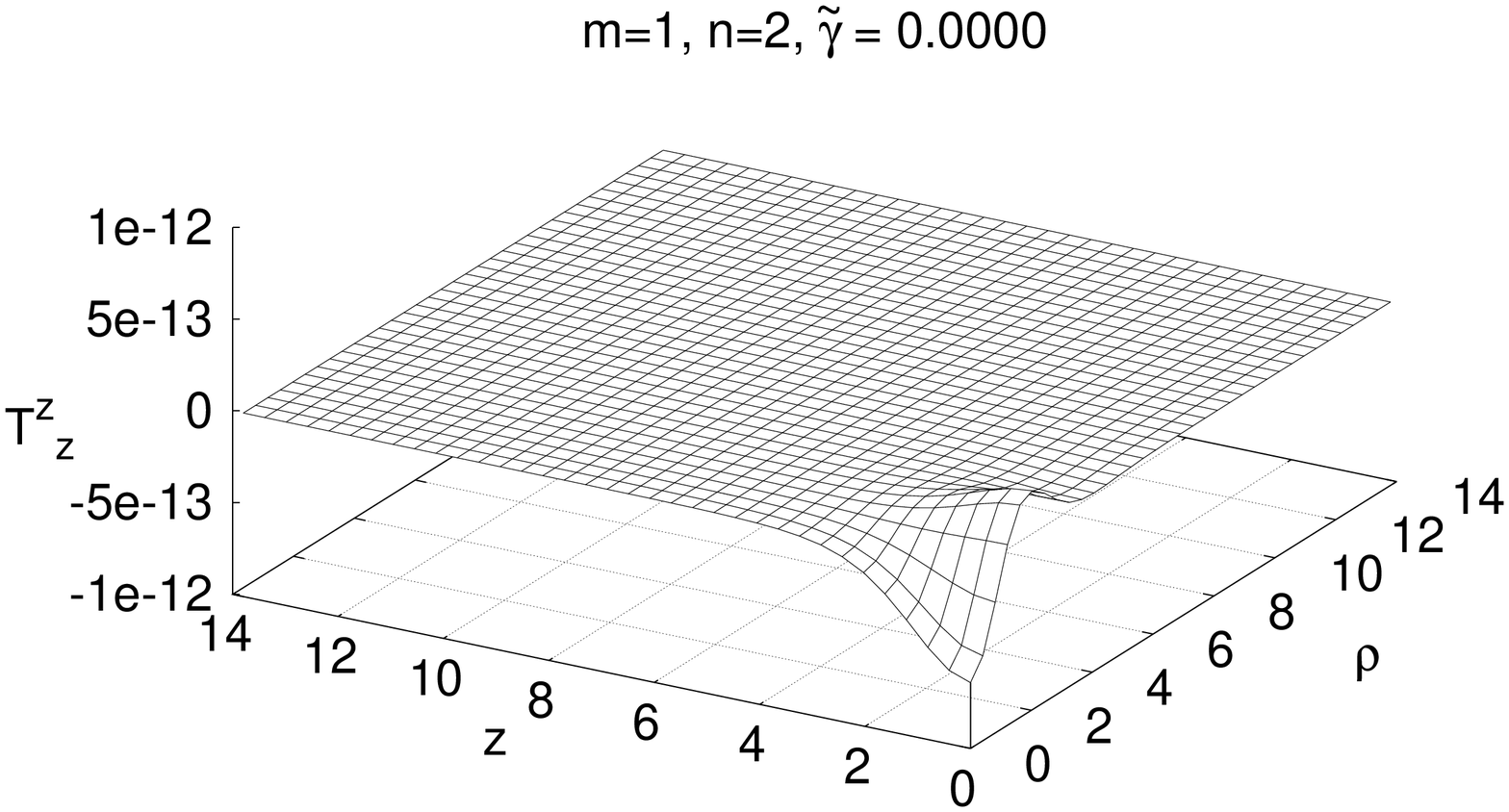}
\hspace{0.0cm} (d2)\hspace{-0.6cm}
\includegraphics[width=.45\textwidth, angle =0]{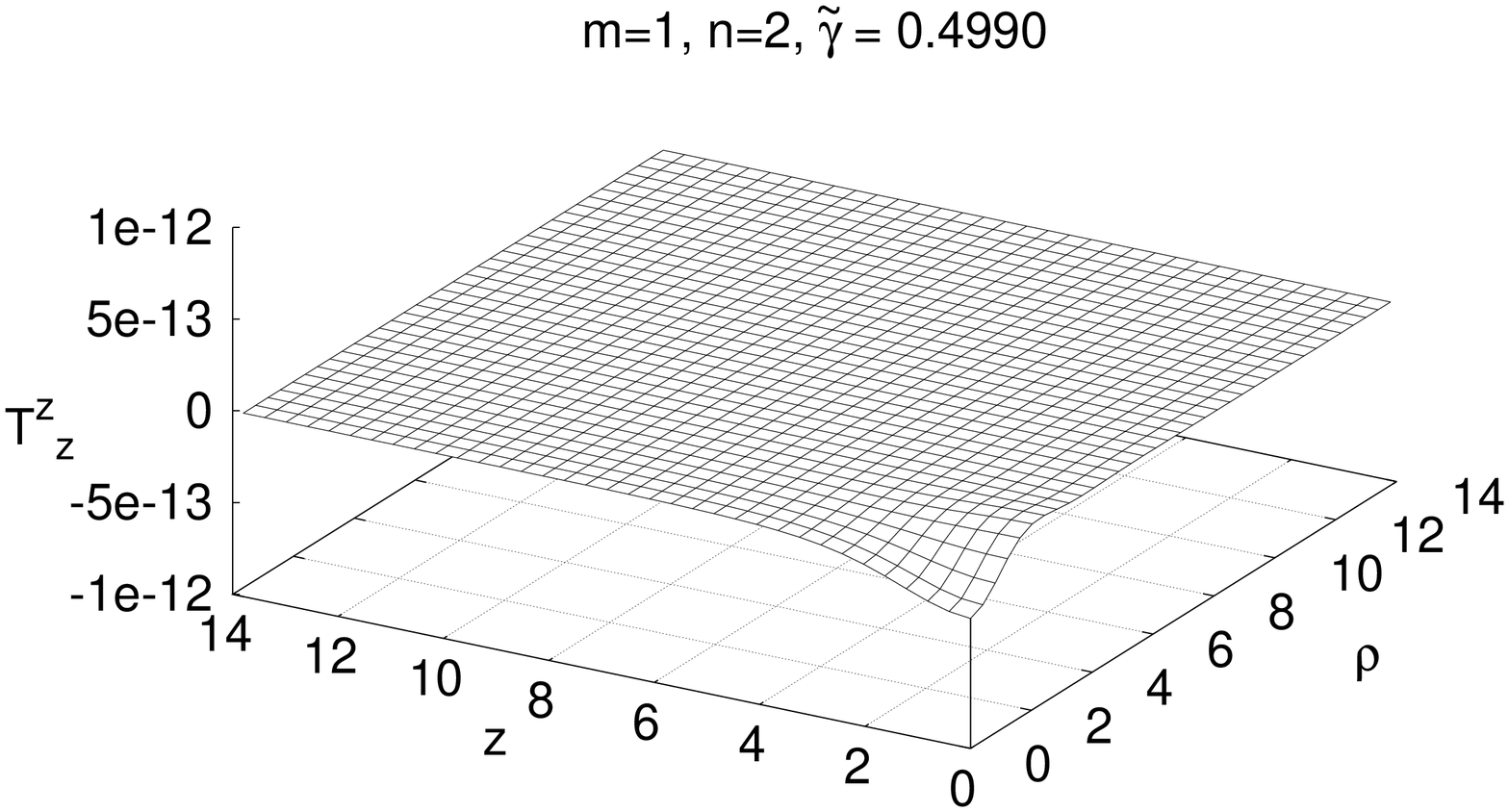}
\end{center}
\vspace{-0.5cm}
\caption{\small
The energy density $-T^t_t$ (a),
the angular momentum density $T^t_\varphi$ (b),
the modulus of the Higgs field $|\Phi|$ (c),
and the stress energy density $T^z_{z}$ (d)
are exhibited for $m=1$, $n=2$ solutions 
with $\tilde\gamma=0$ (left) and $\tilde\gamma \approx 0.5$ (right).
}
\end{figure}

\begin{figure}[h!]
\lbfig{f-6}
\begin{center}
\hspace{0.0cm} (a1)\hspace{-0.6cm}
\includegraphics[width=.45\textwidth, angle =0]{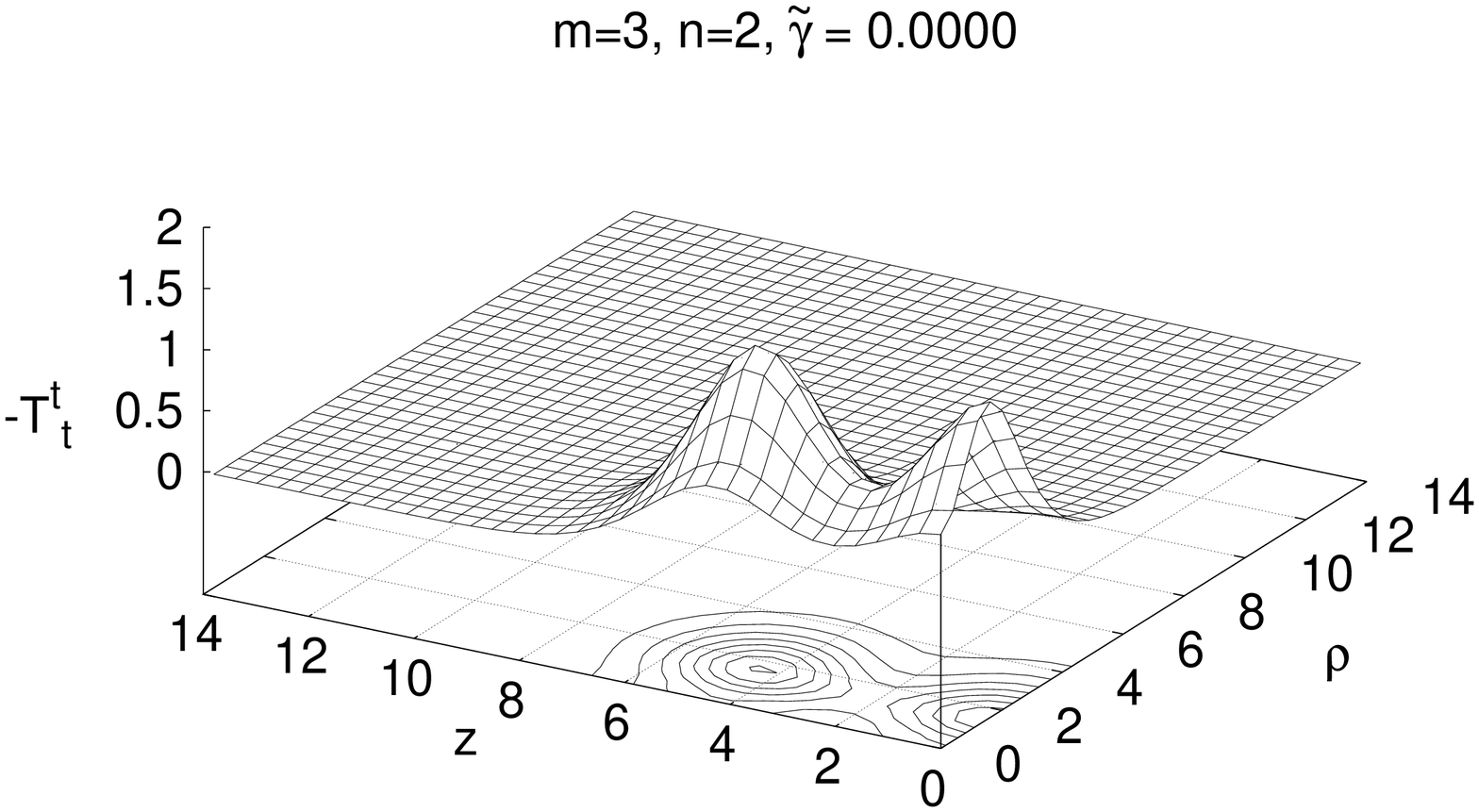}
\hspace{0.0cm} (a2)\hspace{-0.6cm}
\includegraphics[width=.45\textwidth, angle =0]{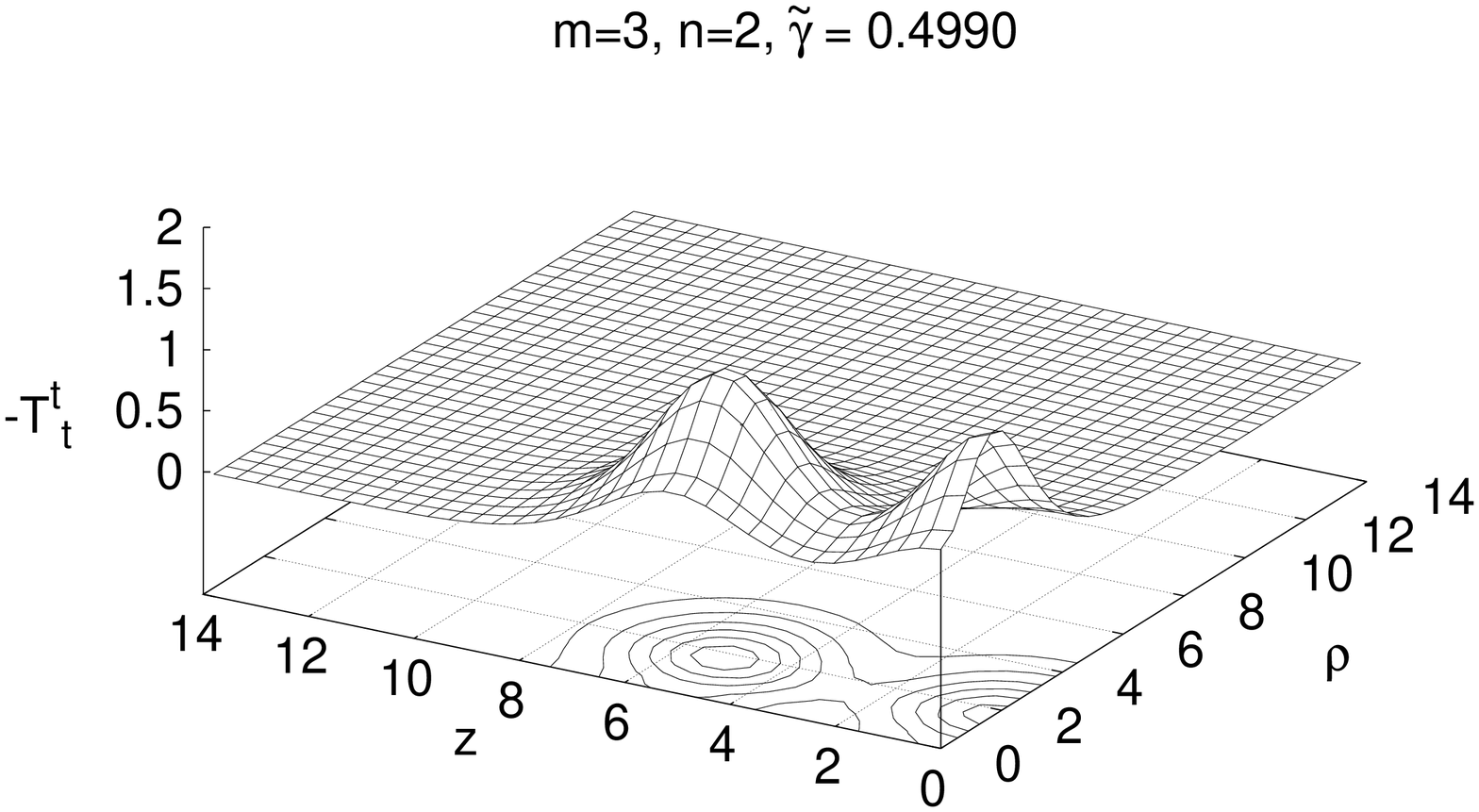}\\
\hspace{0.0cm} (b1)\hspace{-0.6cm}
\includegraphics[width=.45\textwidth, angle =0]{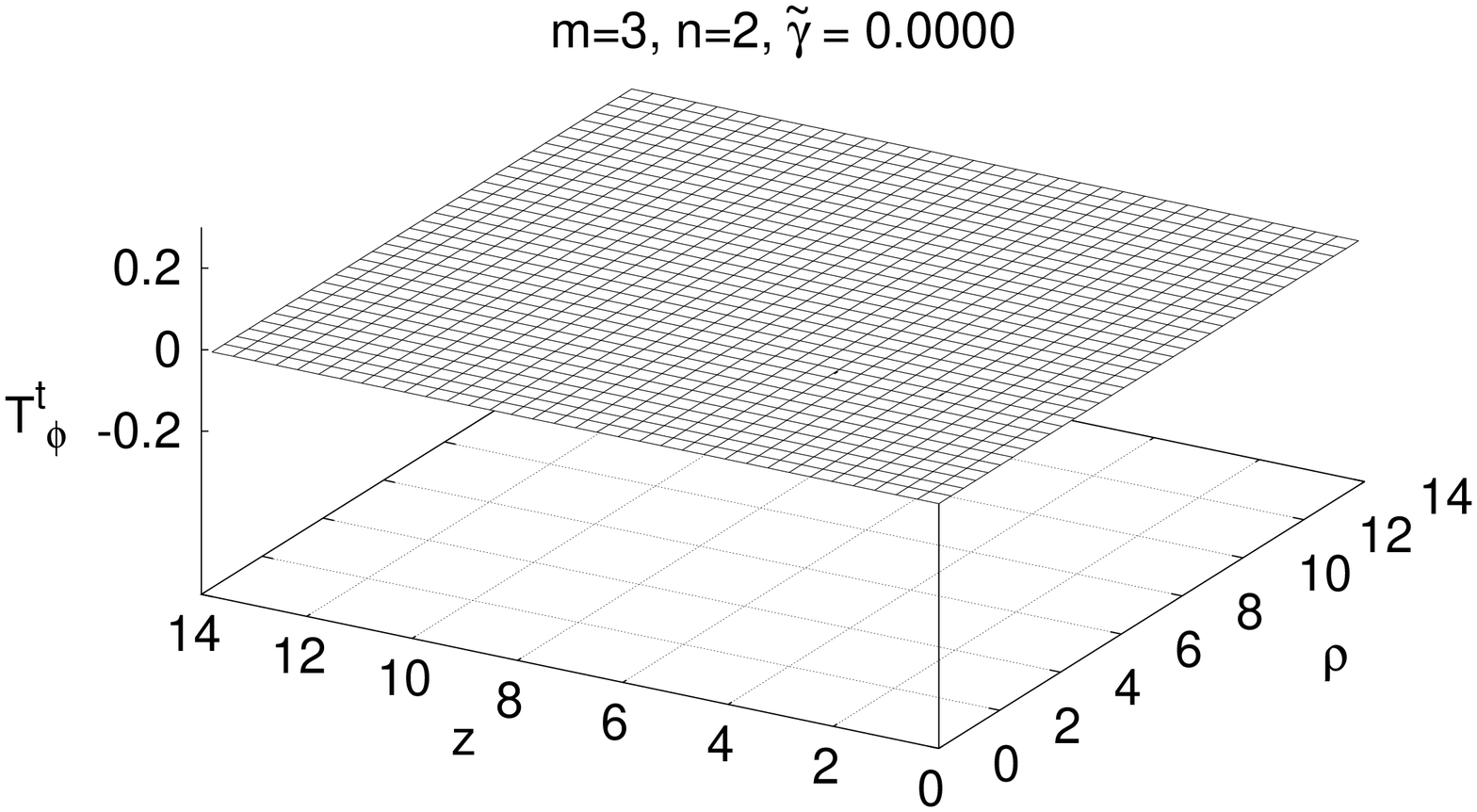}
\hspace{0.0cm} (b2)\hspace{-0.6cm}
\includegraphics[width=.45\textwidth, angle =0]{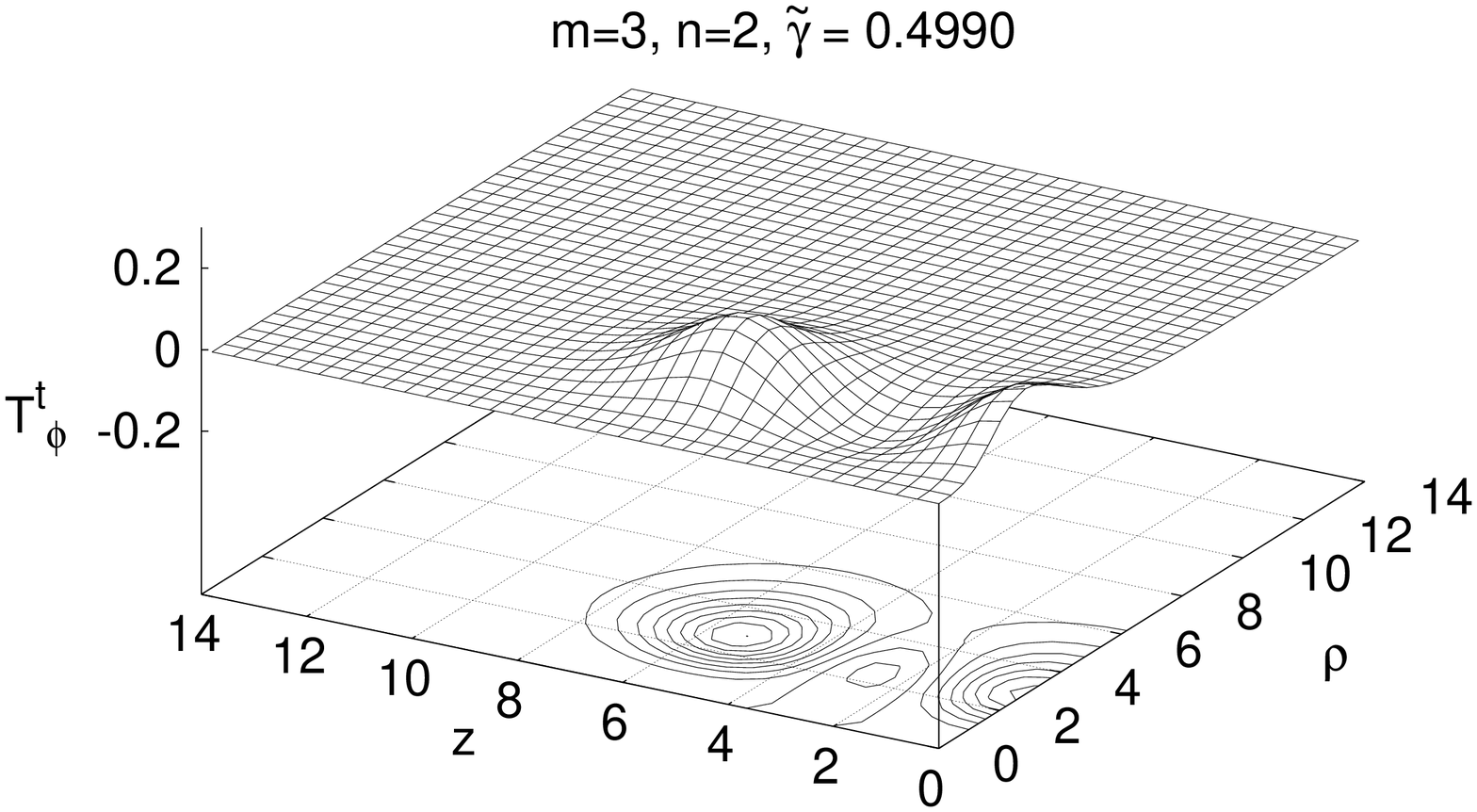}\\
\hspace{0.0cm} (c1)\hspace{-0.6cm}
\includegraphics[width=.45\textwidth, angle =0]{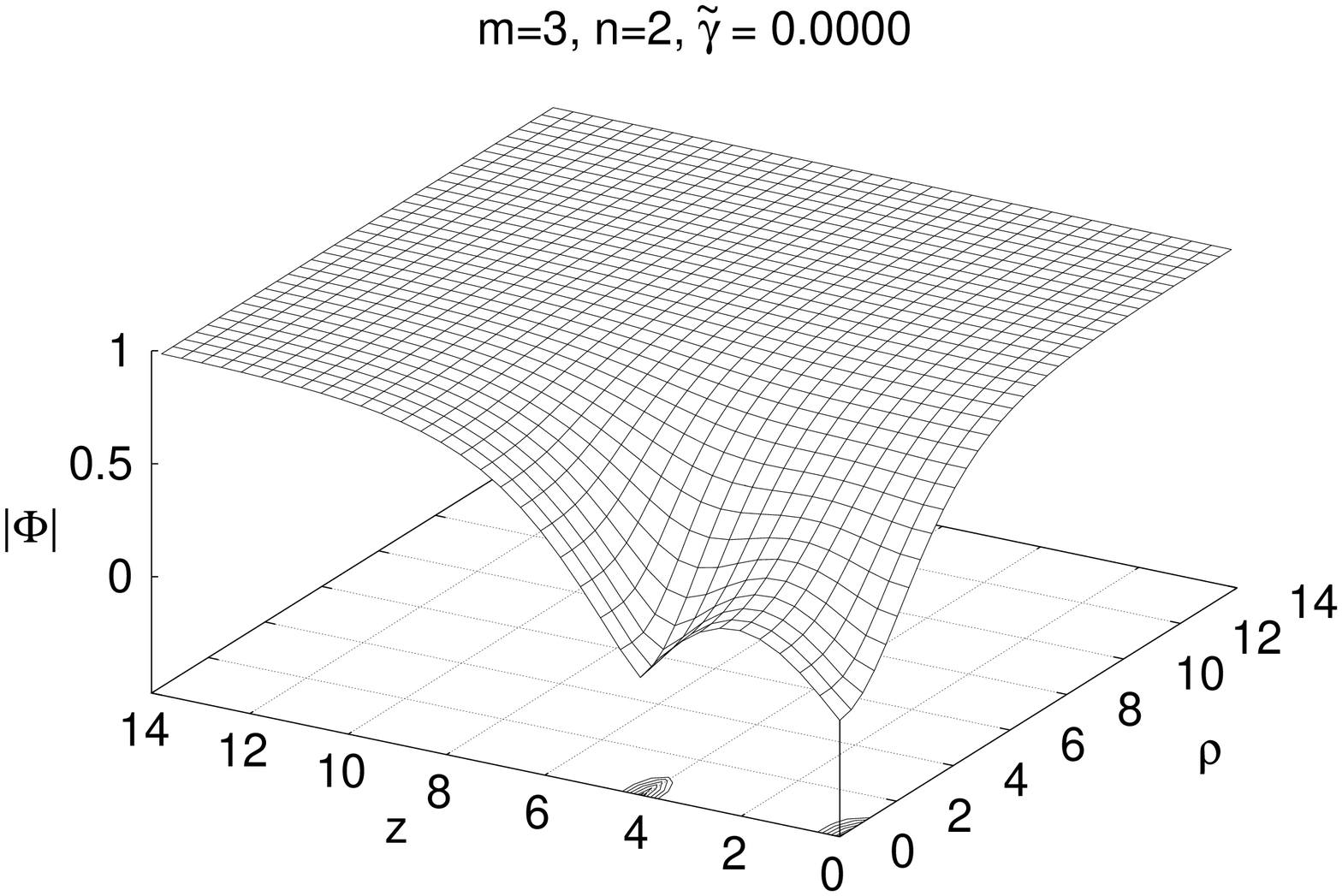}
\hspace{0.0cm} (c2)\hspace{-0.6cm}
\includegraphics[width=.45\textwidth, angle =0]{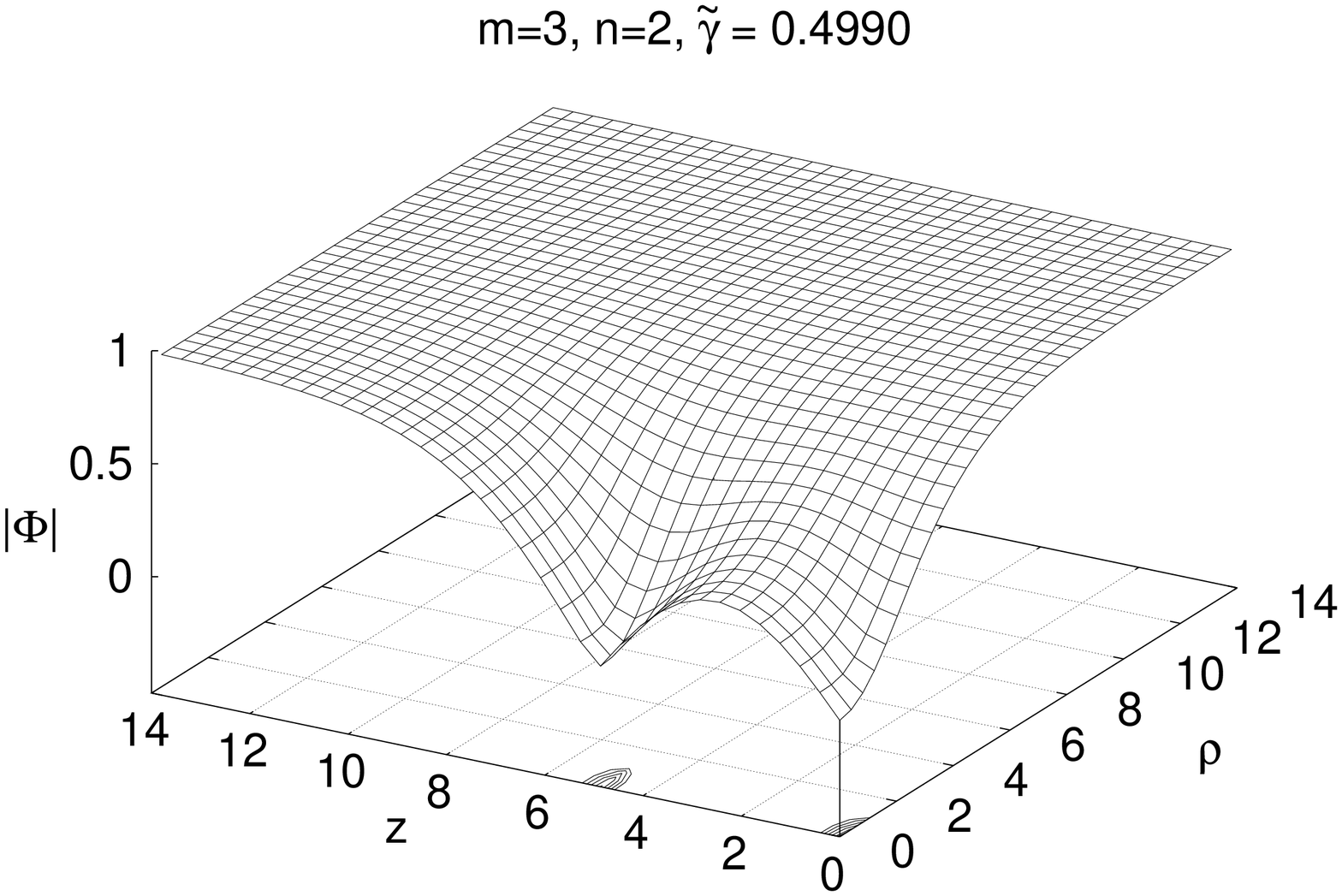}\\
\hspace{0.0cm} (d1)\hspace{-0.6cm}
\includegraphics[width=.45\textwidth, angle =0]{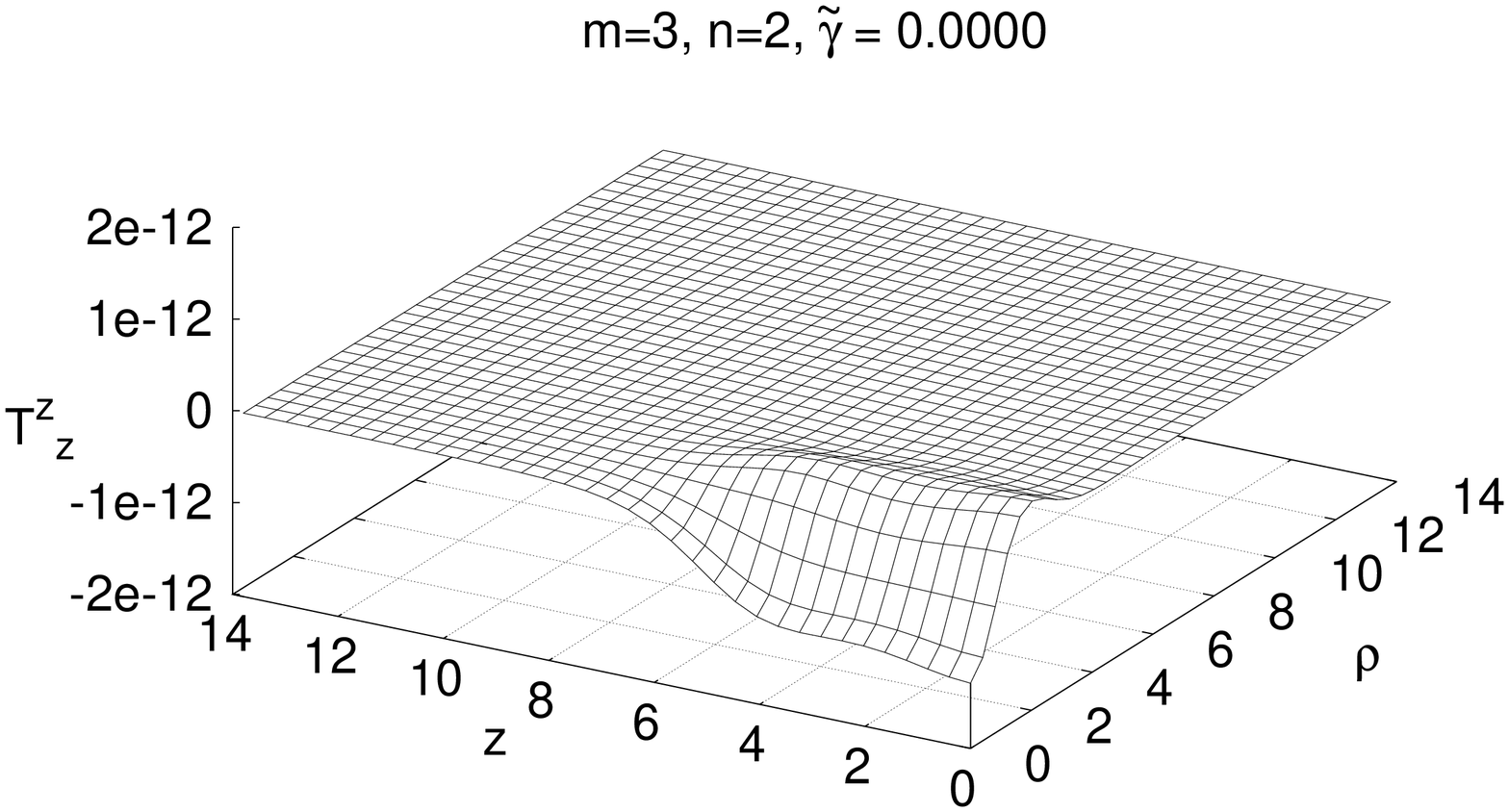}
\hspace{0.0cm} (d2)\hspace{-0.6cm}
\includegraphics[width=.45\textwidth, angle =0]{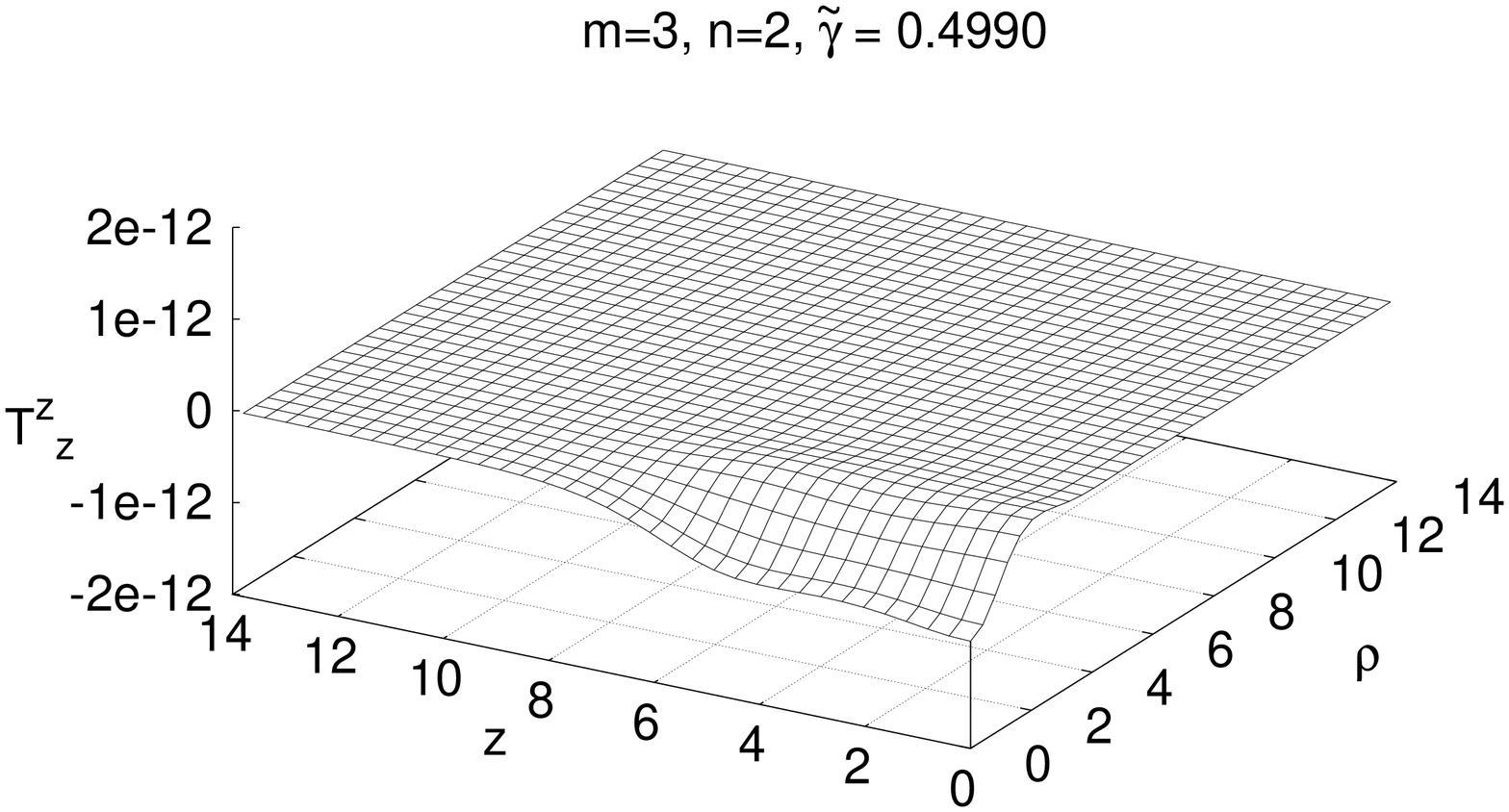}
\end{center}
\vspace{-0.5cm}
\caption{\small
The energy density $-T^t_t$ (a),
the angular momentum density $T^t_\varphi$ (b),
the modulus of the Higgs field $|\Phi|$ (c),
and the stress energy density $T^z_{z}$ (d)
are exhibited for $m=3$, $n=2$ solutions 
with $\tilde\gamma=0$ (left) and $\tilde\gamma \approx 0.5$ (right).
}
\end{figure}

\begin{figure}[h!]
\lbfig{f-7}
\begin{center}
\hspace{0.0cm} (a1)\hspace{-0.6cm}
\includegraphics[width=.45\textwidth, angle =0]{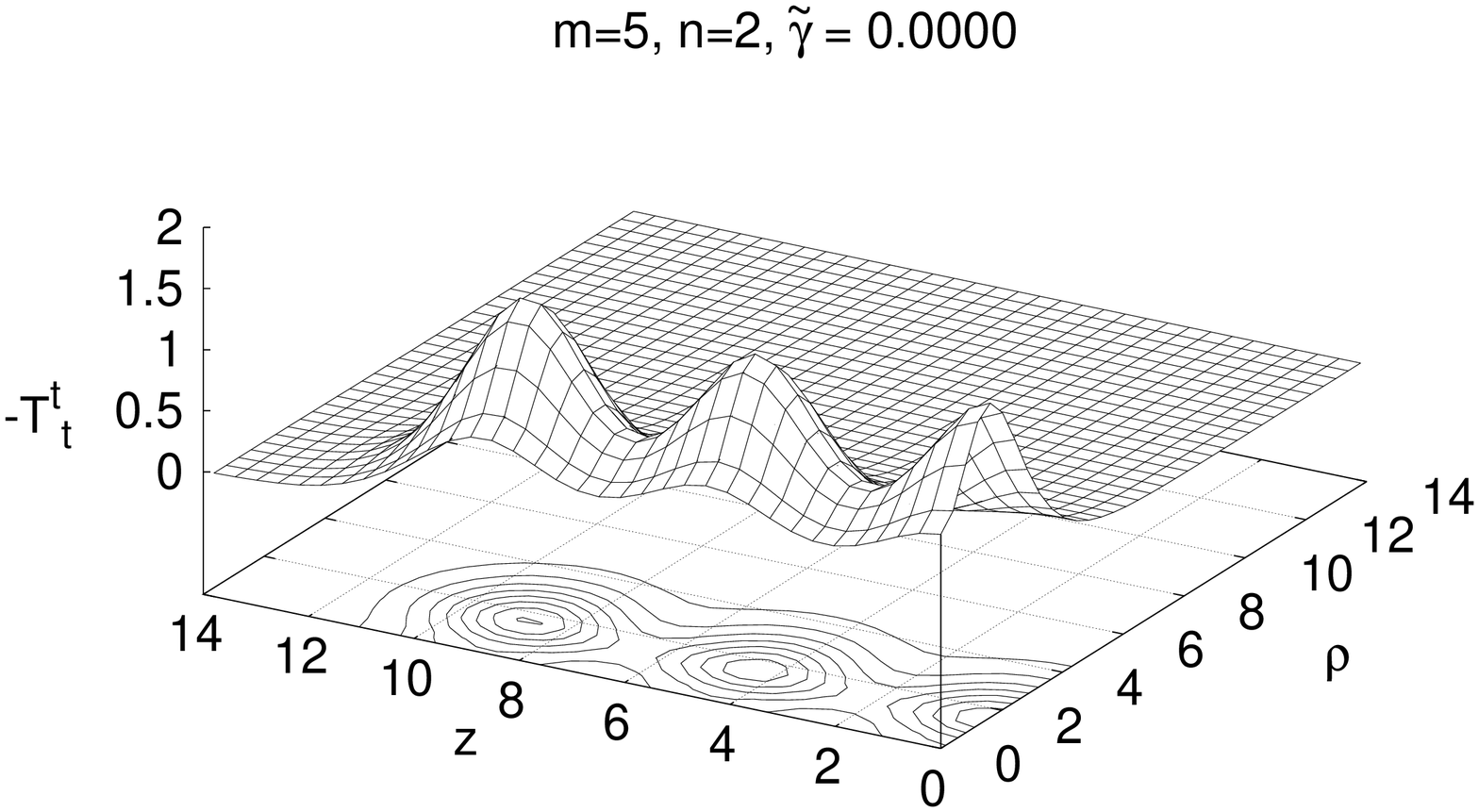}
\hspace{0.0cm} (a2)\hspace{-0.6cm}
\includegraphics[width=.45\textwidth, angle =0]{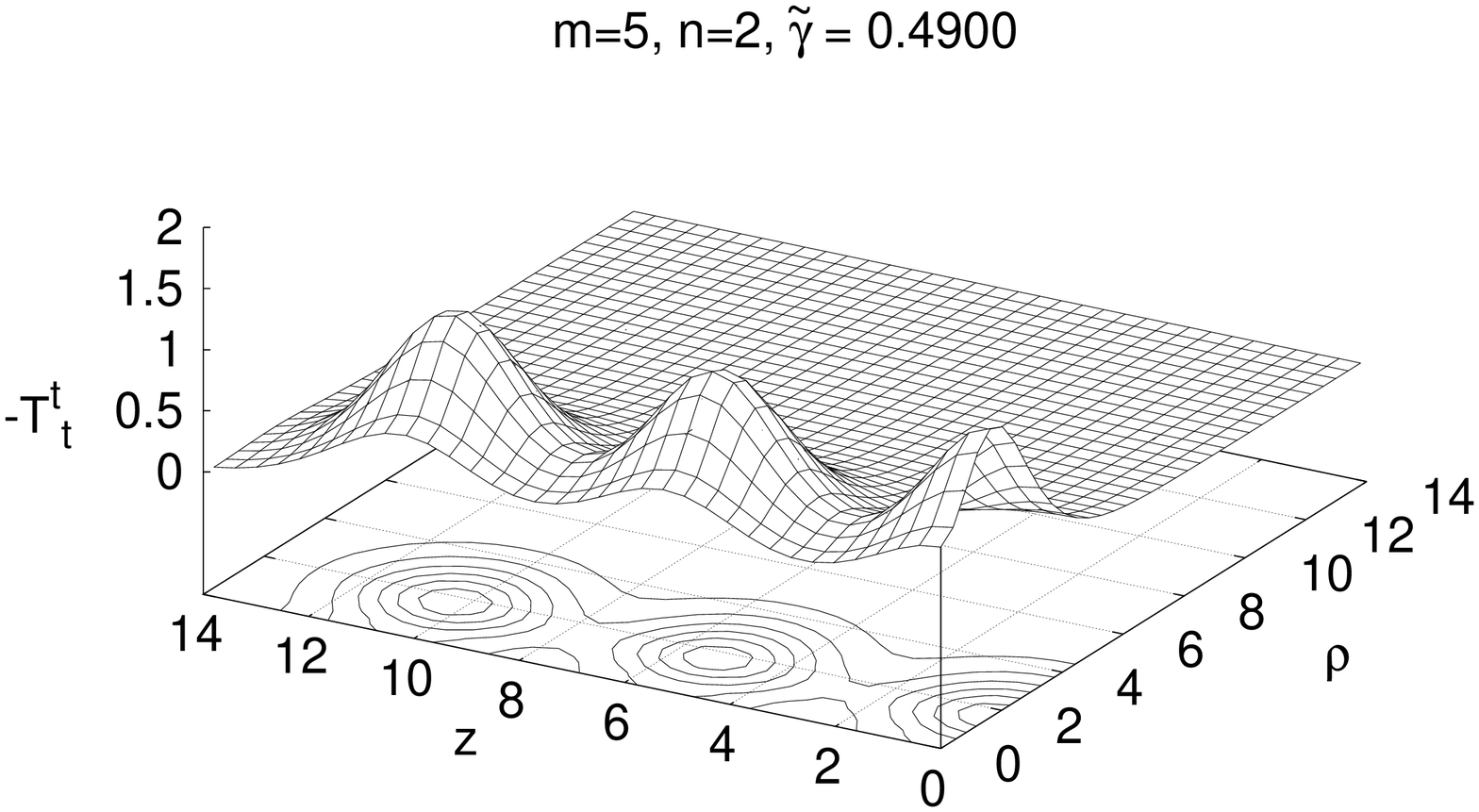}\\
\hspace{0.0cm} (b1)\hspace{-0.6cm}
\includegraphics[width=.45\textwidth, angle =0]{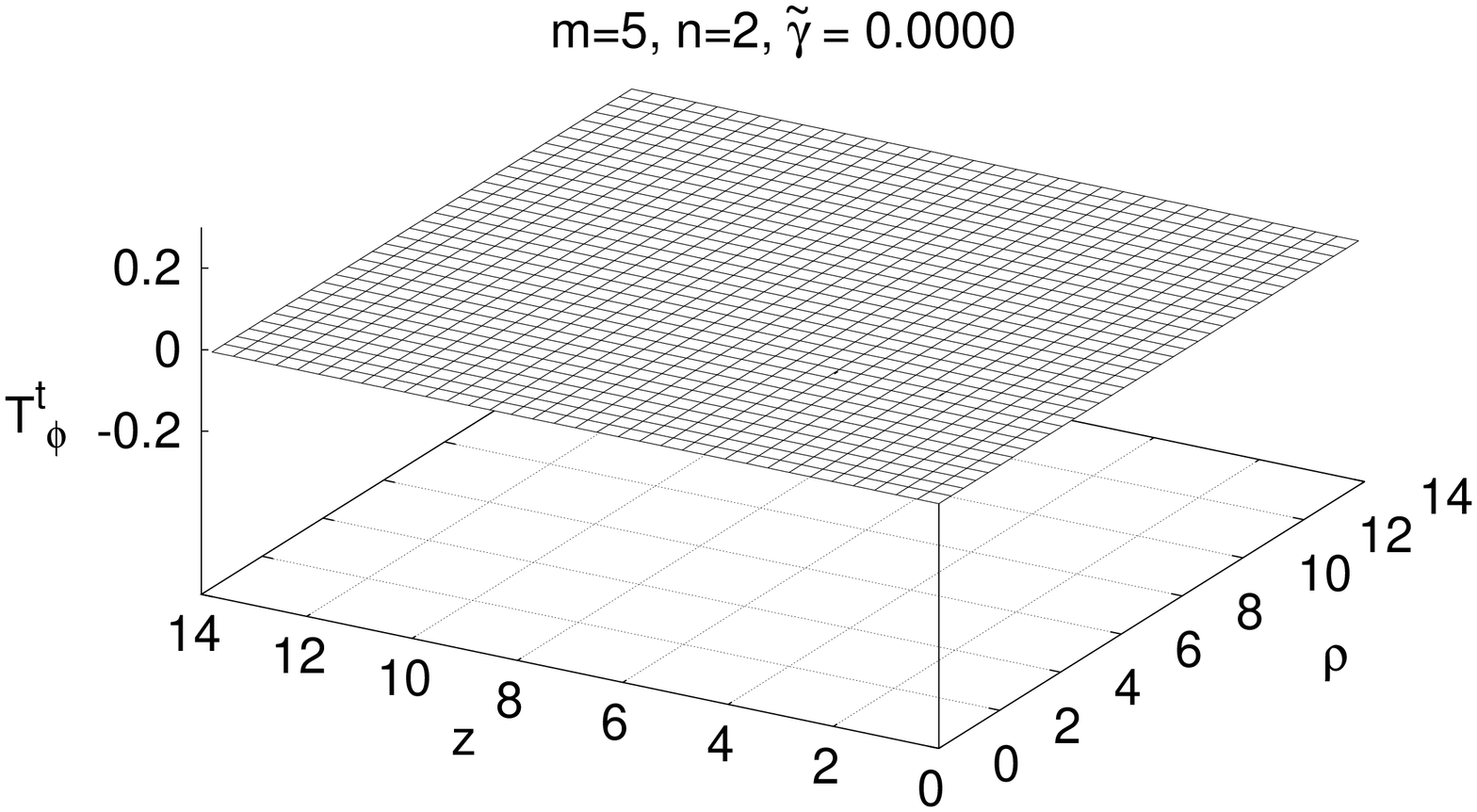}
\hspace{0.0cm} (b2)\hspace{-0.6cm}
\includegraphics[width=.45\textwidth, angle =0]{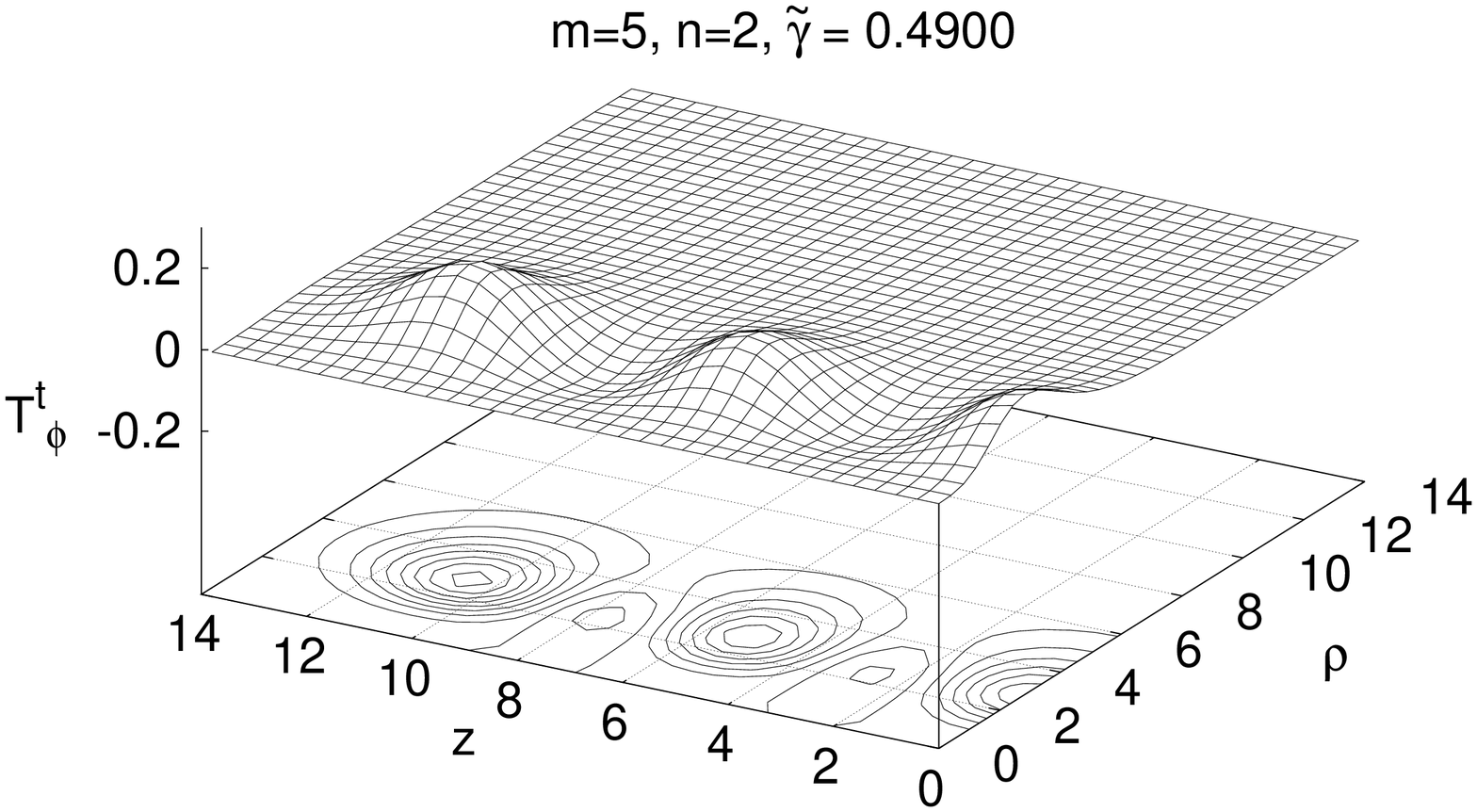}\\
\hspace{0.0cm} (c1)\hspace{-0.6cm}
\includegraphics[width=.45\textwidth, angle =0]{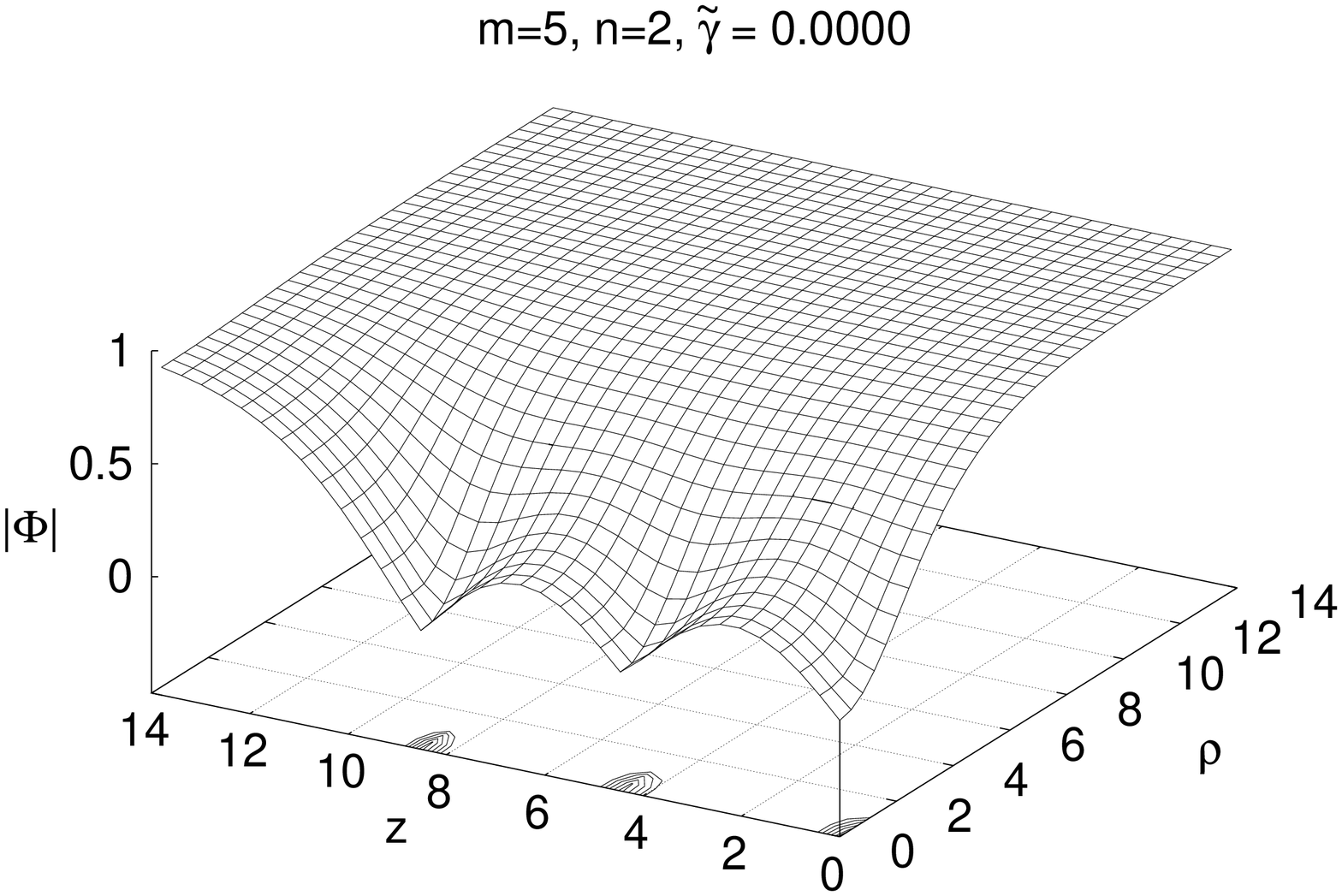}
\hspace{0.0cm} (c2)\hspace{-0.6cm}
\includegraphics[width=.45\textwidth, angle =0]{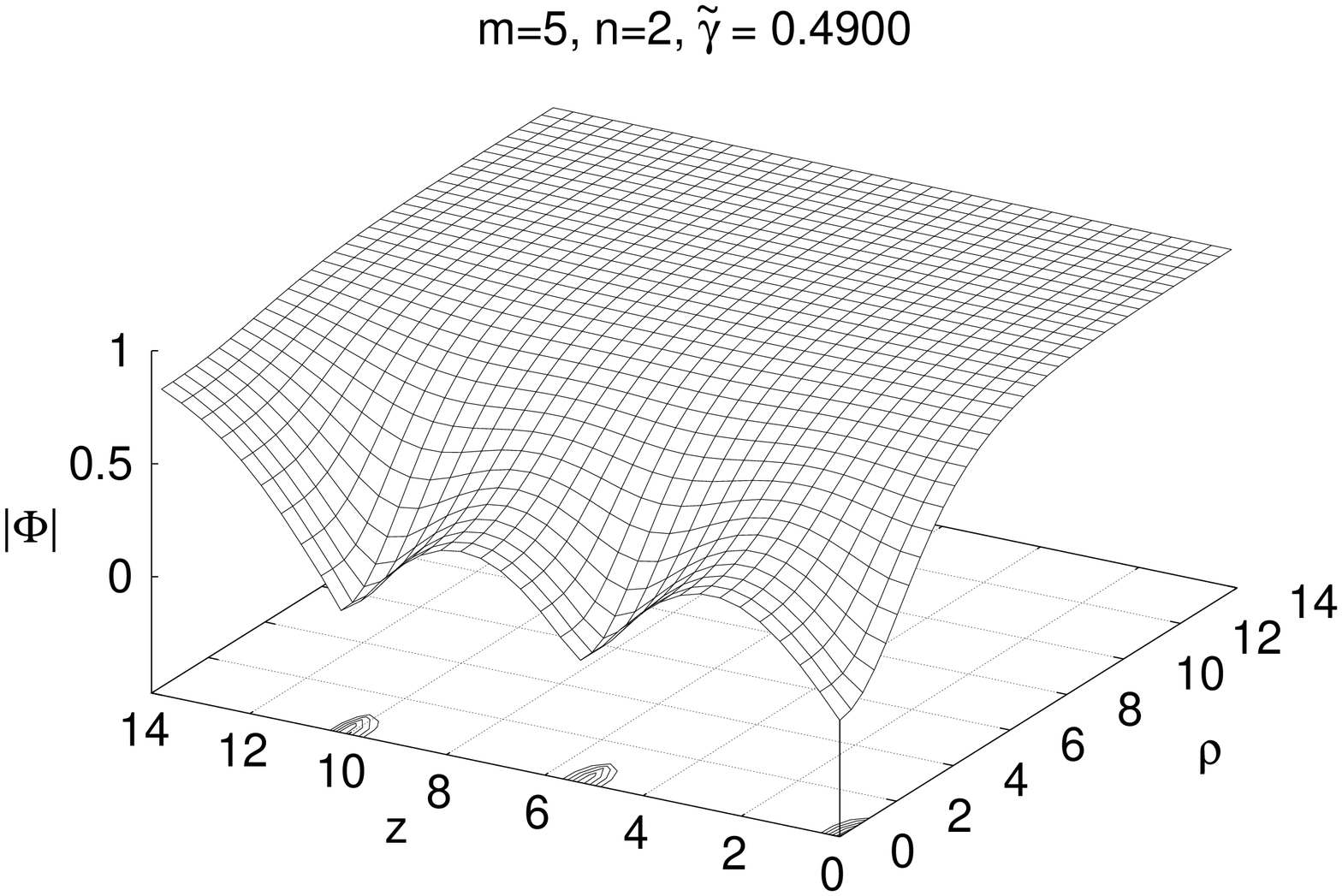}\\
\hspace{0.0cm} (d1)\hspace{-0.6cm}
\includegraphics[width=.45\textwidth, angle =0]{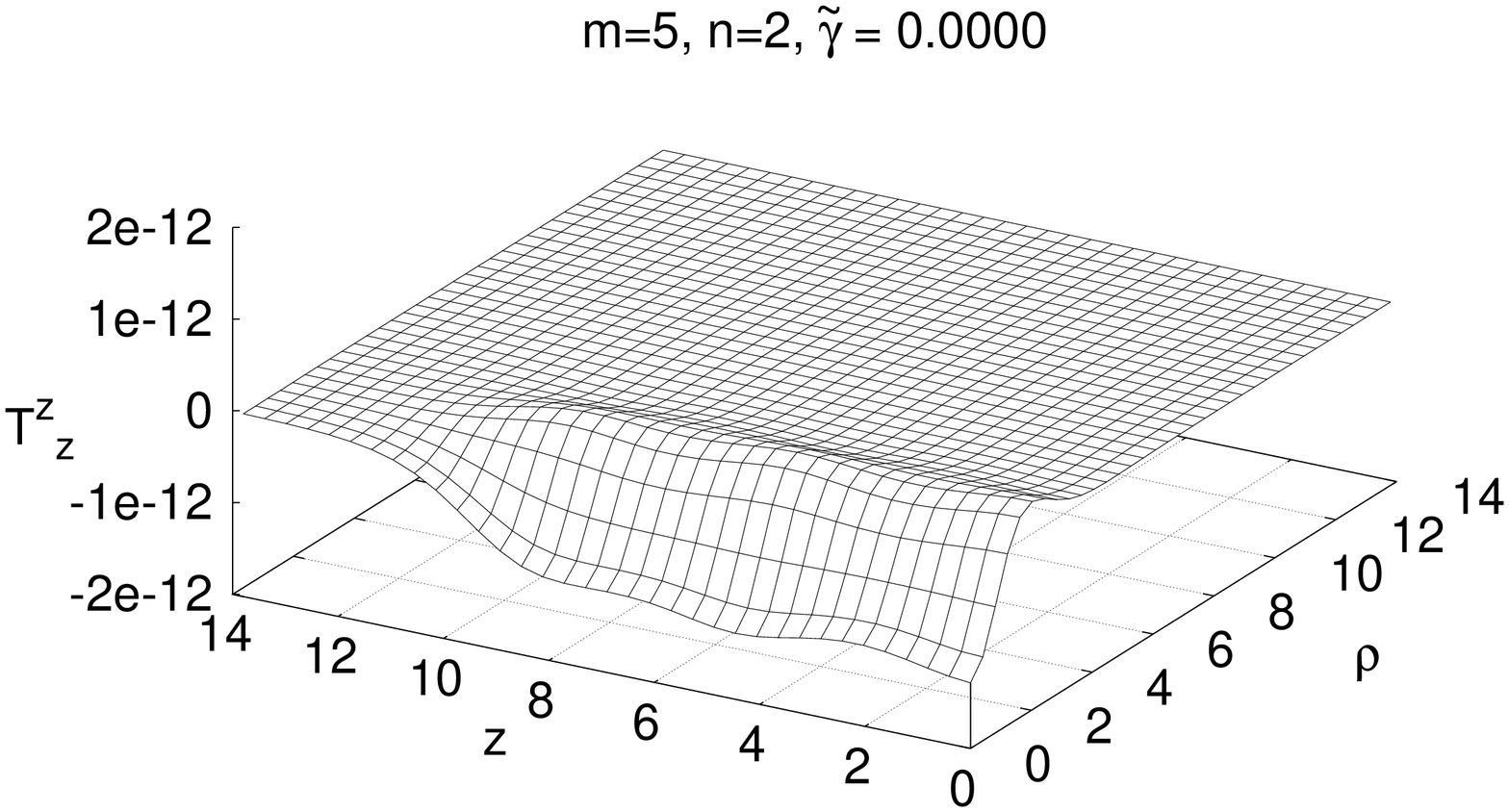}
\hspace{0.0cm} (d2)\hspace{-0.6cm}
\includegraphics[width=.45\textwidth, angle =0]{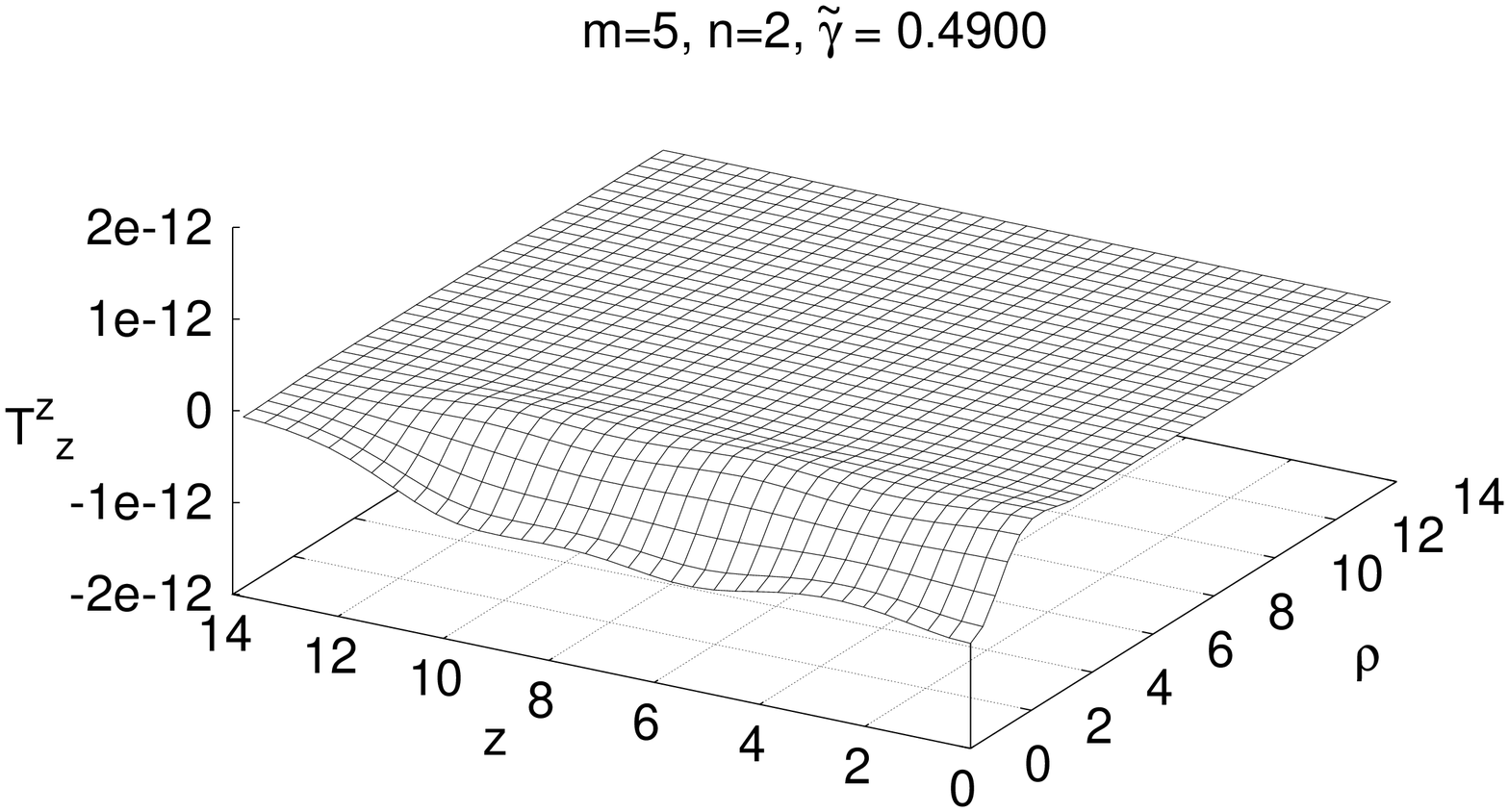}
\end{center}
\vspace{-0.5cm}
\caption{\small
The energy density $-T^t_t$ (a),
the angular momentum density $T^t_\varphi$ (b),
the modulus of the Higgs field $|\Phi|$ (c),
and the stress energy density $T^z_{z}$ (d)
are exhibited for $m=5$, $n=2$ solutions 
with $\tilde\gamma=0$ (left) and $\tilde\gamma \approx 0.5$ (right).
}
\end{figure}

In multisphalerons ($m=1$, $n>1$), the region with large energy density 
is torus-like and the maximum is forming a ring in 
the equatorial plane.
When we add electric charge and angular momentum
to the static configuration, 
we observe that
the energy density is spreading further out,
while at the same time its overall 
magnitude is reduced.
Such a spreading of the energy density with increasing charge is
also seen in dyons, for instance.
We therefore attribute this effect to the presence of charge
and the associated repulsion.
Indeed, this spreading
becomes quite pronounced for large values of the charge.
The expected effect of the presence of angular momentum,
on the other hand,
is a centrifugal shift of the energy density.
Indeed, we observe, that with increasing angular momentum
the torus-like region of large energy density
moves further outward to larger values of $\rho$.
These effects are seen in Fig.~\ref{f-5}a,
where we exhibit the energy density $-T^t_t$
for a multisphaleron solution ($m=1$, $n=2$)
in the static case ($\tilde\gamma=0$)
and for the almost maximally rotating case ($\tilde\gamma \approx 0.5$).
In Fig.~\ref{f-5}b-d we also exhibit
the magnitude of the Higgs field $|\Phi|$,
the angular momentum density $T^t_\varphi$,
and the stress energy density component $T^z_z$
for
these two solutions.
%
We note, that in contrast to our previous paper \cite{Ibadov:2010ei} we decided to change the definition of $T^z_z$ by a constant factor of $1/(gv)^5$, so it refers to a dimensionless $z=\tilde r\cos\theta$.

The modulus of the Higgs field of the multisphaleron
solutions has a single node at the origin,
from where it starts to increase linearly 
in the direction of the symmetry axis,
to reach its vacuum expectation value at infinity.
In the equatorial plane, in contrast,
the modulus of the Higgs field starts to
increase from the origin much more slowly (i.e., only with power $\rho^n$).
As the configurations are endowed with
charge and rotation, the Higgs field changes only slightly.
Indeed, the effect of charge and rotation on
the modulus of the Higgs field is barely noticeable in Fig.~\ref{f-5}c
even at the maximal strength.

The angular momentum density for the multisphaleron solutions
is torus-like and centered in
the equatorial plane analogous to the energy density.
However, the region of large angular momentum density
is located further outwards at larger values of $\rho$,
while it vanishes on the symmetry axis.
The angular momentum density $T^t_\varphi$
for the multisphaleron solution ($m=1$, $n=2$)
for the maximally rotating case ($\gamma \approx 0.5$) is seen in 
Fig.~\ref{f-5}b2.


Let us now turn to sphaleron-antisphaleron systems.
For $n=1$ they represent sphaleron-antisphaleron chains,
where $m$ sphalerons and antisphalerons are located
on the symmetry axis, in static equilibrium.
For $n=2$ the chain is formed
from  $m$ multisphalerons and -antisphalerons,
thus the modulus of the Higgs field
still possesses only isolated nodes on the symmetry axis.

We demonstrate the effect of charge
and rotation on these sphaleron-antisphaleron chains
with $n=2$, exhibiting the configurations with $m=3$ and
$m=5$ in Fig.~\ref{f-6} and Fig.~\ref{f-7}, respectively.
Associated with each multisphaleron and multiantisphaleron
is a torus-like part of the energy density.
Thus a configuration with $m$ multi(anti)sphalerons
has $m$ tori, located symmetrically with respect to 
the equatorial plane.
As for the single multisphaleron,
we observe that
when we add electric charge and angular momentum
to the static configuration,
the energy density is spreading further out,
while its overall magnitude is reduced.
Indeed, we observe, that 
the torus-like regions of large energy density
move further outward to larger values of $\rho$
and further from each other to larger values of $|z|$.
These effects are seen in Fig.~\ref{f-6}a and Fig.~\ref{f-7}a.
As clearly observable in Fig.~\ref{f-6}c and Fig.~\ref{f-7}c,
the nodes of the Higgs field move further apart,
as the chains become charged and the rotation sets in.

Let us turn now to more complicated 
sphaleron-antisphaleron systems.
As $n$ increases beyond the value two, the character of the solutions
changes, and new types of configurations appear,
where the modulus of the Higgs field vanishes on rings
centered around the symmetry axis.
Therefore we refer to these solutions as vortex ring solutions.
We note, that the precise evolution of the isolated nodes 
on the symmetry axis and the vortex rings in the bulk
with increasing $n$ is somewhat sensitive
to the value of the Higgs mass \cite{Kleihaus:2008gn}.

The nodes of the sphaleron-antisphaleron systems
in the range $m=2-6$ and $n=3-6$ are exhibited in Fig.~\ref{f-11}
for vanishing charge and rotation.
For the chosen parameters, the modulus of the Higgs field
of the simplest vortex ring
configuration, i.e., the system with $m=2$, $n=3$,
vanishes on a single ring located in the equatorial plane.
As the winding number $n$ increases, this single ring merely 
increases in size, as seen in the figure.
For $n\ge5$ this increase is to a high degree linear.

\begin{figure}[h!]
\lbfig{f-11}
\begin{center}
\includegraphics[width=.245\textwidth, angle =0]{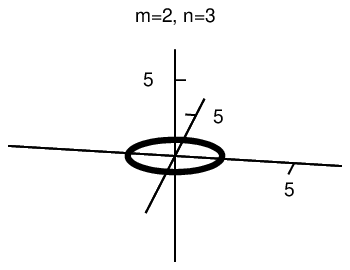}
\includegraphics[width=.245\textwidth, angle =0]{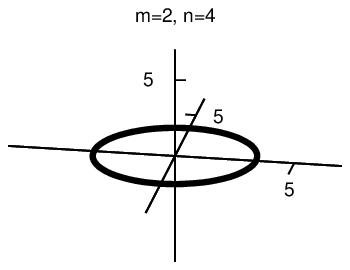}
\includegraphics[width=.245\textwidth, angle =0]{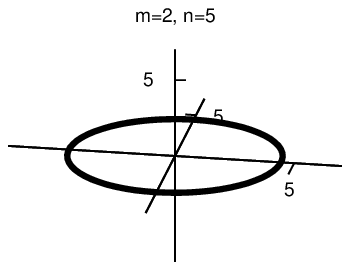}
\includegraphics[width=.245\textwidth, angle =0]{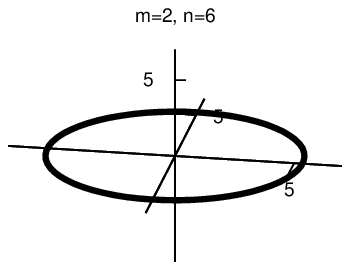} \\
\includegraphics[width=.245\textwidth, angle =0]{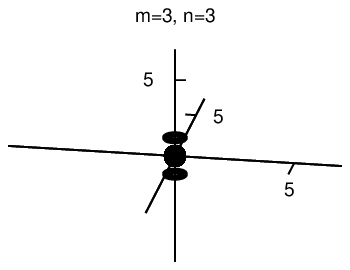}
\includegraphics[width=.245\textwidth, angle =0]{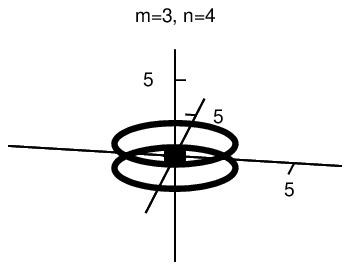}
\includegraphics[width=.245\textwidth, angle =0]{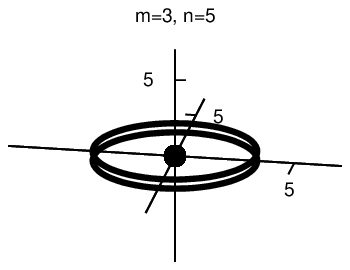}
\includegraphics[width=.245\textwidth, angle =0]{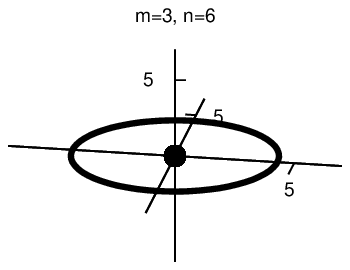} \\
\includegraphics[width=.245\textwidth, angle =0]{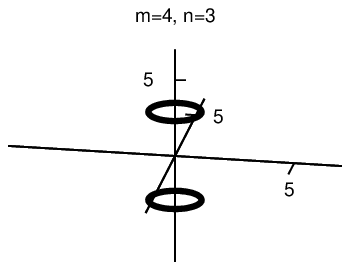}
\includegraphics[width=.245\textwidth, angle =0]{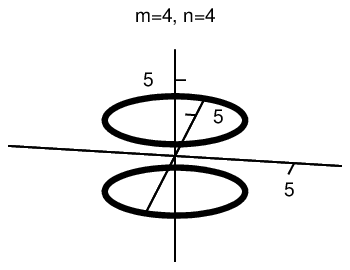}
\includegraphics[width=.245\textwidth, angle =0]{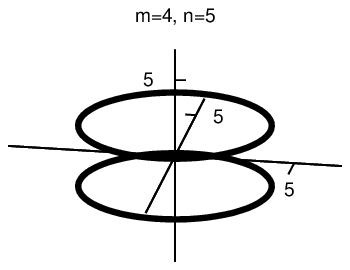}
\includegraphics[width=.245\textwidth, angle =0]{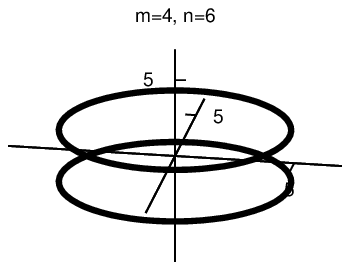} \\
\includegraphics[width=.245\textwidth, angle =0]{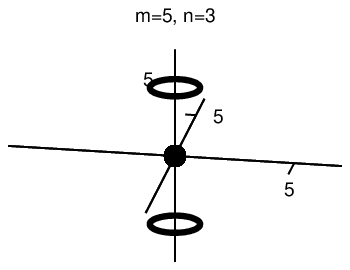}
\includegraphics[width=.245\textwidth, angle =0]{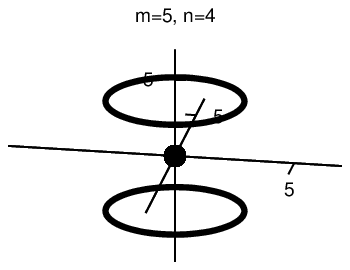}
\includegraphics[width=.245\textwidth, angle =0]{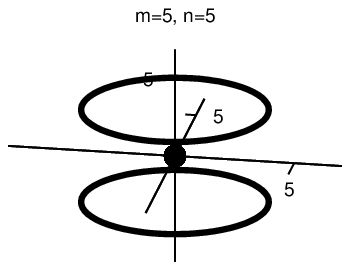}
\includegraphics[width=.245\textwidth, angle =0]{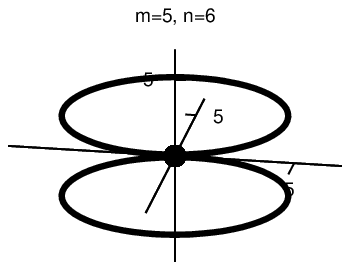} \\
\includegraphics[width=.245\textwidth, angle =0]{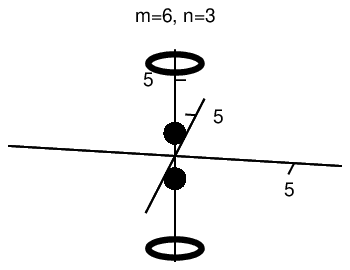}
\includegraphics[width=.245\textwidth, angle =0]{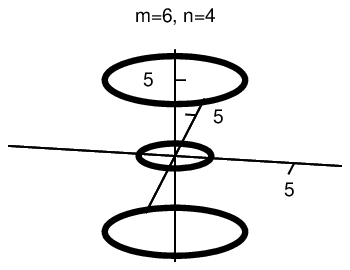}
\includegraphics[width=.245\textwidth, angle =0]{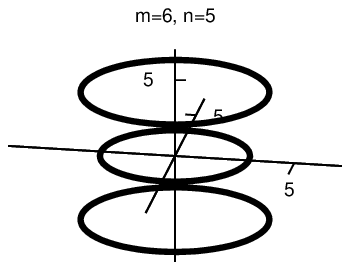}
\includegraphics[width=.245\textwidth, angle =0]{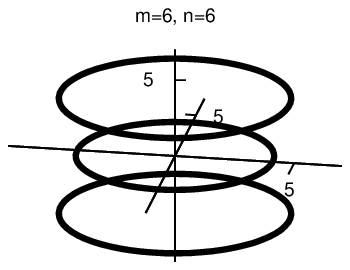}
\end{center}
\caption{\small
The nodes of the Higgs field
for the static sphaleron-antisphaleron systems
with $m=2-6$, $n=3-6$.
Rows: fixed $m$, columns: fixed $n$.
}
\end{figure}

The $m=3$, $n=3$ configuration
has one node at the origin and in addition two tiny rings,
located symmetrically above and below the $xy$-plane.
As $n$ increases, the rings grow in size and move towards each other.
For $n=6$ the rings have merged into a single ring in the
equatorial plane. 
This ring then grows in size as $n$ increases further,
while the central node is always retained.

The static $m=4$, $n=3$ configuration has two vortex rings,
located symmetrically with respect to the $xy$-plane.
While the rings increase in size with increasing $n$,
they hardly change their mutual distance for intermediate
values of $n$, as depicted in the figure.
To see the further evolution of the nodes and be able to
decide, whether they merge into a single ring in the
equatorial plane, we have continued the calculations
up to $n=72$. Here the rings are already rather close,
but they have still not merged. Extrapolating 
the curve $z_{\rm ring}(n)$ indicates, that the merging 
may happen only beyond $n=100$.

For $m=5$, $n=3$ there are two symmetrically located vortex rings
supplemented by a node at the origin.
As $n$ increases, the rings again increase in size,
while retaining at first roughly their distance.
However, they slowly move towards each other and
merge into a single ring at $n=37$.

For $m=6$, $n=3$ there are two symmetrical vortex rings and two inner nodes
on the symmetry axis.
For $n=4$ the inner nodes have already formed a ring in the
equatorial plane.
With increasing $n$ all 3 rings then grow in size.
For $n=83$ the outer rings have come closer to the equatorial plane,
but merging has not yet taken place.
An extrapolation of $z_{\rm ring}(n)$ and $\rho_{\rm ring}(n)$
for the 3 rings shows, that the merging of the rings
will most likely not happen until $n$ is well above $100$.
We display $z_{\rm ring}(n)$ and $\rho_{\rm ring}(n)$ in Fig.\ref{f-14}
for the solutions with $m=4,5,6$.

\begin{figure}[h!]
\lbfig{f-14}
\begin{center}
\hspace{0.0cm} (a)\hspace{-0.6cm}
\includegraphics[width=.45\textwidth, angle =0]{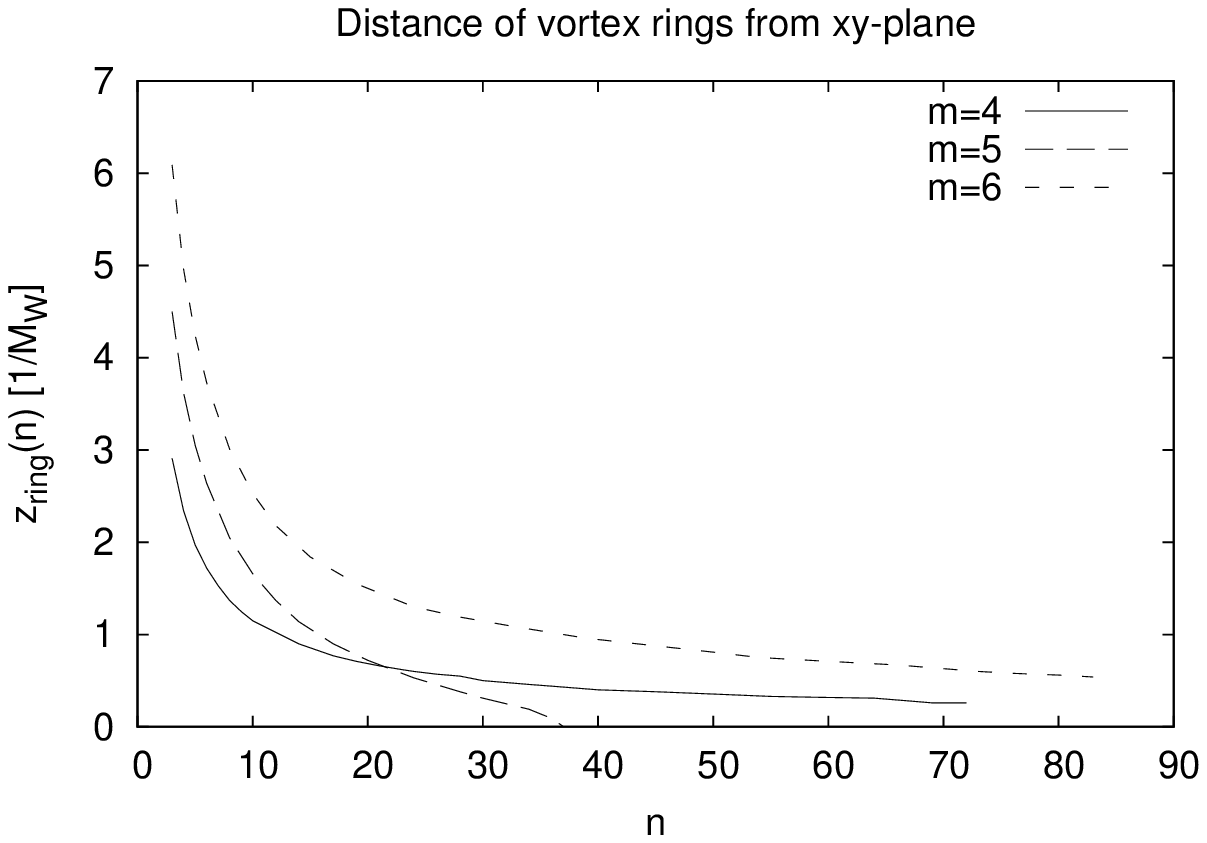}
\hspace{0.0cm} (b)\hspace{-0.6cm}
\includegraphics[width=.45\textwidth, angle =0]{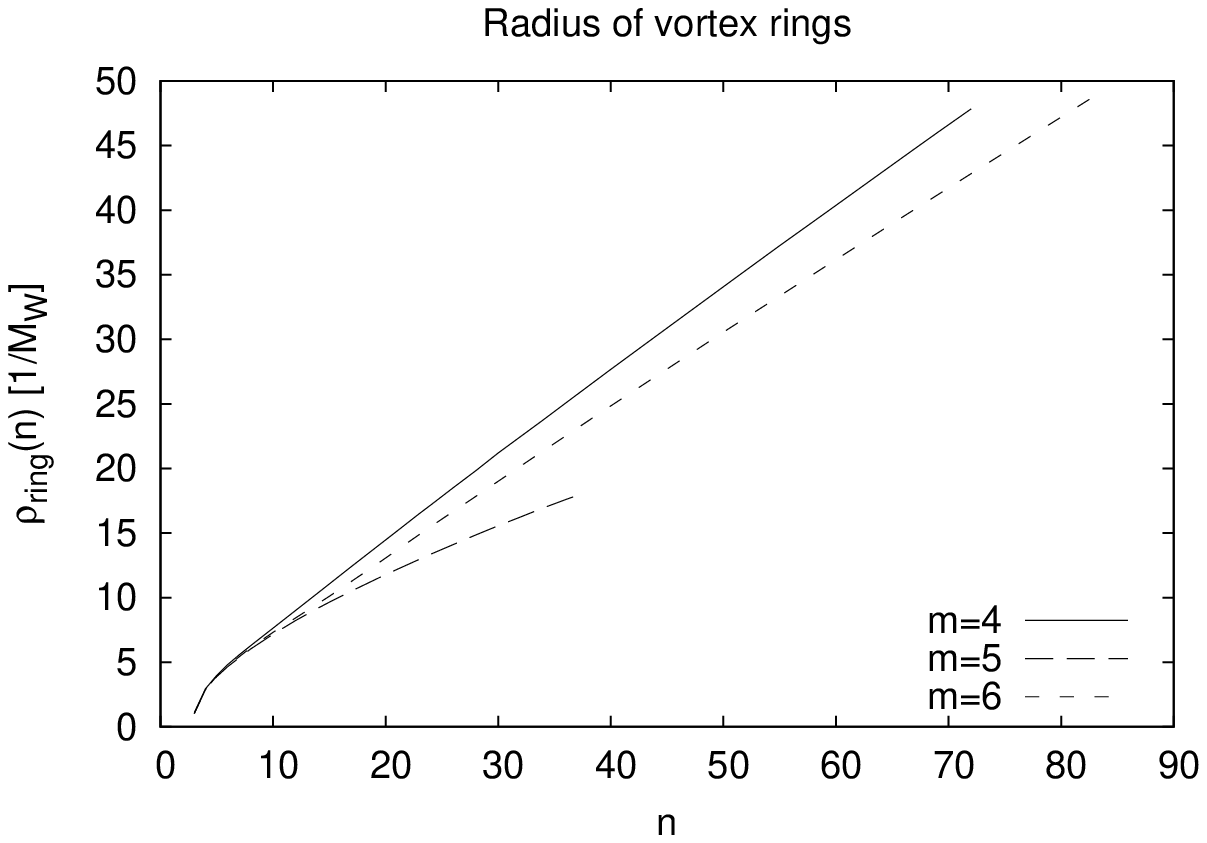}
\end{center}
\vspace{-0.5cm}
\caption{\small
Distance of vortex rings from $xy$-plane (a) and their radius (b) for $m=4,5,6$, $\tilde\gamma=0$ as a function of winding number $n$. The radius of the central ring of the $m=6$ solutions is not shown.
}
\end{figure}

In the following we demonstrate the effect of charge
and rotation on these sphaleron-antisphaleron systems.
Concerning the location and type of nodes, we observe
only a small effect.
For $m=2$, $n=3$ the location of the rings changes by about 20\%,
between the static case and maximal rotation, but typically the changes are on the order of 10\%.
Only in the vicinity of transitions between the numbers and types 
of nodes, the effects of charge and rotation on the nodes
become rather important.

Let us now illustrate our discussion of the
effect of charge and rotation by considering the configurations 
$m=6$ and $n=3,4,5$. 
These are exhibited in Figs.~\ref{f-8} to \ref{f-10},
where we again display the energy density $-T^t_t$,
the modulus of the Higgs field $|\Phi|$,
angular momentum density $T^t_\varphi$,
and the stress-energy density component $T^z_z$
for static solutions and solutions with maximal rotation.

\begin{figure}[h!]
\lbfig{f-8}
\begin{center}
\hspace{0.0cm} (a1)\hspace{-0.6cm}
\includegraphics[width=.45\textwidth, angle =0]{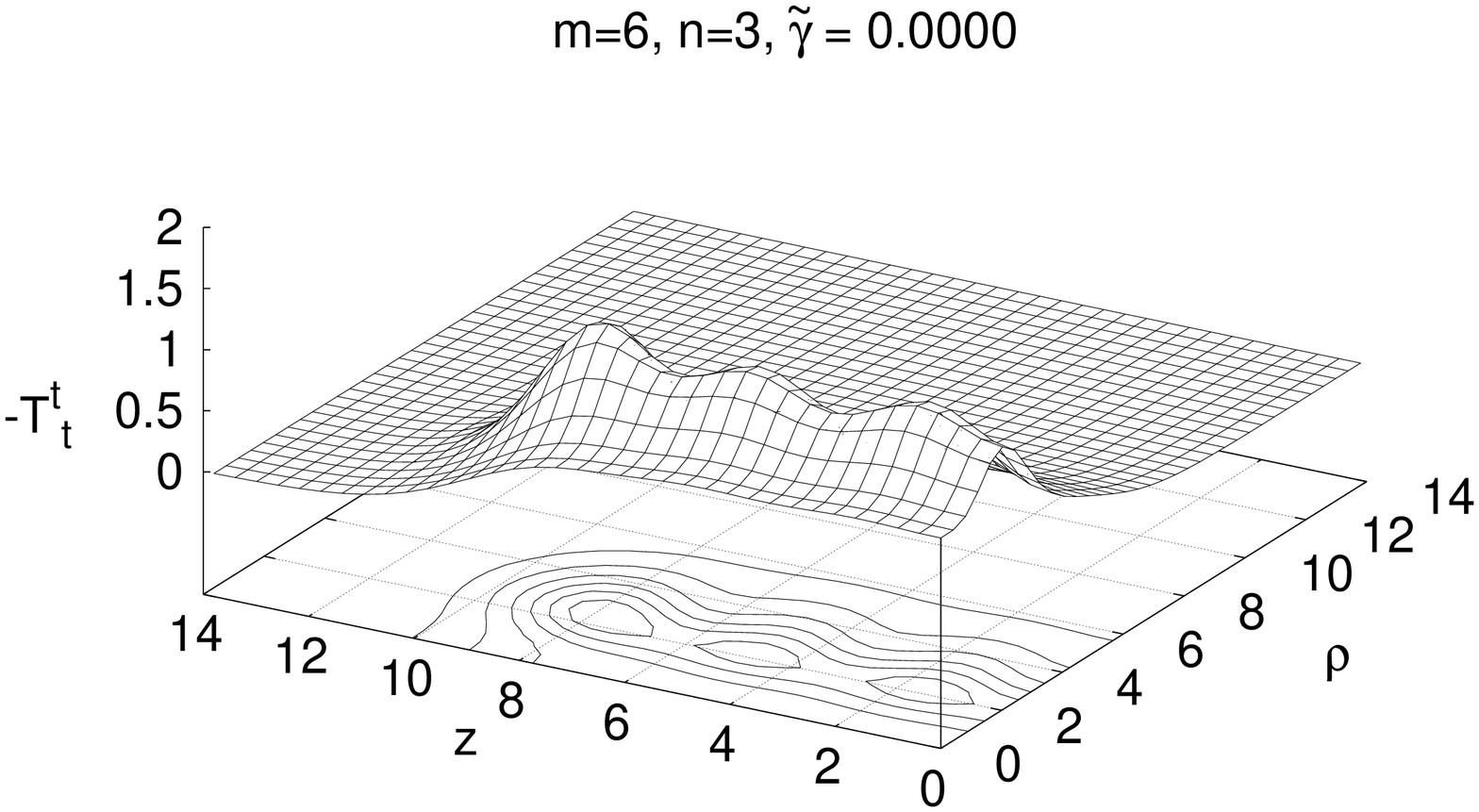}
\hspace{0.0cm} (a2)\hspace{-0.6cm}
\includegraphics[width=.45\textwidth, angle =0]{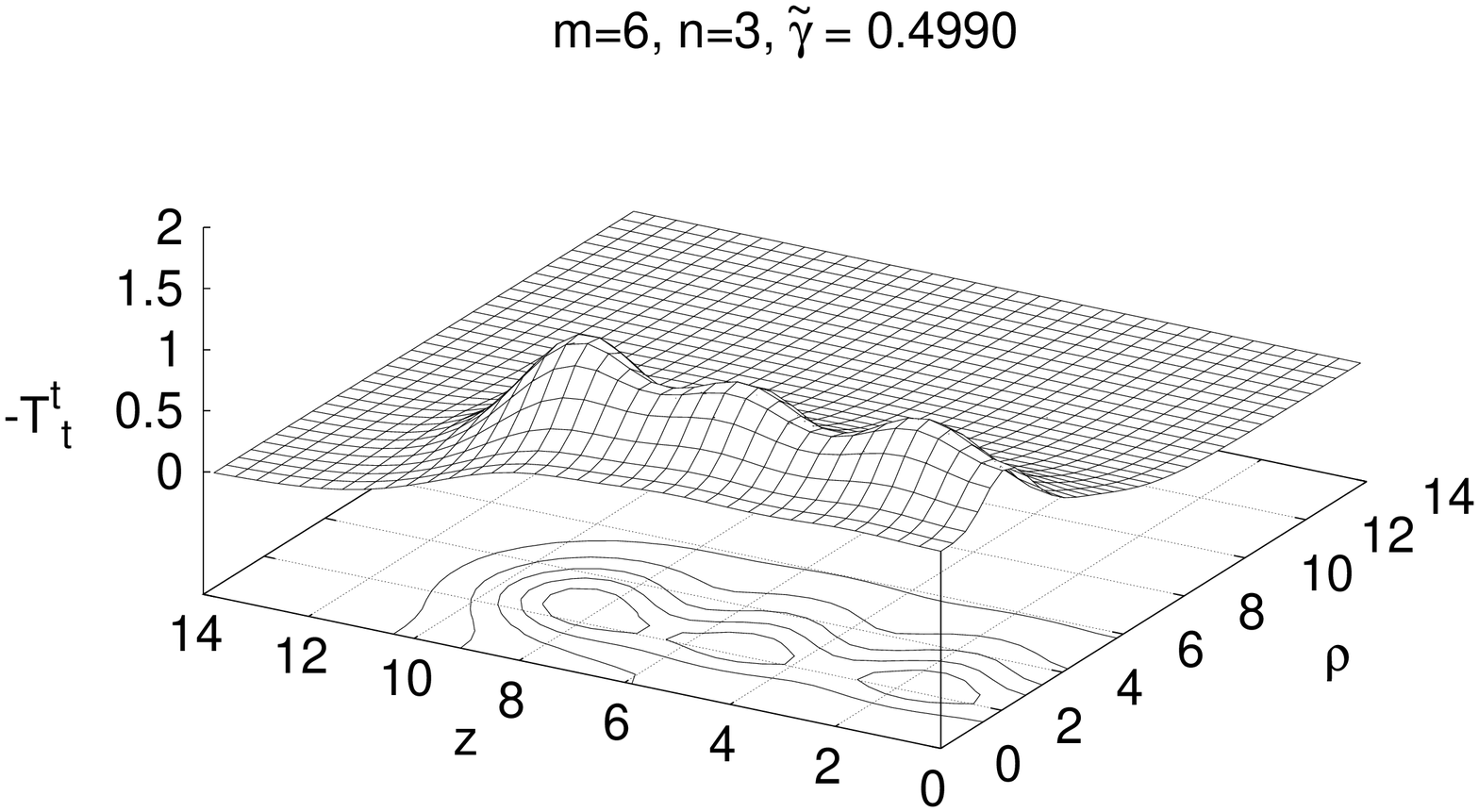}\\
\hspace{0.0cm} (b1)\hspace{-0.6cm}
\includegraphics[width=.45\textwidth, angle =0]{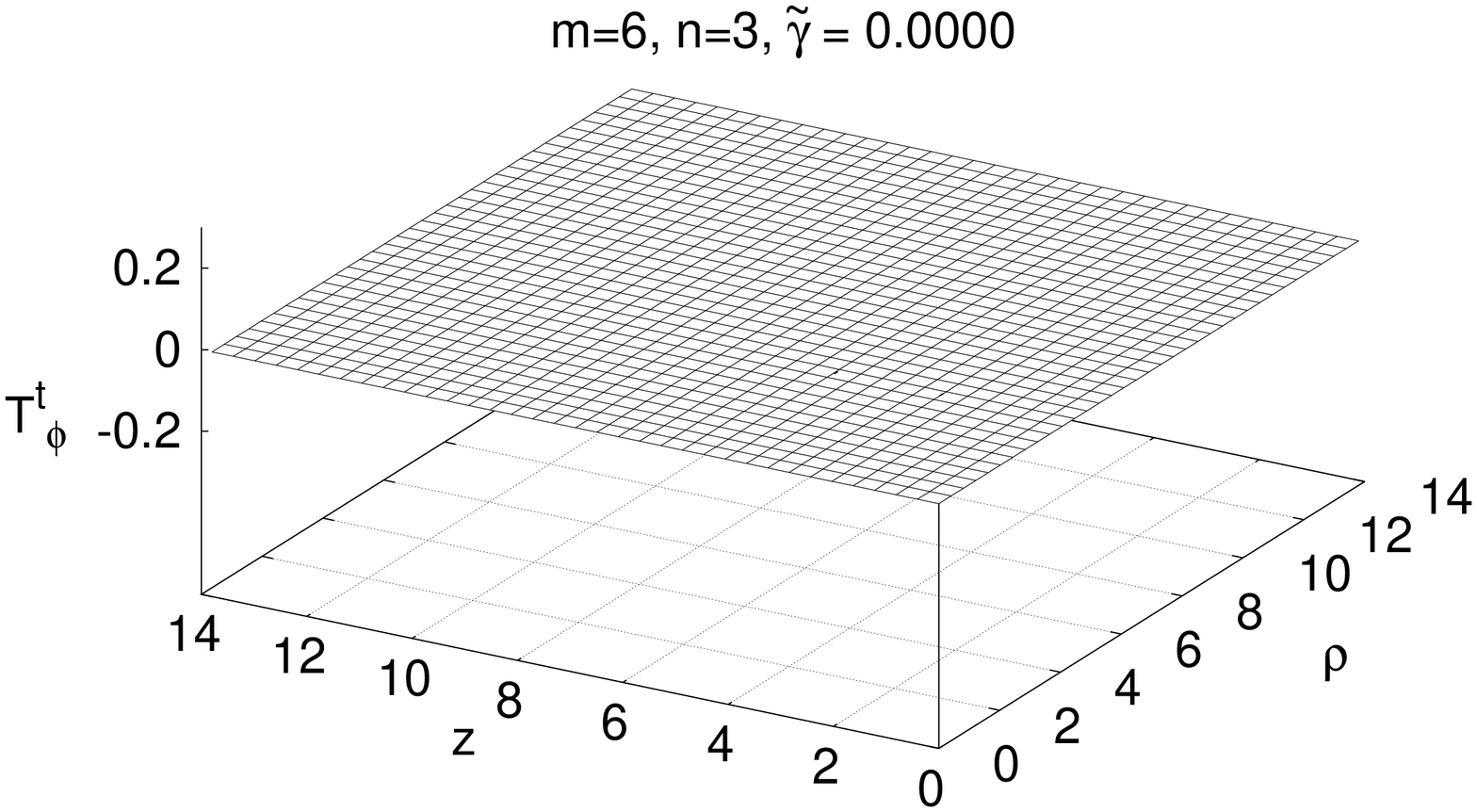}
\hspace{0.0cm} (b2)\hspace{-0.6cm}
\includegraphics[width=.45\textwidth, angle =0]{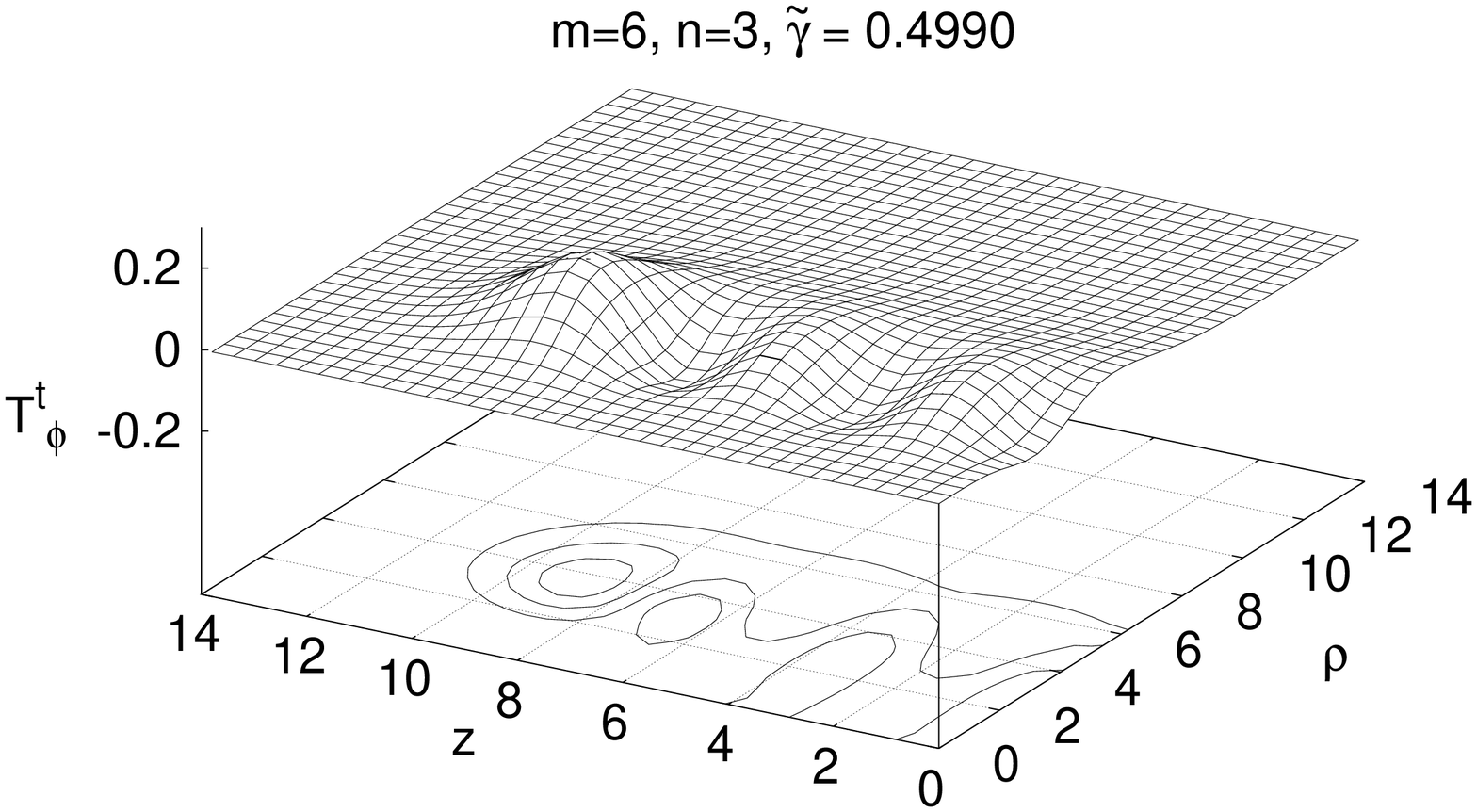}\\
\hspace{0.0cm} (c1)\hspace{-0.6cm}
\includegraphics[width=.45\textwidth, angle =0]{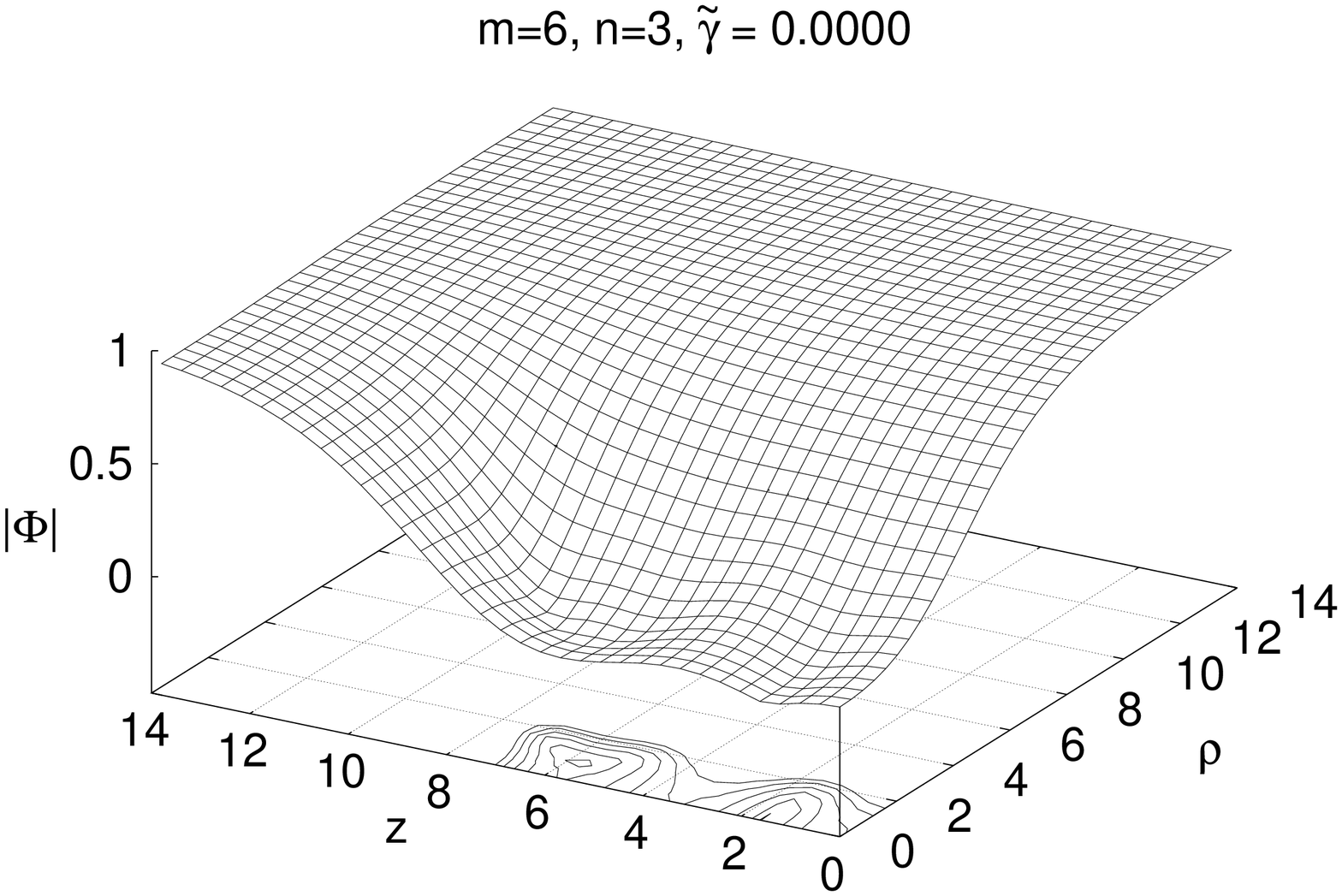}
\hspace{0.0cm} (c2)\hspace{-0.6cm}
\includegraphics[width=.45\textwidth, angle =0]{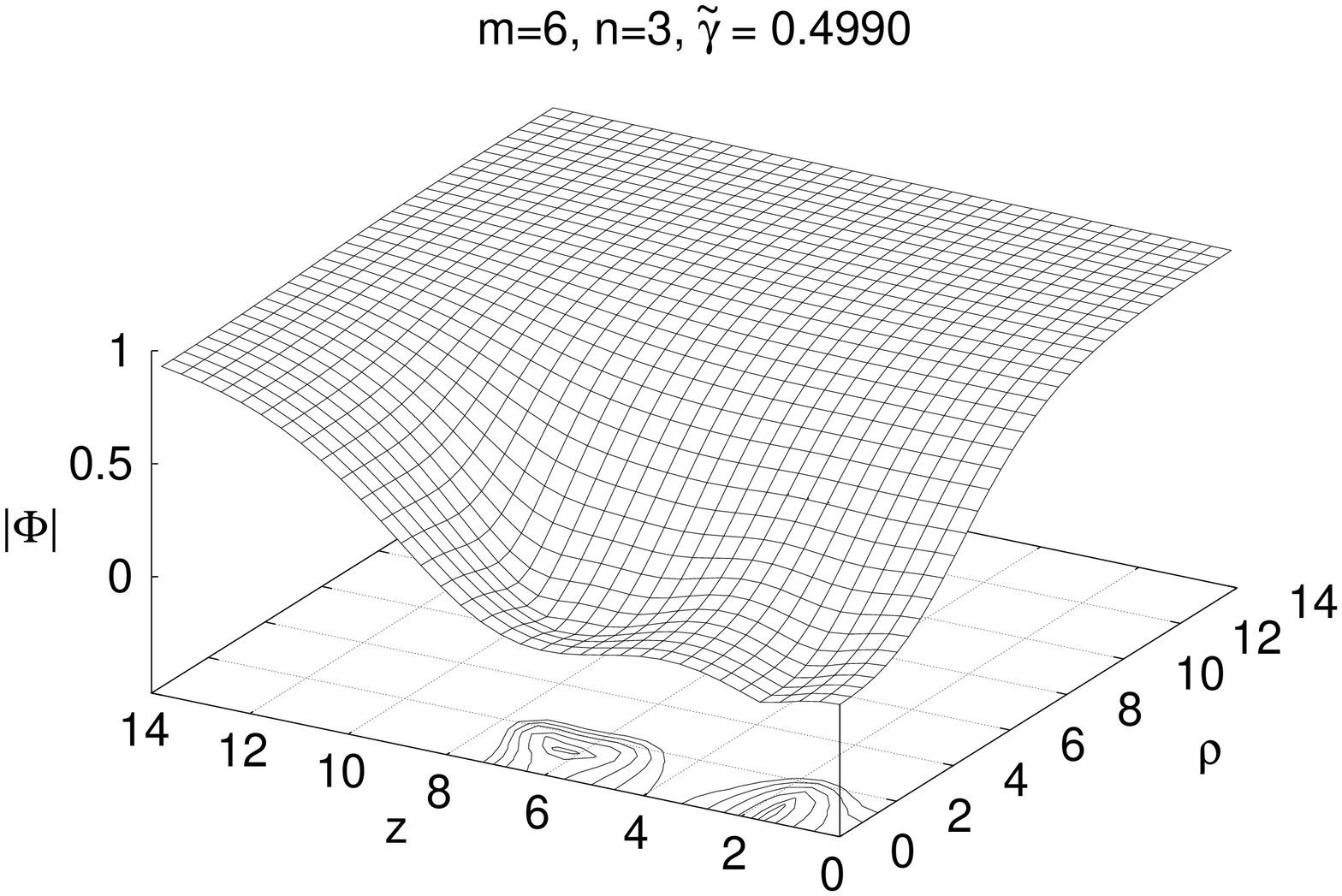}\\
\hspace{0.0cm} (d1)\hspace{-0.6cm}
\includegraphics[width=.45\textwidth, angle =0]{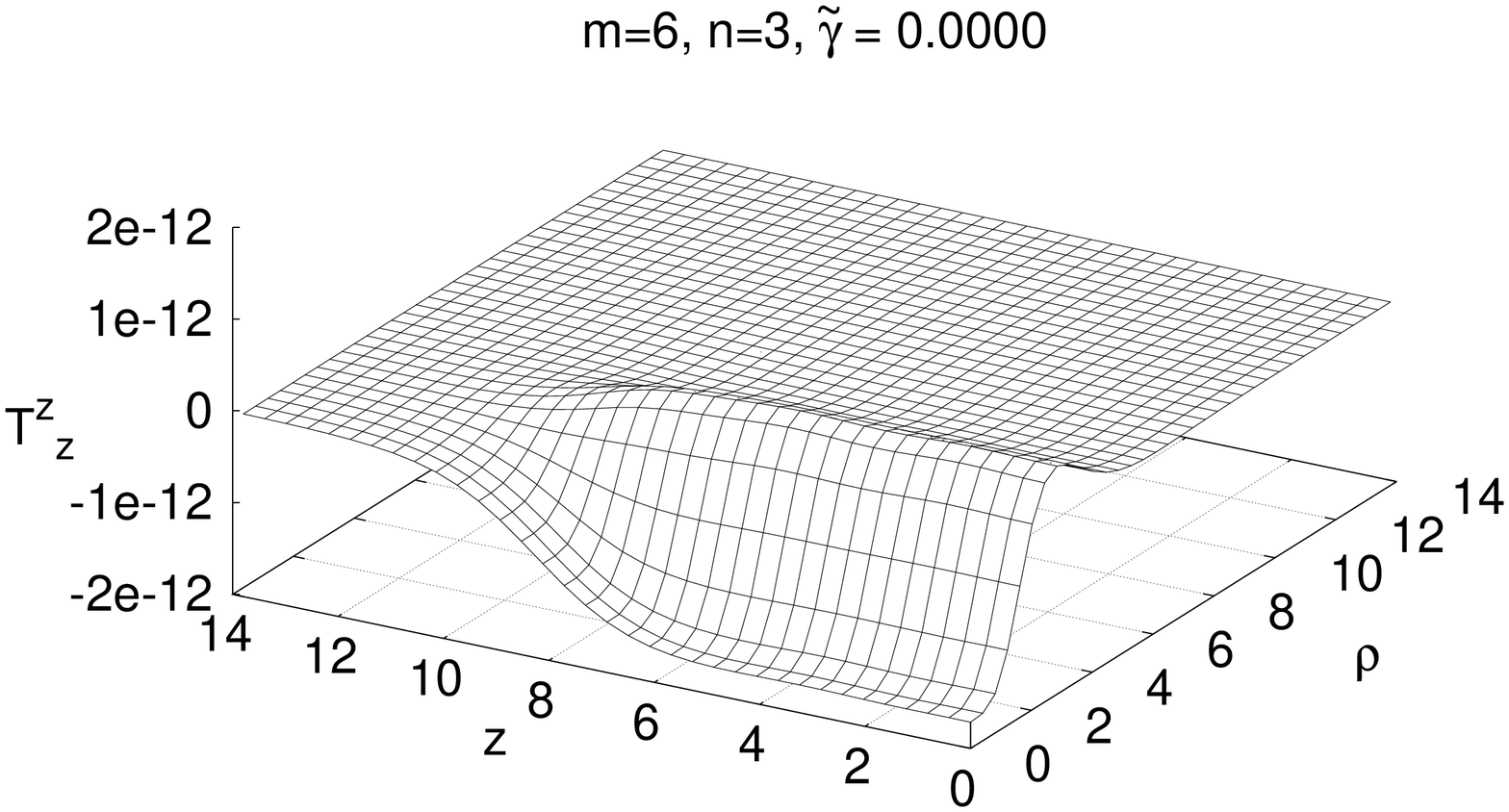}
\hspace{0.0cm} (d2)\hspace{-0.6cm}
\includegraphics[width=.45\textwidth, angle =0]{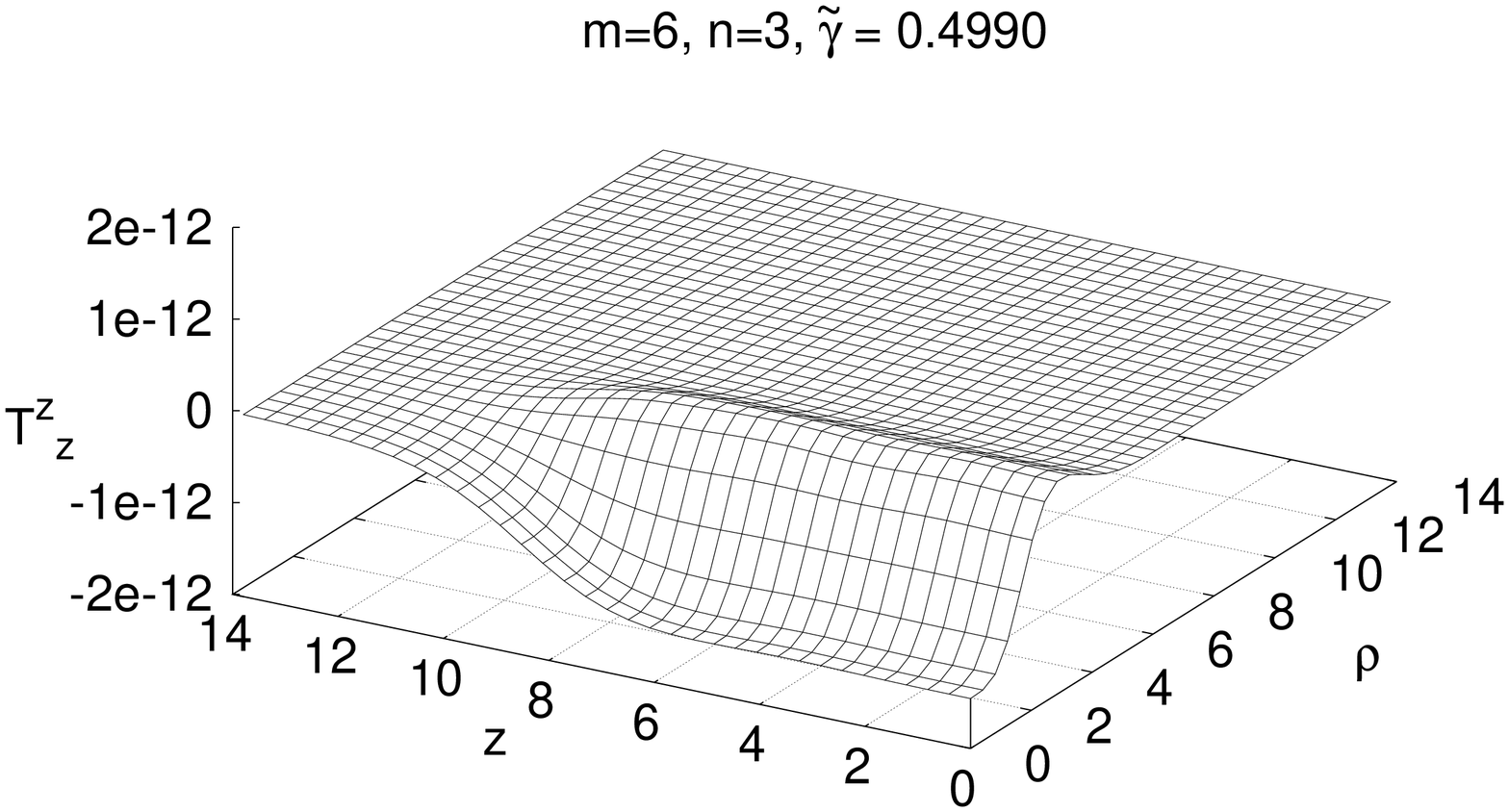}
\end{center}
\vspace{-0.5cm}
\caption{\small
The energy density $-T^t_t$ (a),
the angular momentum density $T^t_\varphi$ (b),
the modulus of the Higgs field $|\Phi|$ (c),
and the stress energy density $T^z_{z}$ (d)
are exhibited for $m=6$, $n=3$ solutions 
with $\tilde\gamma=0$ (left) and $\tilde\gamma \approx 0.5$ (right).
}
\end{figure}

\begin{figure}[h!]
\lbfig{f-9}
\begin{center}
\hspace{0.0cm} (a1)\hspace{-0.6cm}
\includegraphics[width=.45\textwidth, angle =0]{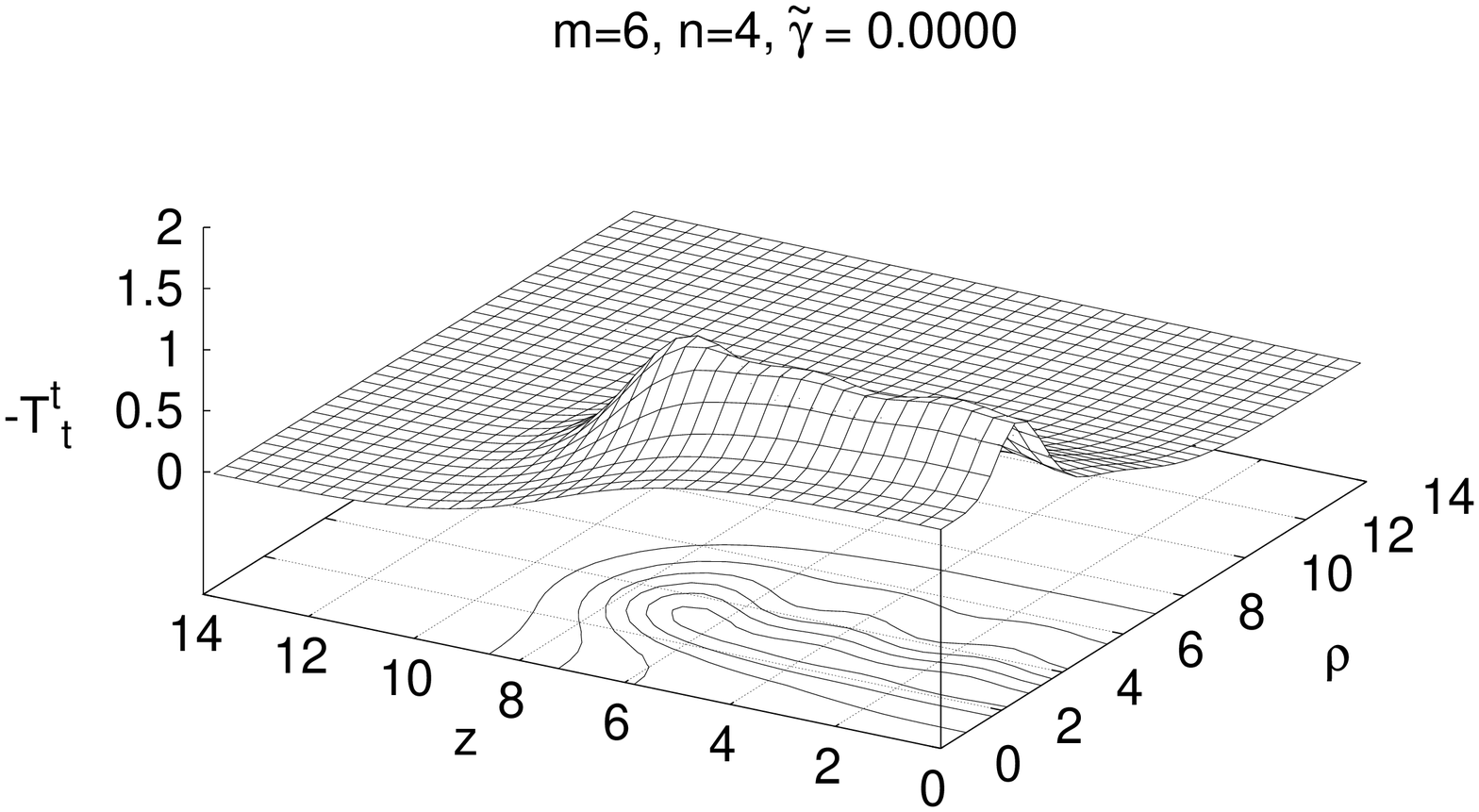}
\hspace{0.0cm} (a2)\hspace{-0.6cm}
\includegraphics[width=.45\textwidth, angle =0]{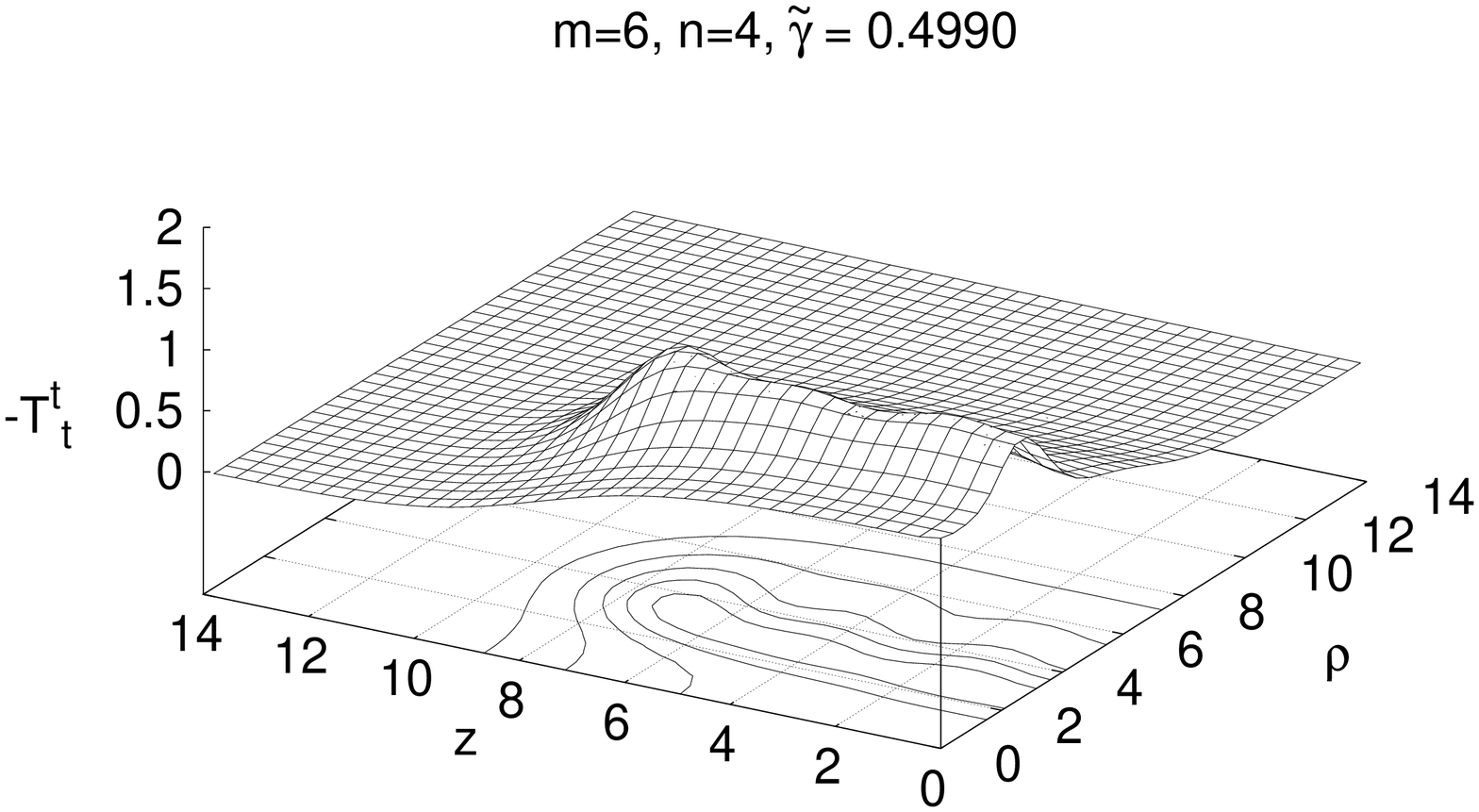}\\
\hspace{0.0cm} (b1)\hspace{-0.6cm}
\includegraphics[width=.45\textwidth, angle =0]{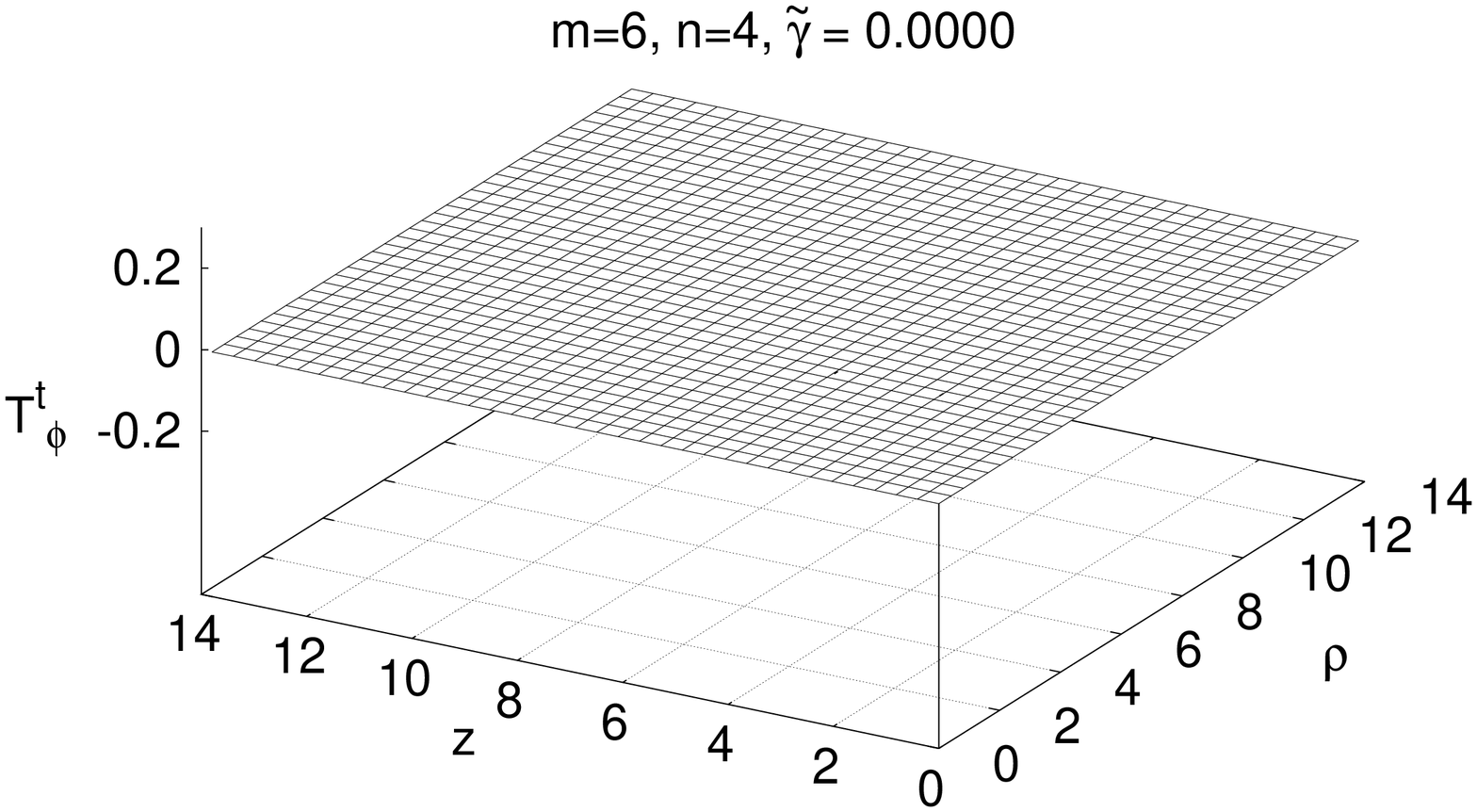}
\hspace{0.0cm} (b2)\hspace{-0.6cm}
\includegraphics[width=.45\textwidth, angle =0]{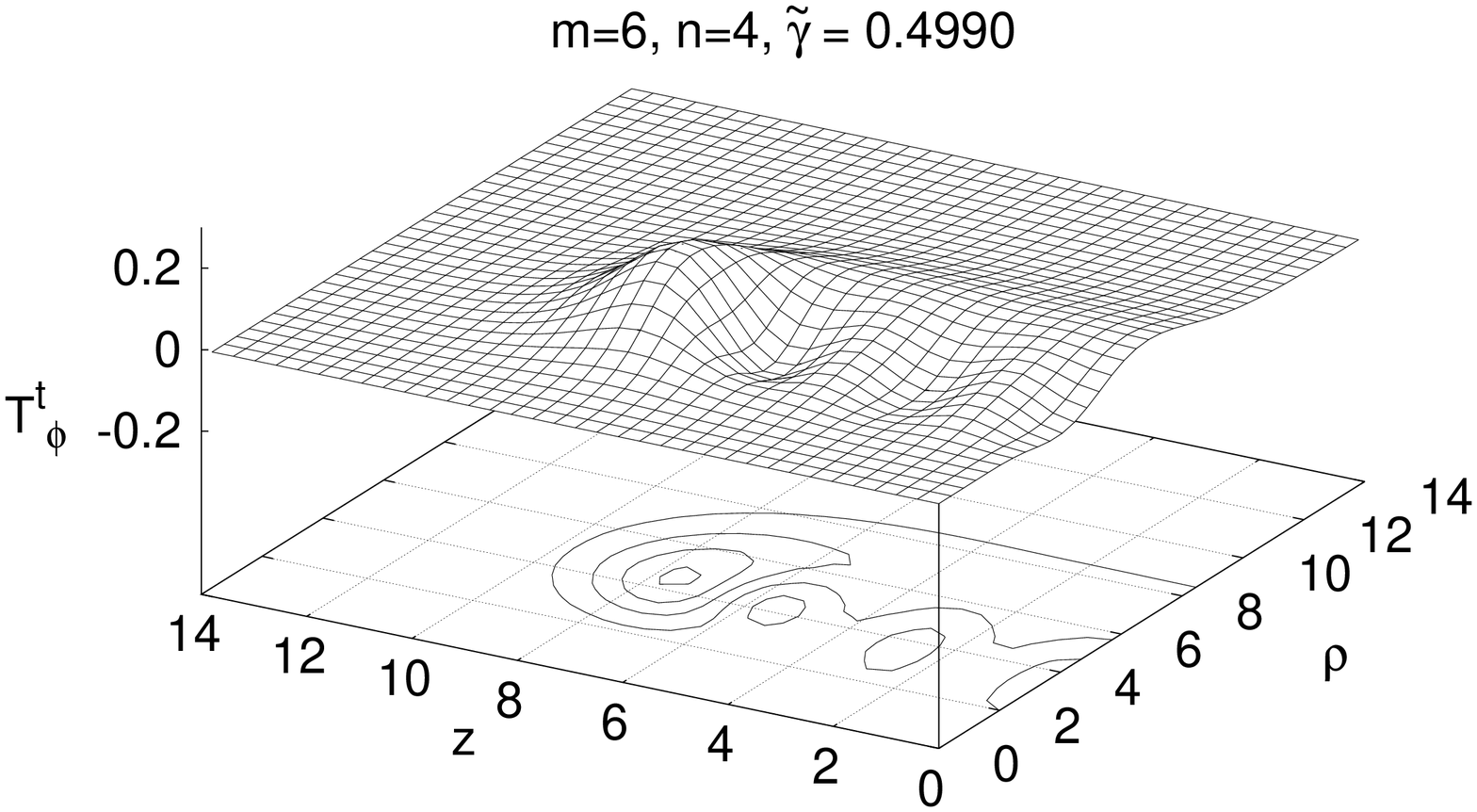}\\
\hspace{0.0cm} (c1)\hspace{-0.6cm}
\includegraphics[width=.45\textwidth, angle =0]{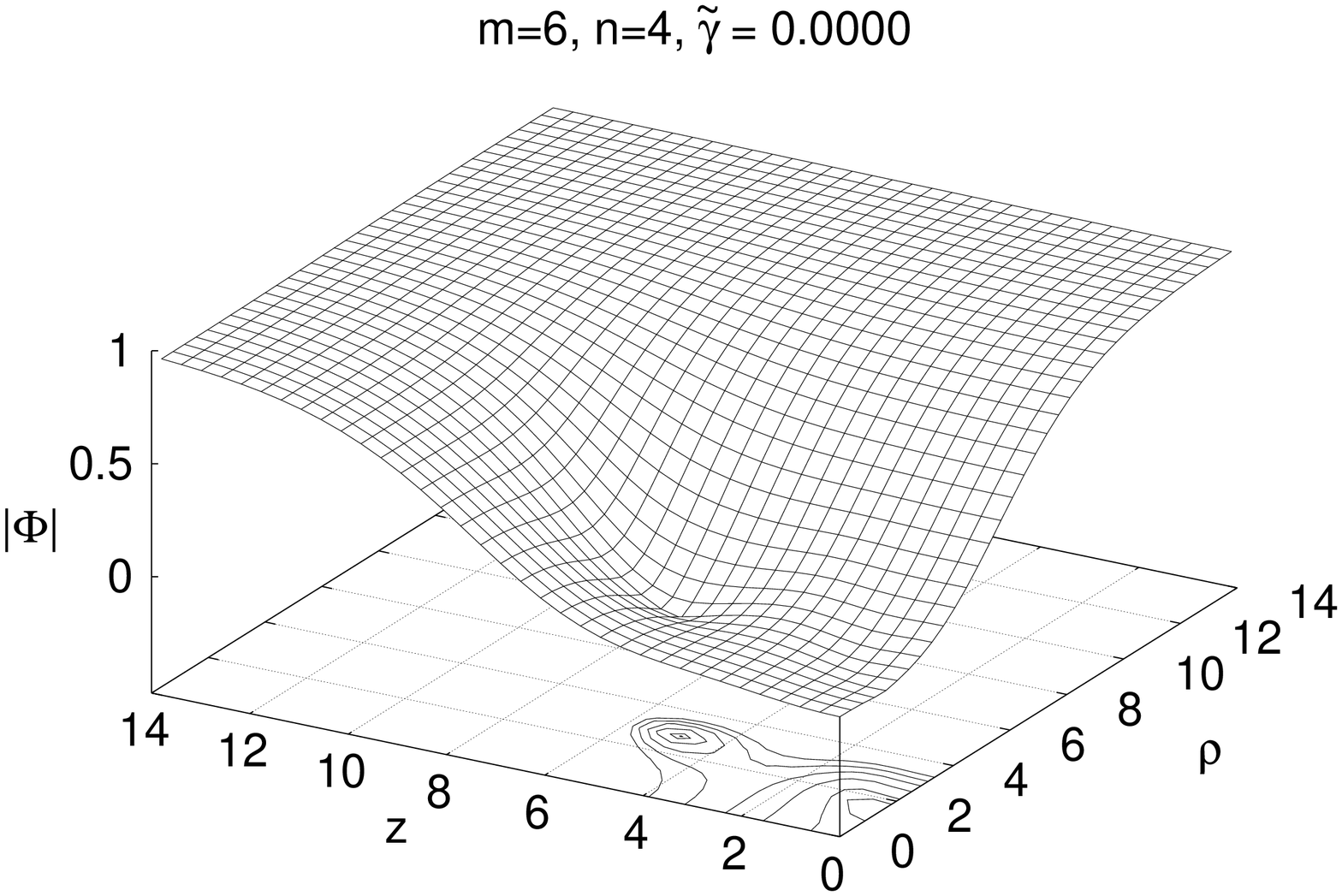}
\hspace{0.0cm} (c2)\hspace{-0.6cm}
\includegraphics[width=.45\textwidth, angle =0]{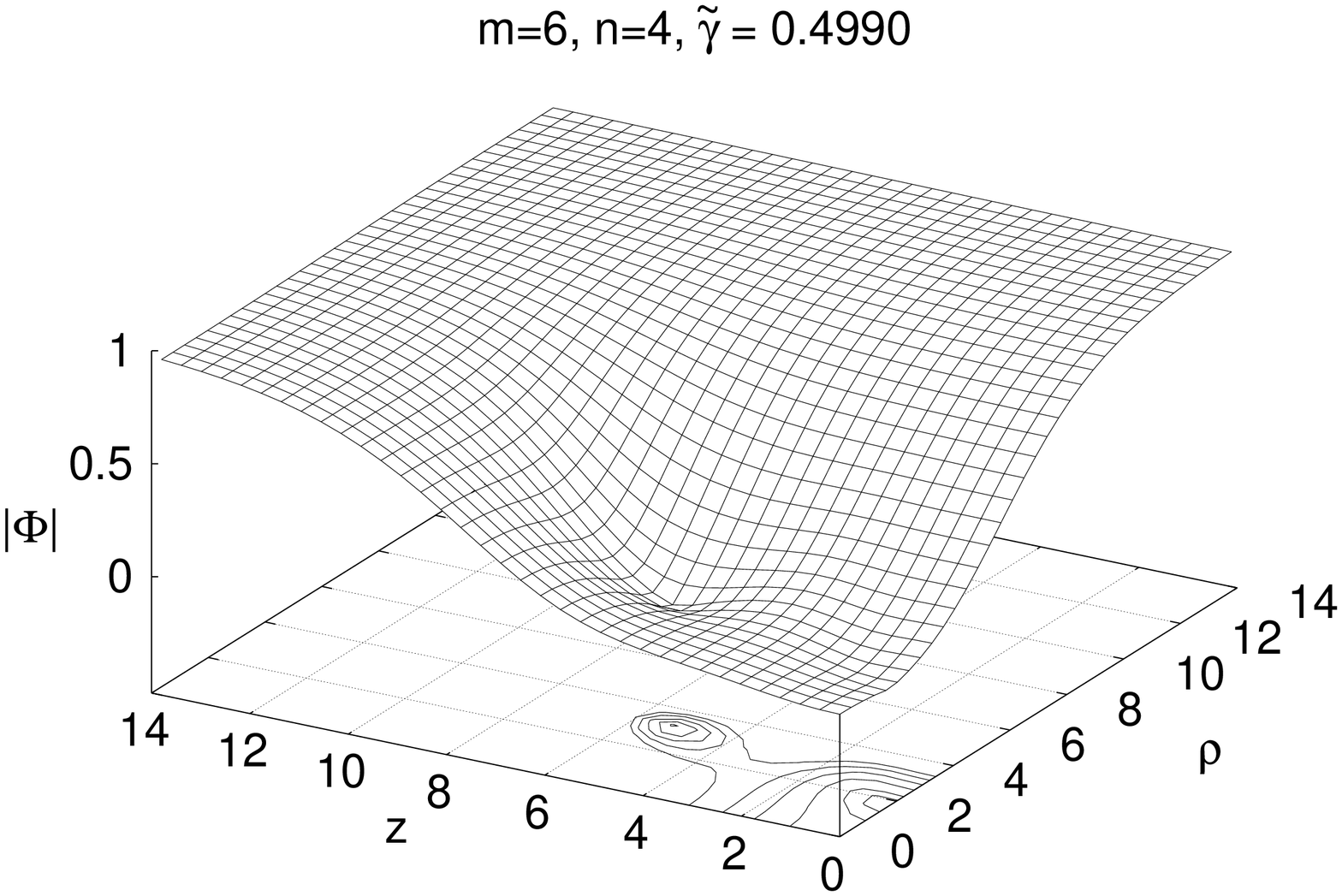}\\
\hspace{0.0cm} (d1)\hspace{-0.6cm}
\includegraphics[width=.45\textwidth, angle =0]{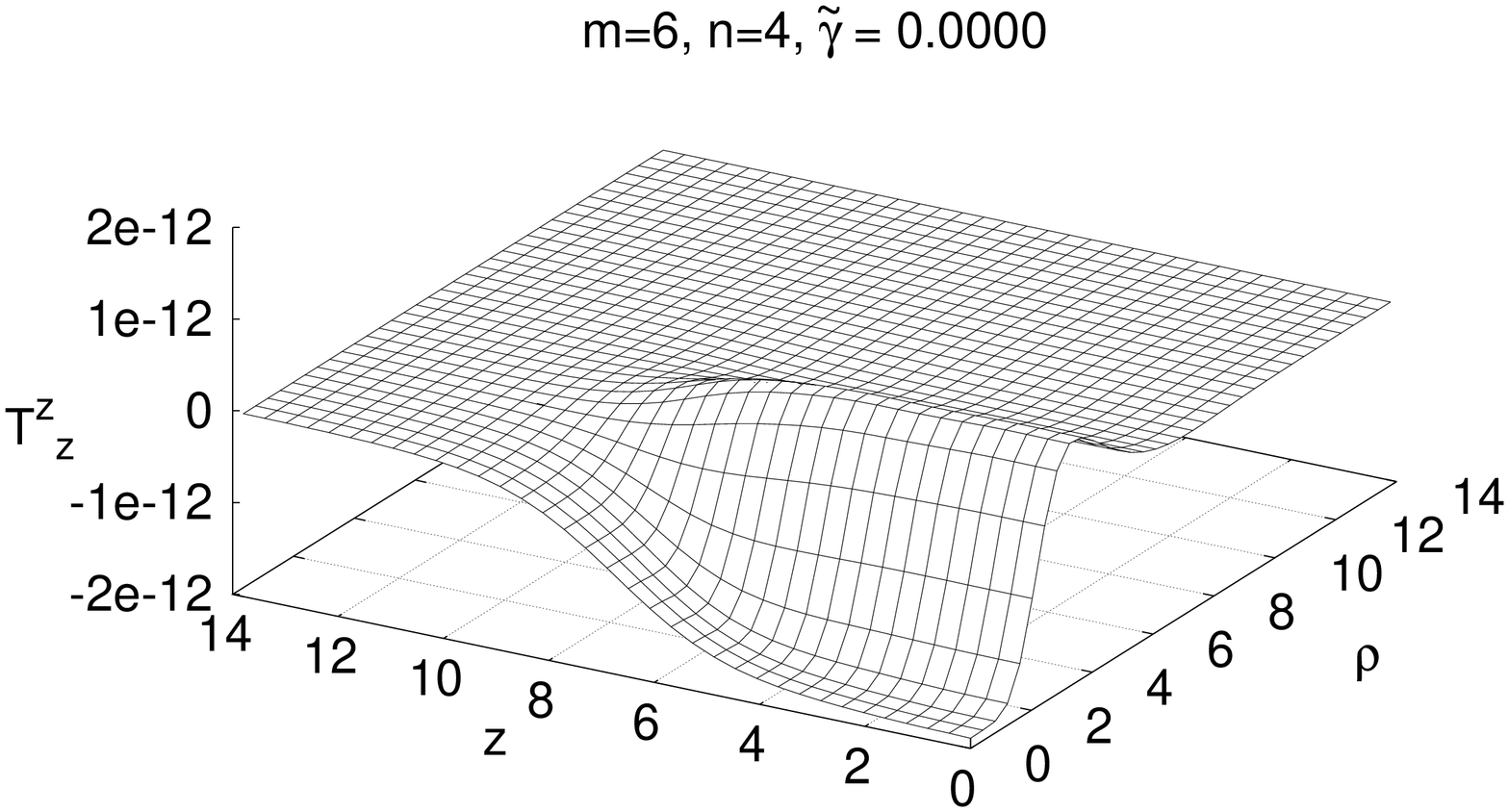}
\hspace{0.0cm} (d2)\hspace{-0.6cm}
\includegraphics[width=.45\textwidth, angle =0]{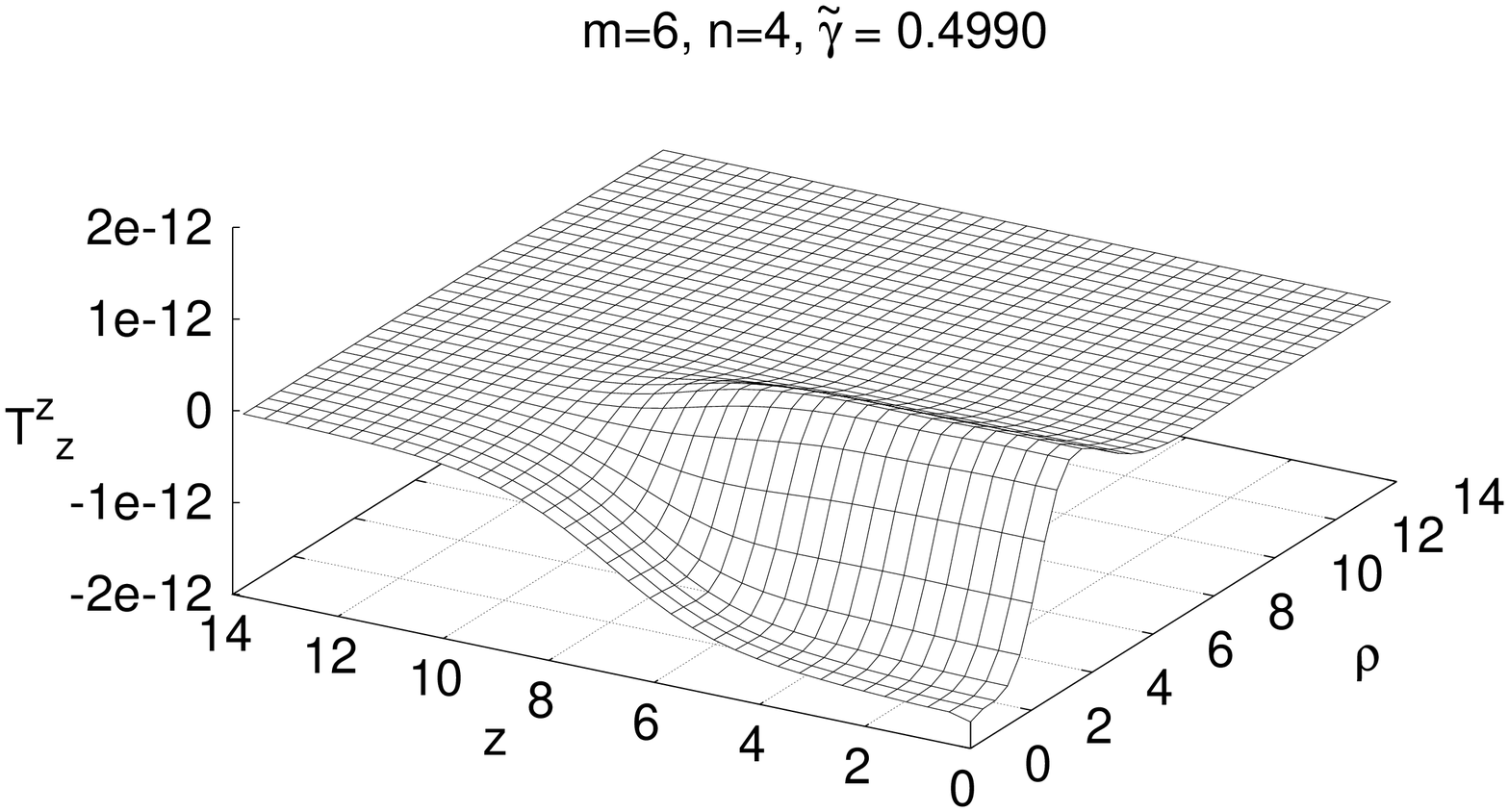}
\end{center}
\vspace{-0.5cm}
\caption{\small
The energy density $-T^t_t$ (a),
the angular momentum density $T^t_\varphi$ (b),
the modulus of the Higgs field $|\Phi|$ (c),
and the stress energy density $T^z_{z}$ (d)
are exhibited for $m=6$, $n=4$ solutions 
with $\tilde\gamma=0$ (left) and $\tilde\gamma \approx 0.5$ (right).
}
\end{figure}

\begin{figure}[h!]
\lbfig{f-10}
\begin{center}
\hspace{0.0cm} (a1)\hspace{-0.6cm}
\includegraphics[width=.45\textwidth, angle =0]{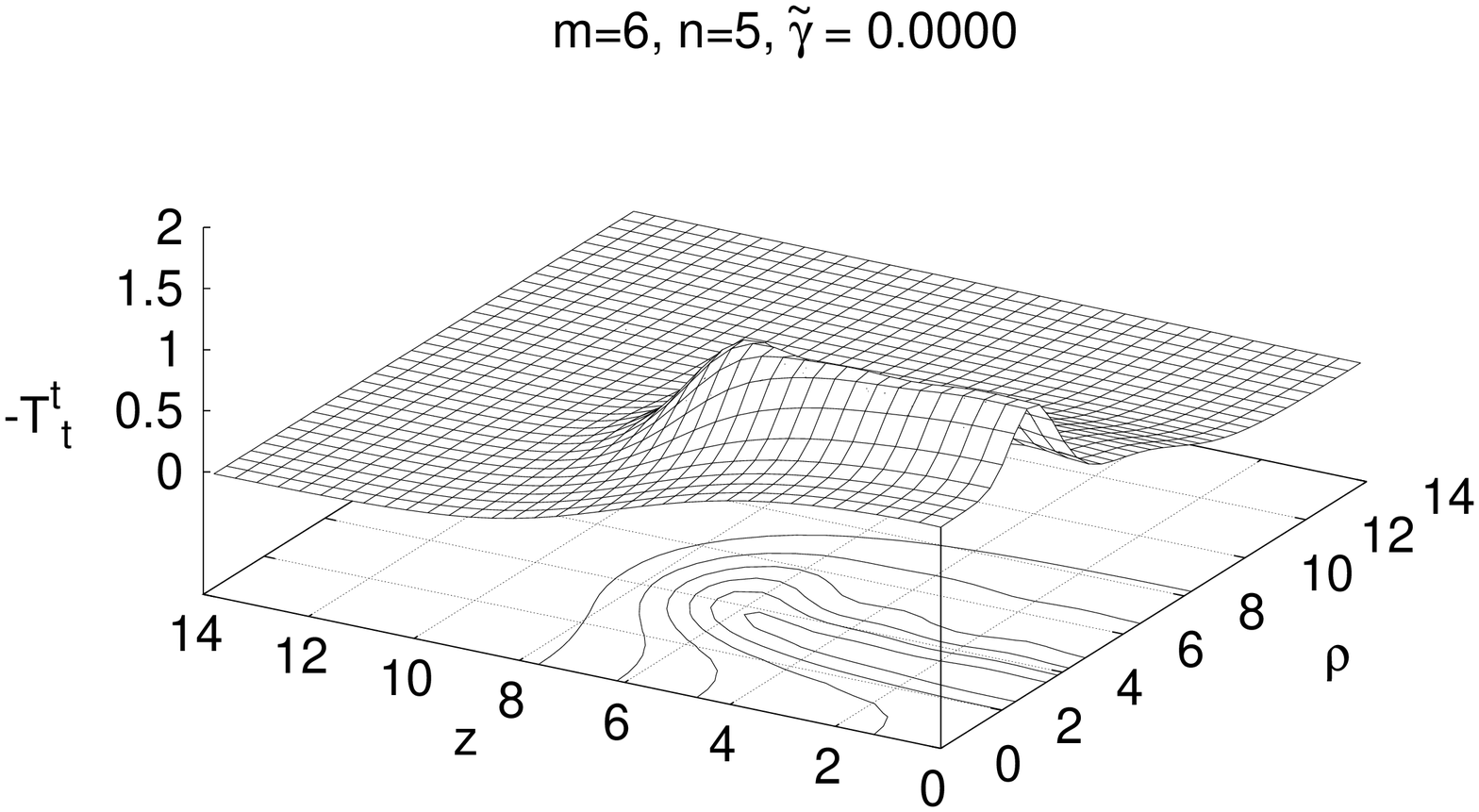}
\hspace{0.0cm} (a2)\hspace{-0.6cm}
\includegraphics[width=.45\textwidth, angle =0]{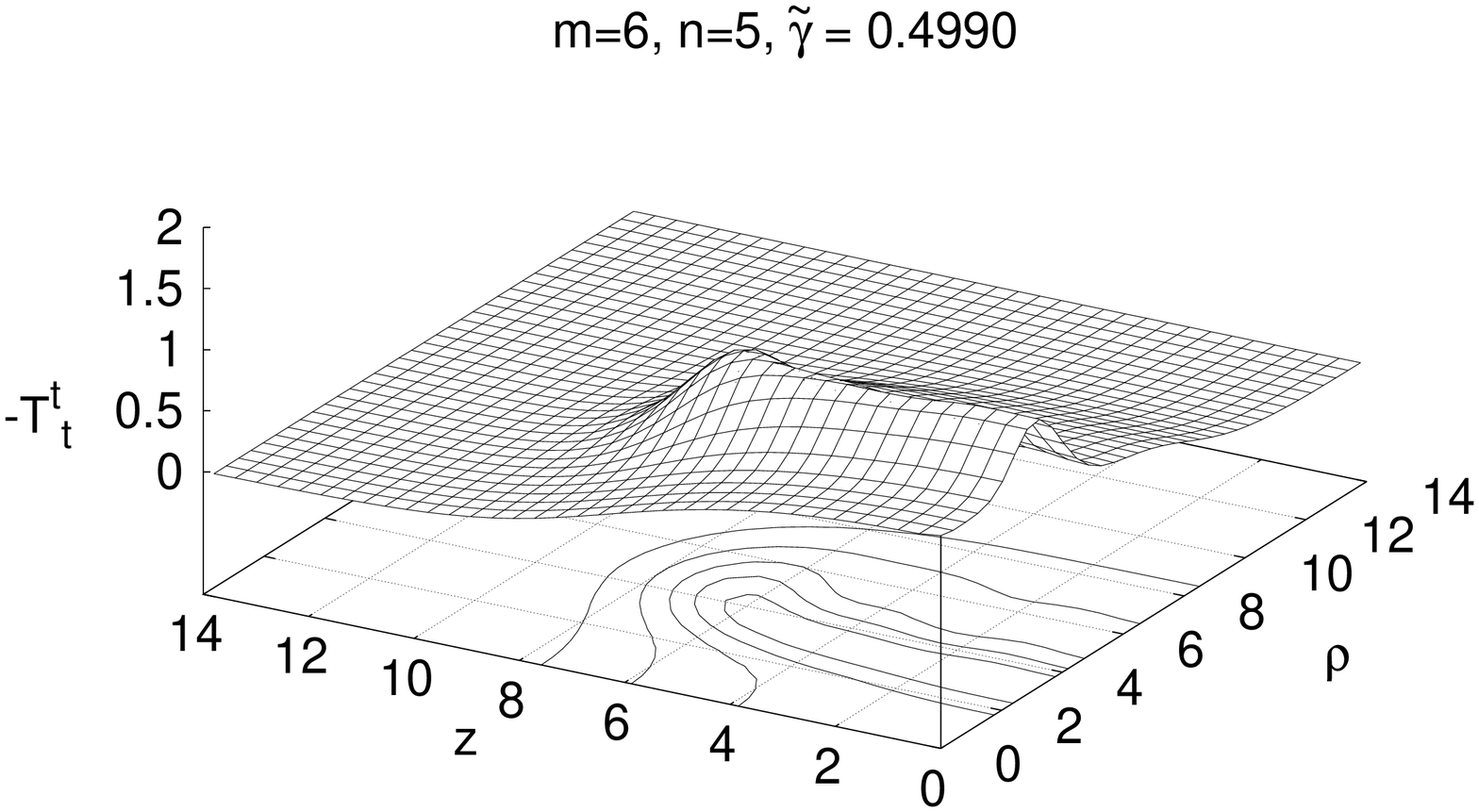}\\
\hspace{0.0cm} (b1)\hspace{-0.6cm}
\includegraphics[width=.45\textwidth, angle =0]{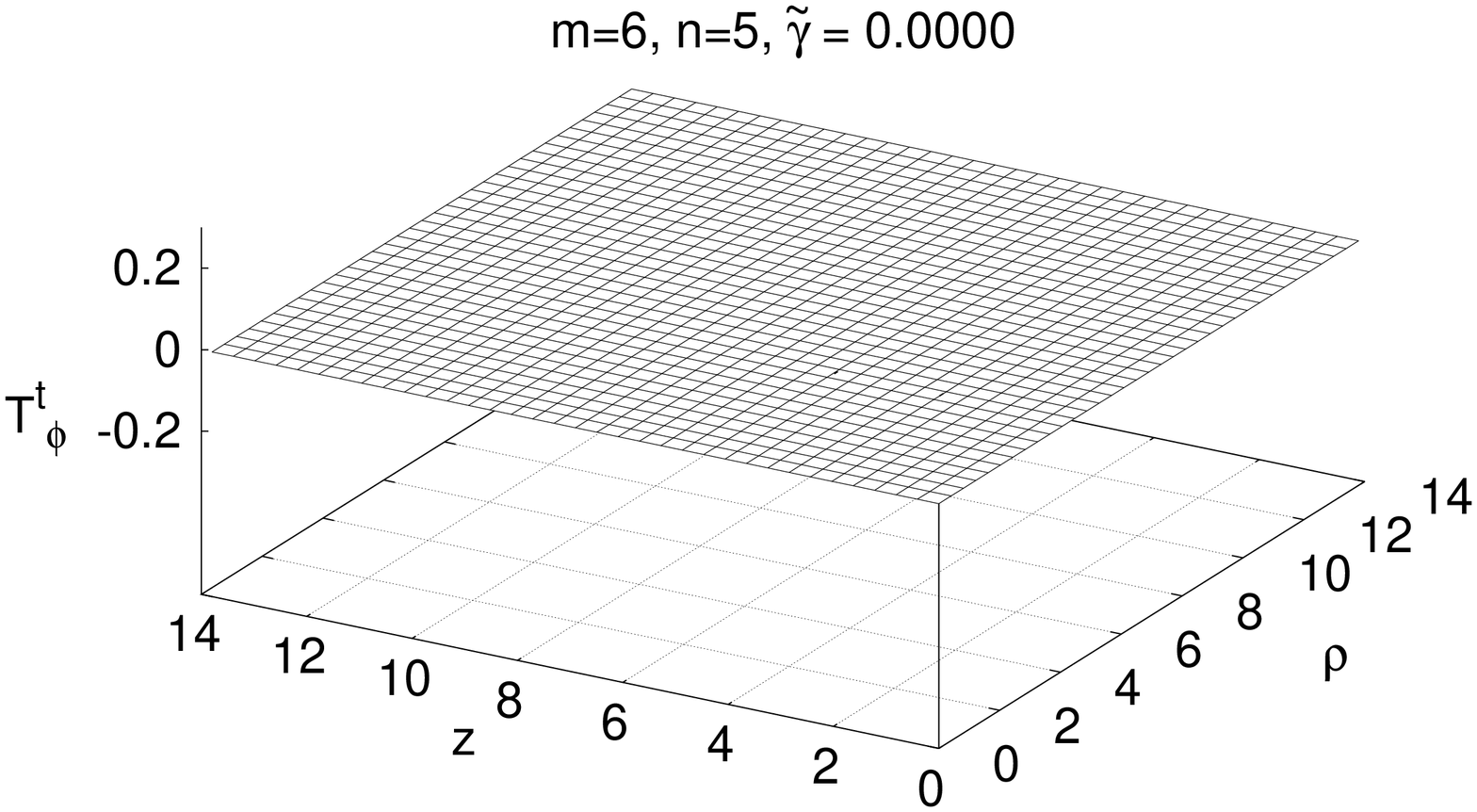}
\hspace{0.0cm} (b2)\hspace{-0.6cm}
\includegraphics[width=.45\textwidth, angle =0]{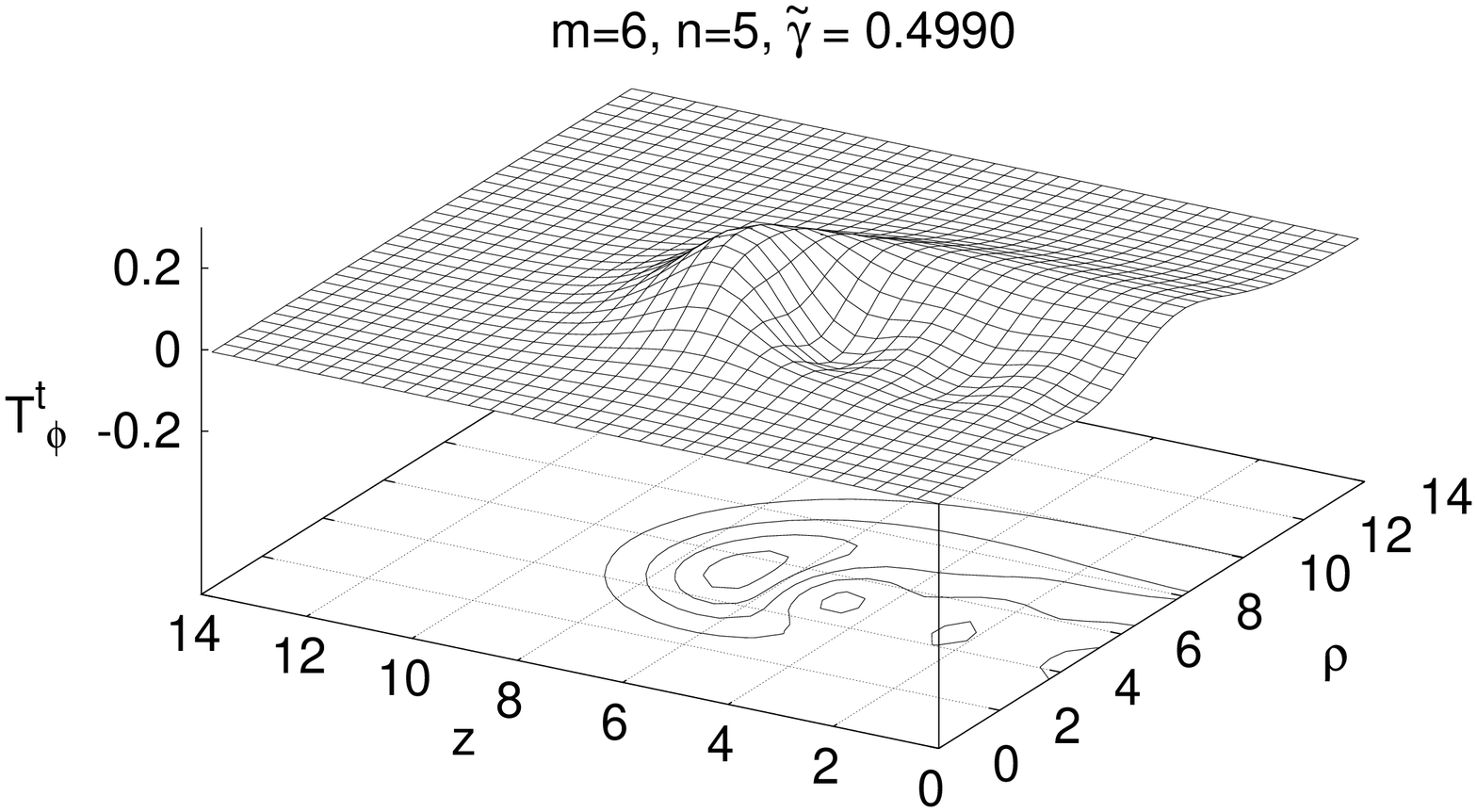}\\
\hspace{0.0cm} (c1)\hspace{-0.6cm}
\includegraphics[width=.45\textwidth, angle =0]{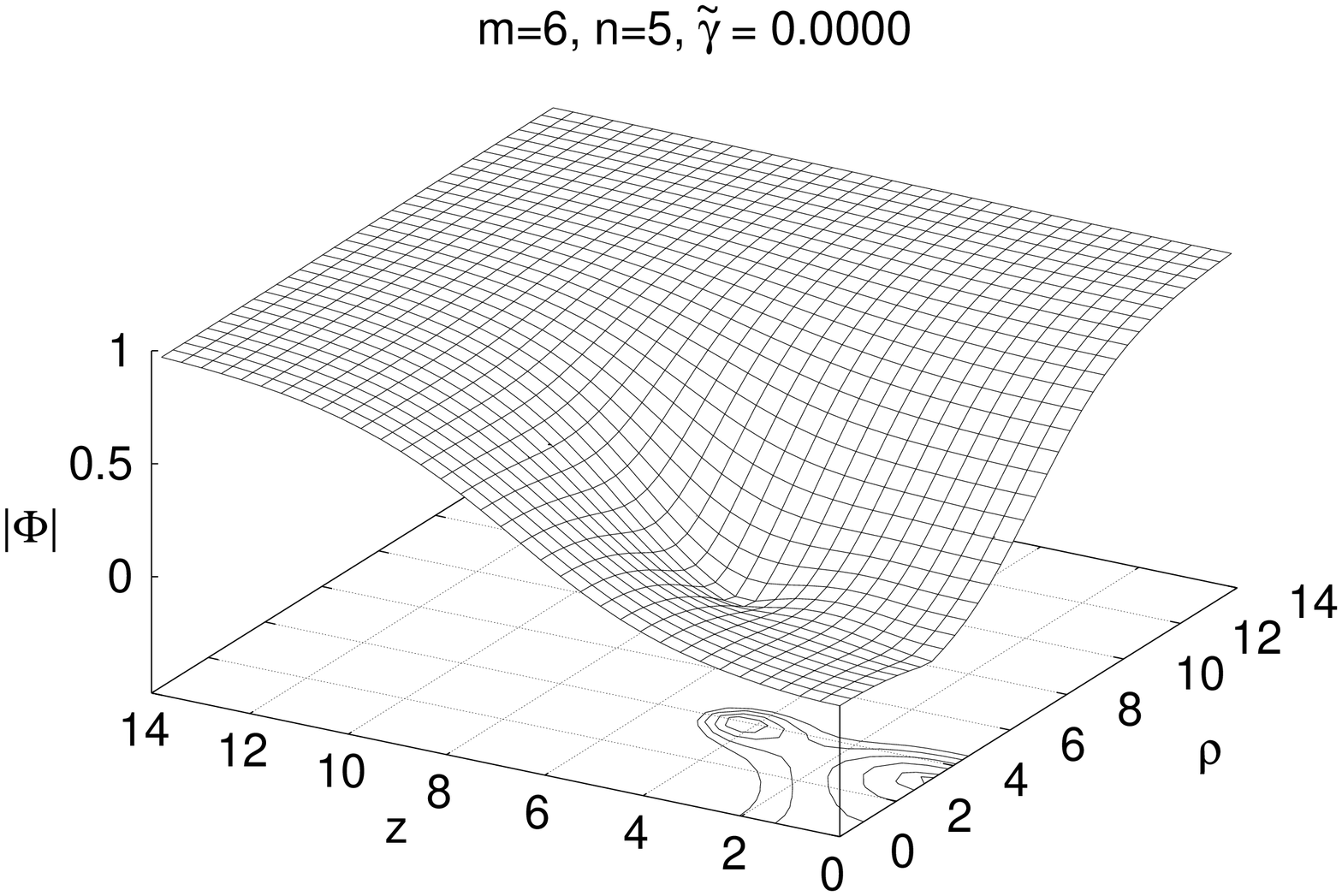}
\hspace{0.0cm} (c2)\hspace{-0.6cm}
\includegraphics[width=.45\textwidth, angle =0]{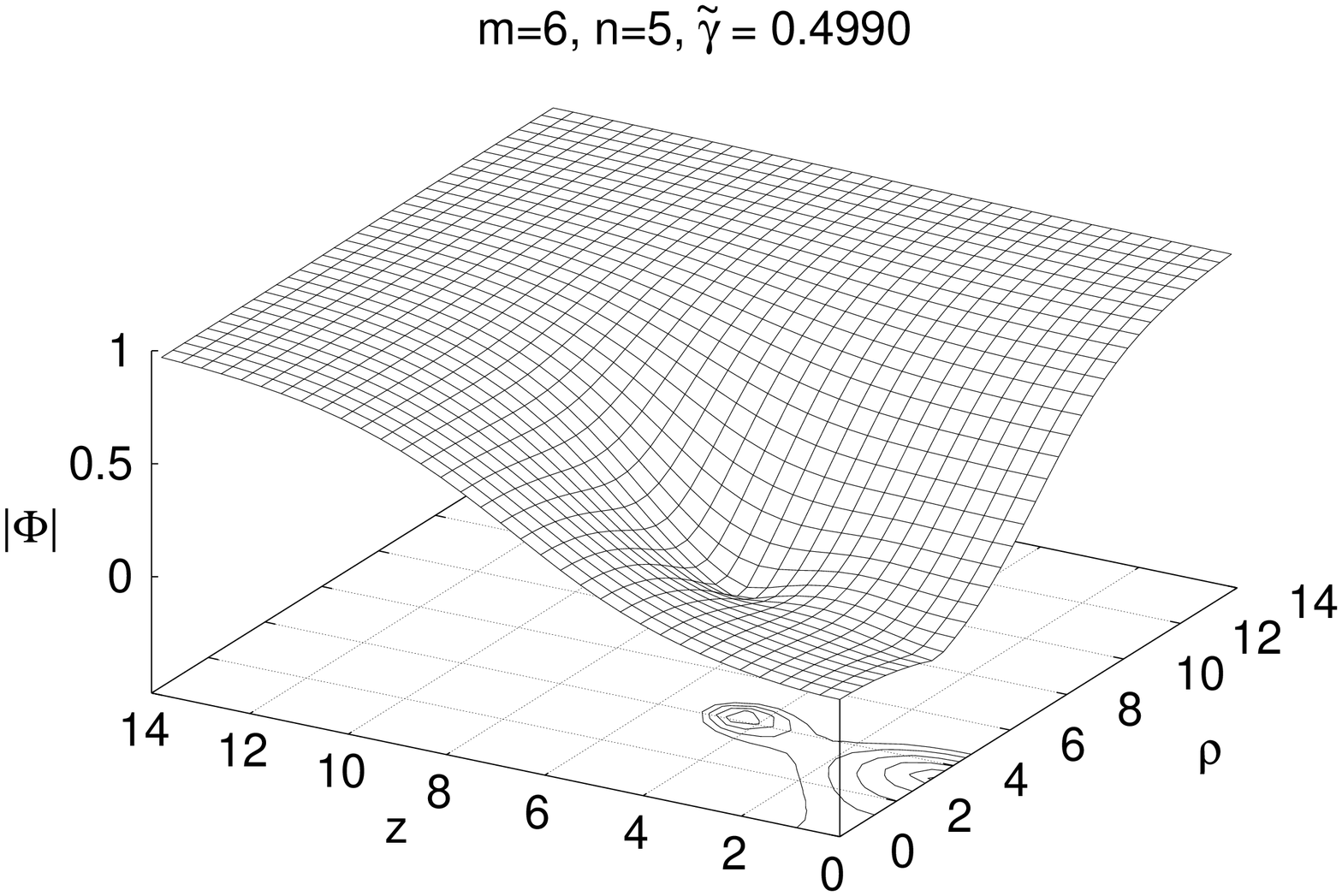}\\
\hspace{0.0cm} (d1)\hspace{-0.6cm}
\includegraphics[width=.45\textwidth, angle =0]{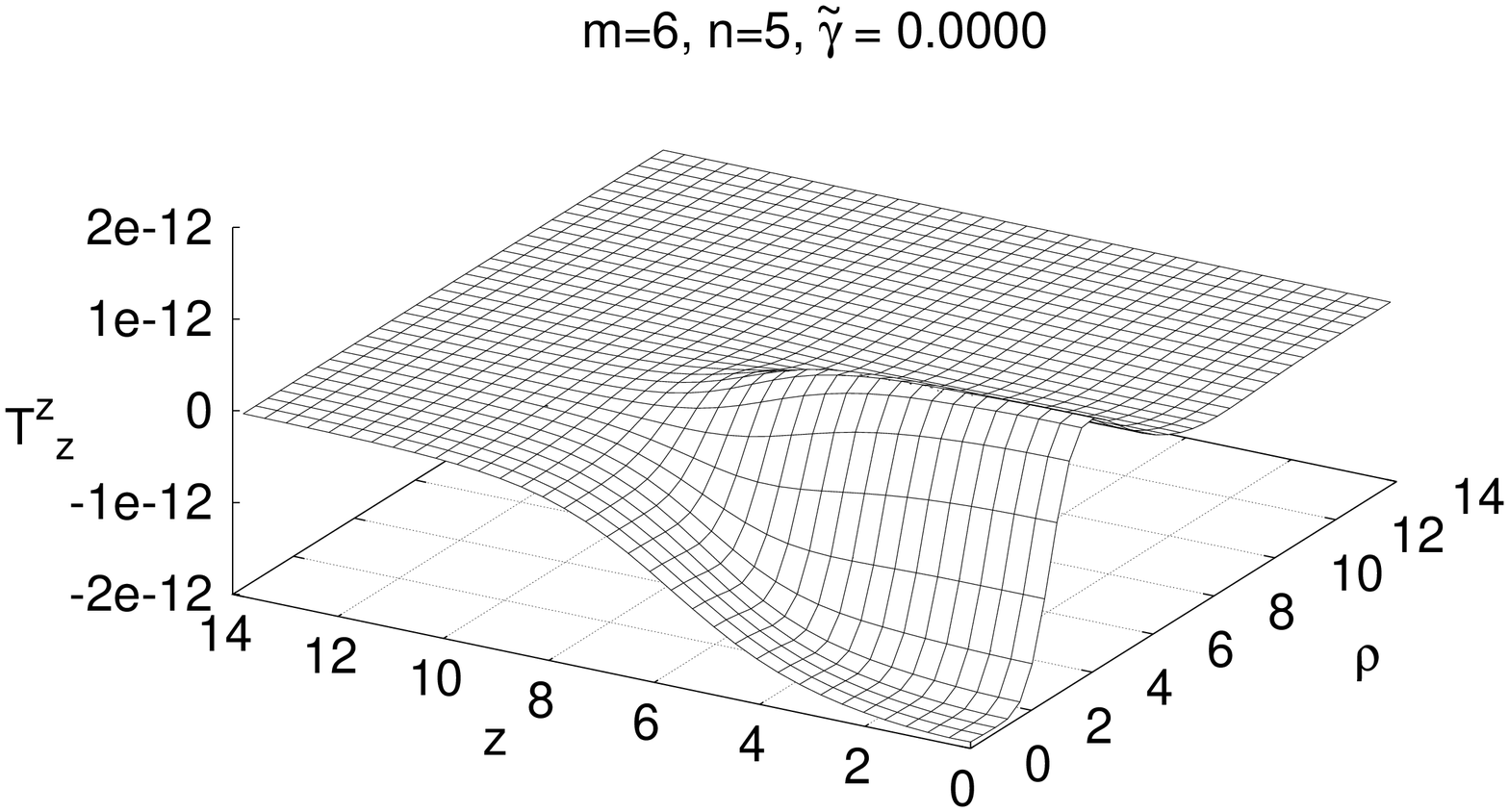}
\hspace{0.0cm} (d2)\hspace{-0.6cm}
\includegraphics[width=.45\textwidth, angle =0]{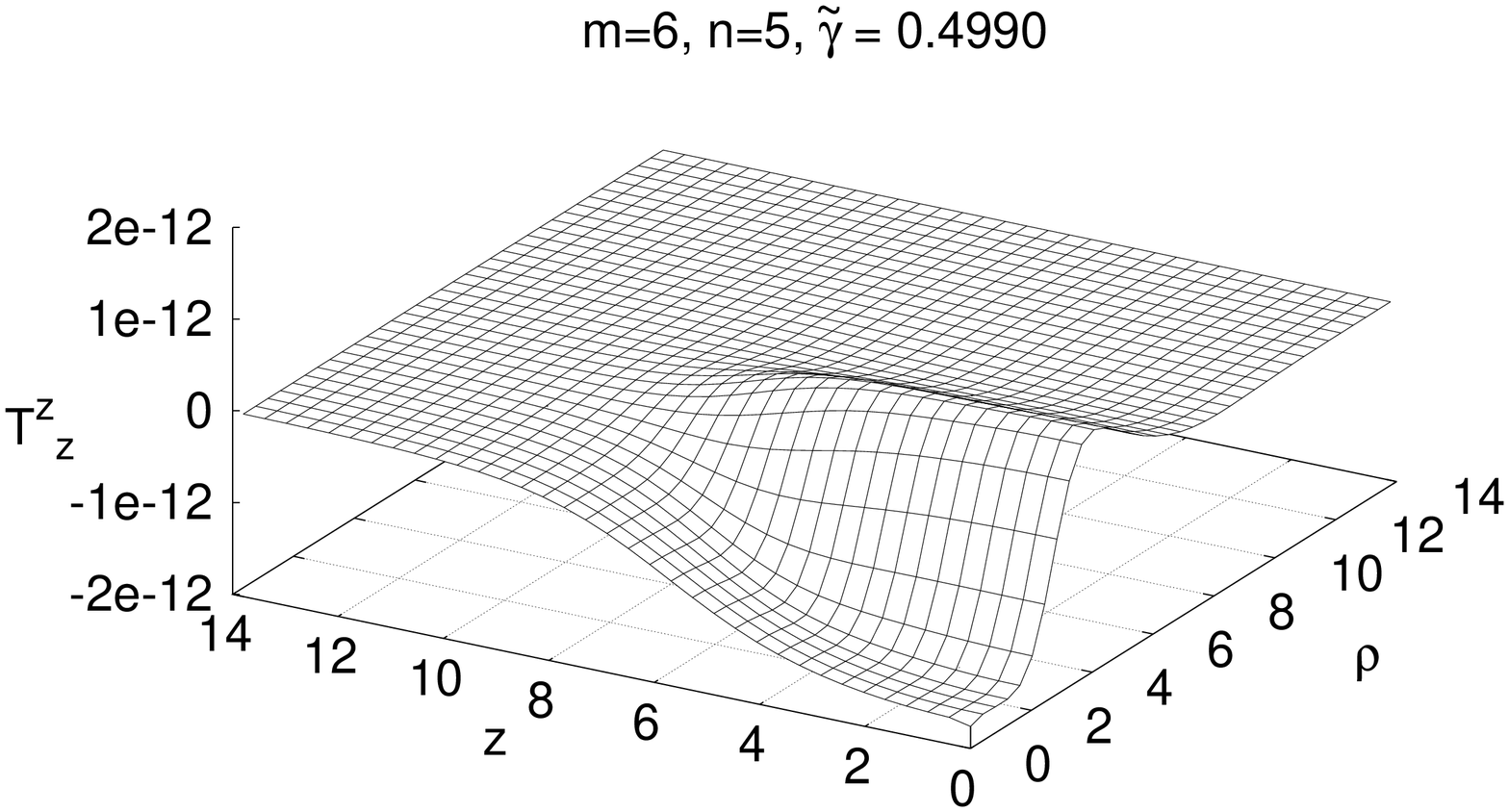}
\end{center}
\vspace{-0.5cm}
\caption{\small
The energy density $-T^t_t$ (a),
the angular momentum density $T^t_\varphi$ (b),
the modulus of the Higgs field $|\Phi|$ (c),
and the stress energy density $T^z_{z}$ (d)
are exhibited for $m=6$, $n=5$ solutions 
with $\tilde\gamma=0$ (left) and $\tilde\gamma \approx 0.5$ (right).
}
\end{figure}

In these sphaleron-antisphaleron systems, 
the regions with large energy density are torus-like,
where the configurations possess six such tori.
Their location depends on $n$ and on the parameters.
With increasing $n$ these tori degenerate
forming basically a single cylinder.
As before,
the effect of the presence of electric charge
is that the energy density spreads further out,
while at the same time its overall 
magnitude reduces.
Likewise, the effect of the presence of angular momentum
is a centrifugal shift of the energy density.
With increasing angular momentum
the torus-like regions of large energy density
move further outward to larger values of $\rho$.

The modulus of the Higgs field of the 
sphaleron-antisphaleron systems changes only little
with increasing charge and angular momentum.
Indeed the change is barely noticeable in these figures,
even though the static systems are compared to those
that carry maximal charge and angular momentum.
We therefore address the effect on the nodes of the
Higgs field separately in Fig.~\ref{f-12},
where we exhibit for the system $m=6$, $n=5$ the
modulus of the Higgs field in the equatorial plane
with increasing charge parameter $\tilde\gamma$.
The effect on the size of the ring is an increase
of roughly 10\% due to charge and rotation.
For comparison we also exhibit the
modulus of the Higgs field on the $z$-axis,
choosing the system $m=5$, $n=2$, where the nodes
are pointlike.
Here we observe an increase of the distance between the nodes
also on the order of 10\% due to charge and rotation.

\begin{figure}[h!]
\lbfig{f-12}
\begin{center}
\includegraphics[width=.45\textwidth, angle =0]{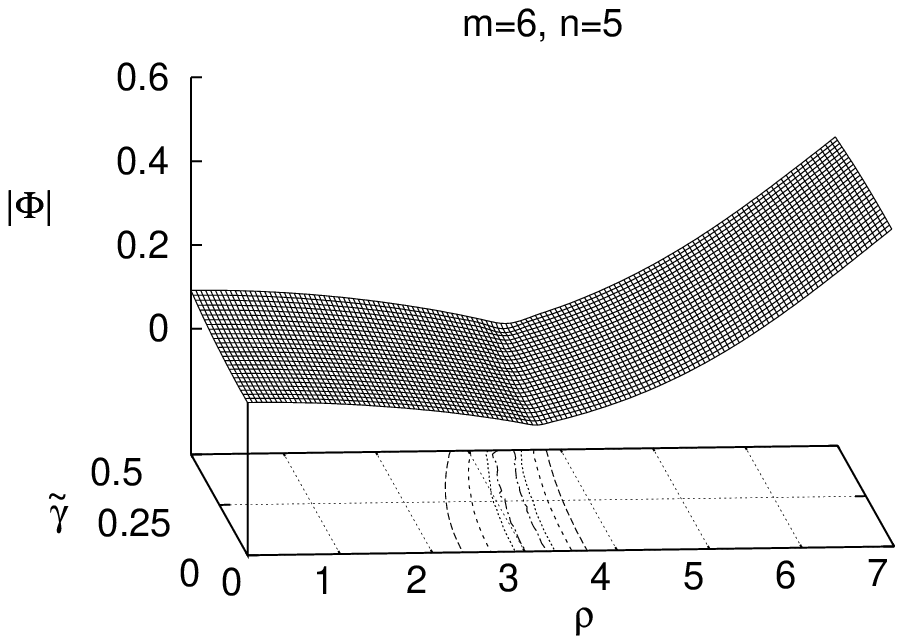}
\includegraphics[width=.45\textwidth, angle =0]{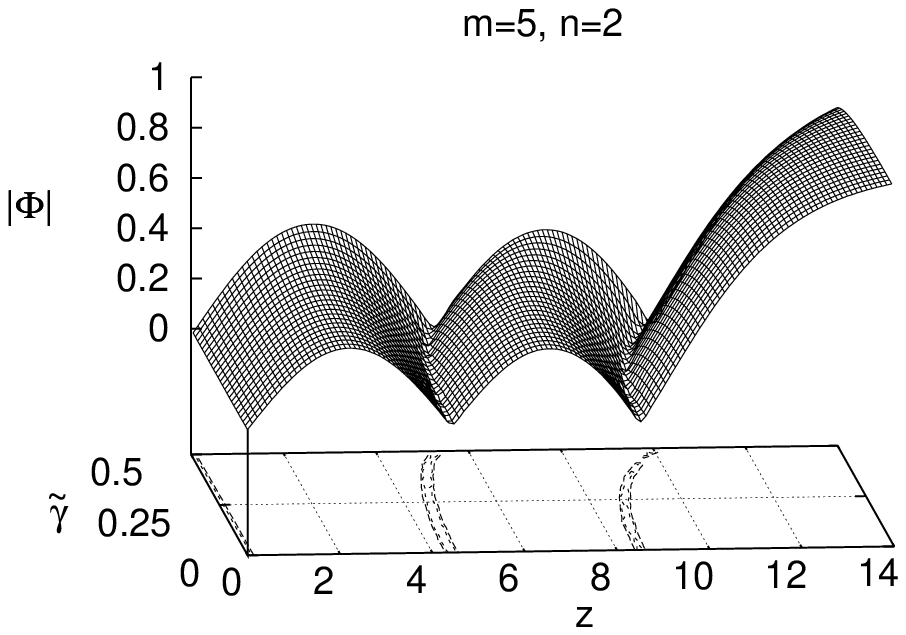}
\end{center}
\vspace{-0.5cm}
\caption{\small
The modulus of the Higgs field $|\Phi|$
versus the charge parameter $\tilde\gamma$
a) in the equatorial plane for the vortex ring
configuration $m=6$, $n=5$,
b) on the $z$-axis
for the sphaleron-antisphaleron chain $m=5$, $n=2$.
}
\end{figure}

The angular momentum density of the sphaleron-antisphaleron systems
is also characterized by the presence of tori.
There are tori of large positive angular momentum density
as well as negative angular momentum density.
The tori of the angular momentum density are spatially related to the tori
of the energy density.
In particular, the location of the positive tori is associated with
the location of the tori of the energy density,
with the negative tori inbetween.

\subsection{Equilibrium condition}

Let us finally address the question of the
equilibrium of such composite configurations
as sphaleron-antisphaleron pairs and more general
sphaleron-antisphaleron systems.
As discussed previously \cite{Aharonov:1992wf,Beig:2008qi,Beig:2009jd},
a necessary condition for the equilibrium
of such axially symmetric configurations is 
\begin{equation}
\int_S T_{zz} dS =0
\label{Tzz}
\end{equation}
where $T_{zz}$ is the respective component
of the stress energy tensor and $S$ is the equatorial plane.
When this condition is satisfied, the net force between
the constituents in the upper and in the lower
hemisphere vanishes, thus yielding equilibrium.
If $T_{zz}$ vanishes everywhere in the equatorial plane,
this condition is met trivially,
if on the other hand $T_{zz}$ does not identically vanish, 
the various contributions to the surface integral 
(\ref{Tzz}) must precisely cancel each other.

To understand how the equilibrium condition is satisfied
in these sphaleron-antisphaleron systems,
we have extracted the $T_{zz}$ component of the
stress energy tensor. 
We illustrate 
$T_{zz}$ for two rather different configurations in Fig.~\ref{f-13}.
In Fig.~\ref{f-13}a 
we display $T_{zz}$
for the static sphaleron-antisphaleron chain with $m=4$, $n=1$
in the upper hemisphere. 
In the equatorial plane $T_{zz}$ appears to almost vanish.
We therefore focus on the equatorial plane in Fig.~\ref{f-13}c. 
Here $T_{zz}$ is small, but finite (except when it changes sign).
To gain further insight into how the equilibrium 
results from the various forces present in the system,
we consider the contributions from the respective parts of the Lagrangian
separately.
We exhibit these also in Fig.~\ref{f-13}c. 
We note, that
the positive contribution from the $SU(2)$ gauge field part
almost cancels the negative contributions from the $U(1)$ and Higgs
parts, yielding in total a $T_{zz}$ which is almost but not quite vanishing
in the equatorial plane.
In the inner region the total $T_{zz}$ is slightly positive, while in the
outer region it is slightly negative, yielding
together a vanishing surface integral (\ref{Tzz}),
within the numerical accuracy.

The situation is similar for other sphaleron-antisphaleron
chains with even $m$ and $n=1$, including the sphaleron-antisphaleron pair.
Also, the inclusion of rotation does not change this
overall behaviour of these types of solutions.
For most other systems, however, $T_{zz}$ does not nearly vanish
in the equatorial plane. 
This is exhibited exemplarily in Fig.~\ref{f-13}b
for the fast rotating sphaleron-antisphaleron system with
$m=6$, $n=5$. The features of $T_{zz}$
seen here, are very typical, and hardly change with rotation,
since the effect of rotation is basically a slight shift in magnitude.
However, while $T_{zz}$ is rather large in the equatorial plane
for these configurations,
its positive and negative contributions to the surface integral 
do cancel as required for equilibrium.

\begin{figure}[h!]
\lbfig{f-13}
\begin{center}
\hspace{0.0cm} (a)\hspace{-0.6cm}
\includegraphics[width=.46\textwidth, angle =0]{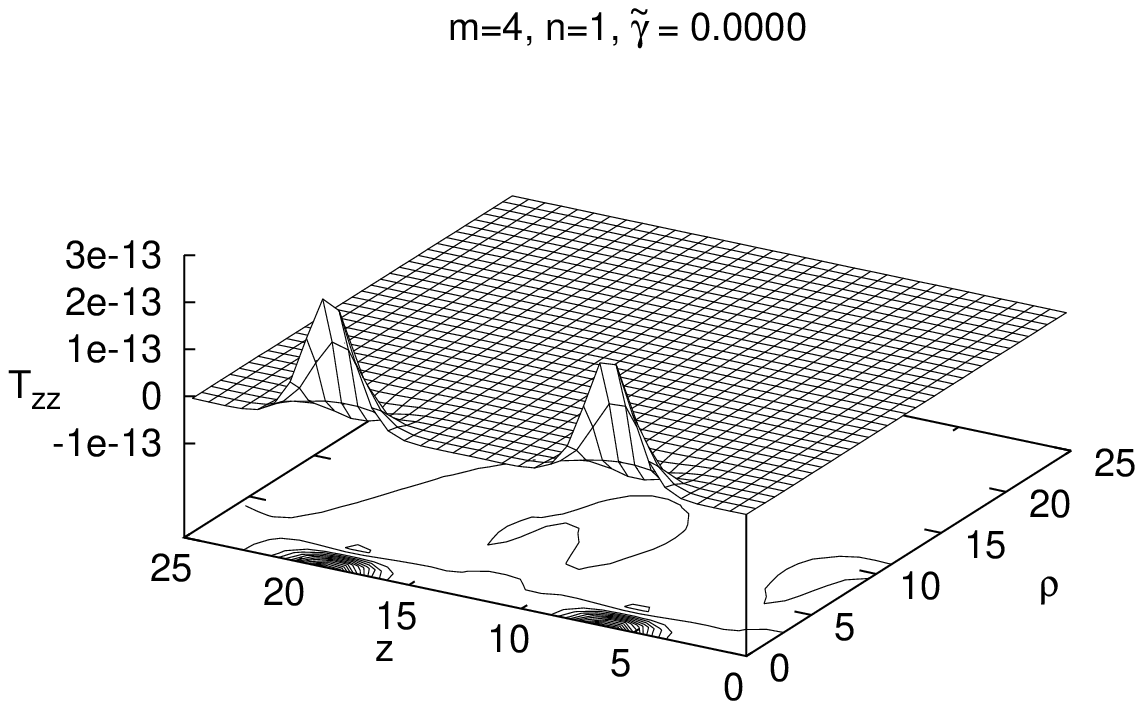}
\hspace{0.5cm} (b)\hspace{-0.6cm}
\includegraphics[width=.46\textwidth, angle =0]{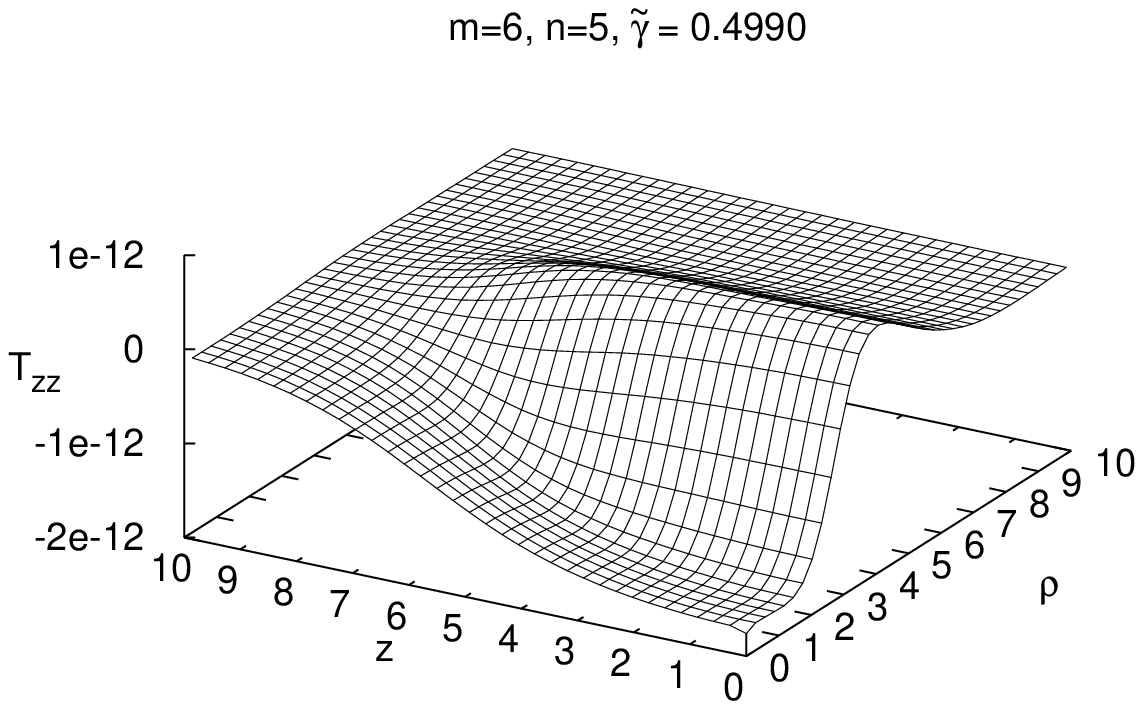}
\\
\hspace{0.0cm} (c)\hspace{-0.6cm}
\includegraphics[width=.46\textwidth, angle =0]{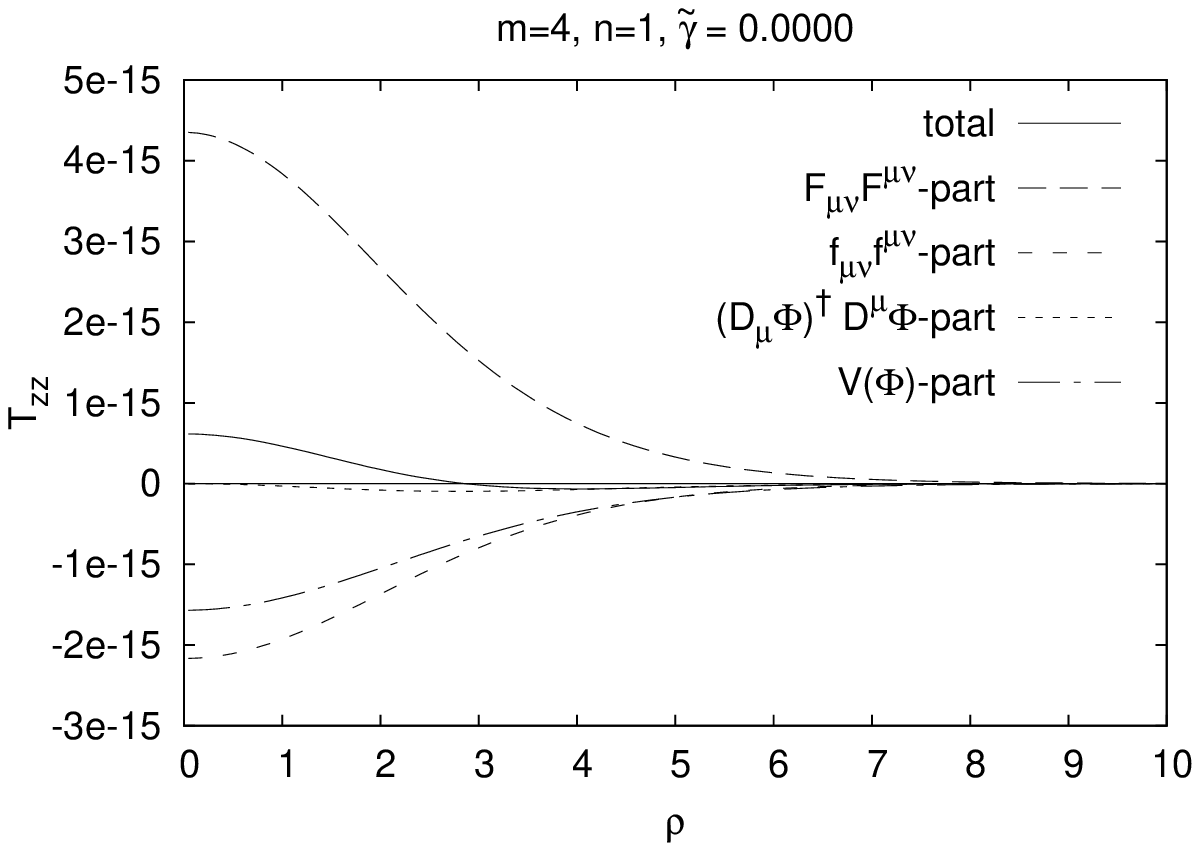}
\hspace{1.0cm} (d)\hspace{-0.6cm}
\includegraphics[width=.46\textwidth, angle =0]{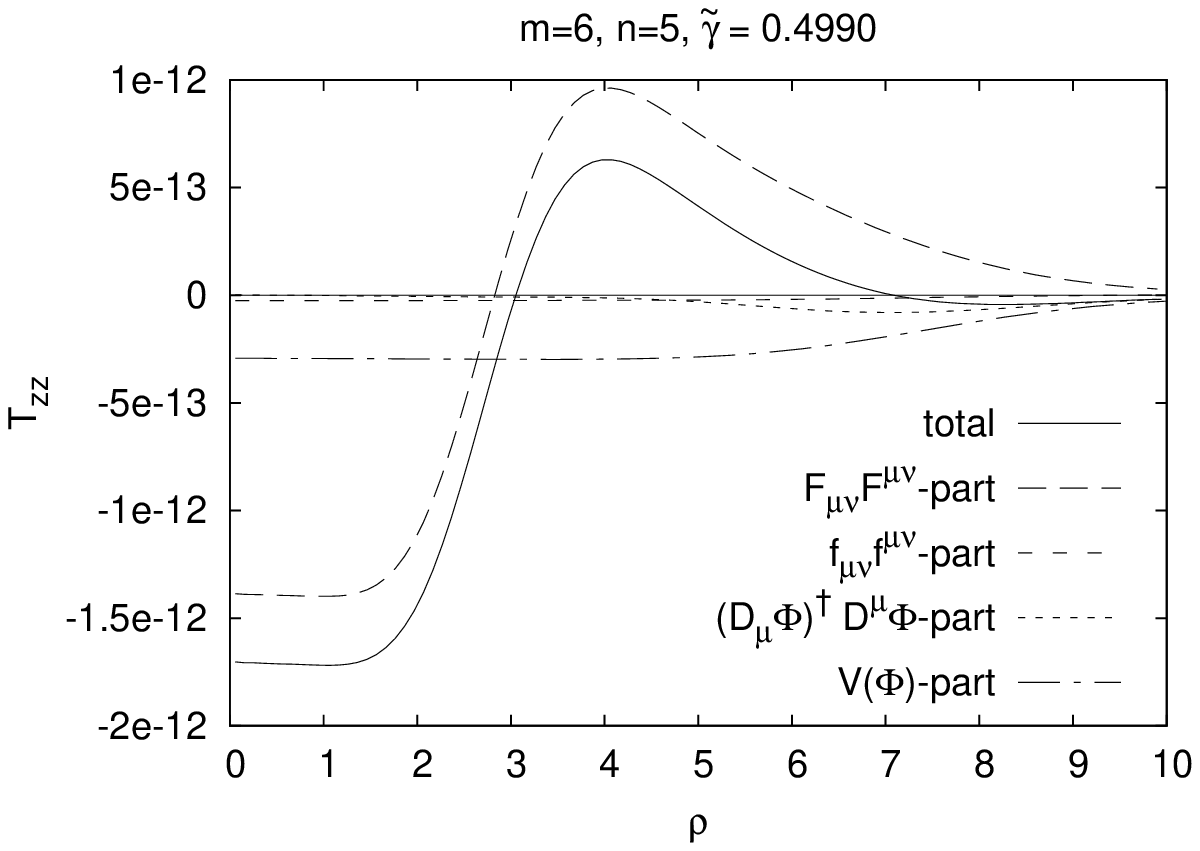}
\end{center}
\vspace{-0.5cm}
\caption{\small
The full stress-energy component $T_{zz}$ 
(upper)
and its $SU(2)$, $U(1)$, Higgs covariant derivative
and Higgs potential parts in
the equatorial plane (lower)
for the sphaleron-antisphaleron systems
$m=4$, $n=1$ $J=0$ (left)
and $m=6$, $n=5$, $J\approx J_{\rm max}$ (right).
}
\end{figure}

\section{Conclusions}

We have considered
sphaleron-antisphaleron pairs, chains and vortex ring solutions
in Weinberg-Salam theory,
which are characterized by two integers $n$ and $m$.
Starting from the respective neutral electroweak configurations,
we have obtained the corresponding 
branches of rotating electrically charged solutions.
These branches exist up to maximal values of the charge
and angular momentum,
beyond which localized solutions are no longer possible.

We have performed a complete study of all
configurations with $m=1-6$ and $n=2-6$,
fixing the weak mixing angle at its physical value
and taking a fixed value of the Higgs mass.
The chain configurations with $m=2-6$ and $n=1$
have only partially been obtained with sufficiently high
accuracy, to include their global properties
such as their energy and magnetic moment into our
systematic survey.
For these chains our efforts are still continuing.

On the other hand, in order to clarify the evolution
of the nodal structure of these sphaleron-antisphaleron systems,
we have gone to rather high values of $n$,
for static configurations, while extrapolating from the
full study that the nodal structure is not much affected
by the presence of charge and rotation.
In particular, we have observed that the various rings
in the vortex ring configurations tend to increase in size
linearly with $n$, while at the same time tending to merge
into a single ring in the equatorial plane.
For $m=3$ this merging transition occurs at $n=6$,
and for $m=5$ at $n=37$, while for the even $m$ cases
4 and 6 the expected merging may occur only beyond $n=100$.

The angular momentum $J$ and the charge $\cal Q$ of these
sphaleron-antisphaleron systems are proportional
$$J = n {\cal Q} /e.$$
Their energy and binding energy increase with increasing rotation,
and so does their magnetic moment.
With increasing charge
the energy density of the configurations spreads further out,
while its overall magnitude reduces.
At the same time the effect of the rotation 
is a centrifugal shift of the energy density tori
to larger radii.

We have also addressed the equilibrium condition (\ref{Tzz})
for these sphaleron-antisphaleron systems. 
In all systems, it is the surface integral that vanishes
to give equilibrium, and not the 
stress-energy tensor component $T_{zz}$ by itself.
However, for the sphaleron-antisphaleron pair
(and other even $m$ chains)
the stress-energy tensor component $T_{zz}$
almost vanishes in the equatorial plane.
In these configurations the positive contribution from the
$SU(2)$ part almost cancels the negative
contributions from the $U(1)$ and Higgs parts,
thus yielding an almost vanishing total $T_{zz}$
in the equatorial plane.

The next step will be to study the fermion modes
in the background of rotating electroweak configurations
to understand their relevance for baryon number violating
processes \cite{Kunz:1993ir}.
Moreover it will be interesting to include the effect of gravitation
\cite{Ibadov:2008hj} to obtain rotating gravitating regular configurations
as well as black hole solutions.
Here a fascinating possibility would be the existence of
a pair of black holes kept apart by the non-Abelian interactions
between the sphaleron and antisphaleron configurations 
without the need for a conical singularity.

{\bf Acknowledgement}:
We gratefully acknowledge discussions with E.~Radu.
R.I.~acknowledges support by the Volkswagen Foundation,
and B.K.~support by the DFG.


\end{document}